\documentclass[showpacs,preprintnumbers,amsmath,amssymb,floatfix,nofootinbib]{revtex4}

\usepackage{amsmath,amsfonts,bbm,euscript,hyperref}
\usepackage{graphicx}
\usepackage{dcolumn}
\usepackage{bm}
\usepackage{epsfig}
\epsfclipon

\newcommand{\nco}{\newcommand}
\nco{\beq}{\begin{equation}} \nco{\eeq}{\end{equation}}
\nco{\beqa}{\begin{eqnarray}} \nco{\eeqa}{\end{eqnarray}}
\def\be{\begin{equation}}
\def\ee{\end{equation}}    
\def\baray{\begin{eqnarray}}
\def\earay{\end{eqnarray}}

\nco{\lra}{\leftrightarrow}

\nco{\sss}{\scriptscriptstyle} \nco{\dphi}{\varphi}
\nco{\lsim}{\mbox{\raisebox{-.6ex}{~$\stackrel{<}{\sim}$~}}}
\nco{\gsim}{\mbox{\raisebox{-.6ex}{~$\stackrel{>}{\sim}$~}}}

\def\IK{\relax{\rm I\kern-.20em K}}
\def\IM{\relax{\rm I\kern-.20em M}}

\def\lsim{\mbox{\raisebox{-.6ex}{~$\stackrel{<}{\sim}$~}}}
\def\gsim{\mbox{\raisebox{-.6ex}{~$\stackrel{>}{\sim}$~}}}
\def\sss{\scriptscriptstyle}

\def\sH{\mathcal{H}}

\def\grad{\vec{\nabla}}

\begin{document}

\title{Nongaussianity from Particle Production During Inflation}

\author{Neil Barnaby}

\affiliation{%
\centerline{Canadian Institute for Theoretical Astrophysics,}
\centerline{University of Toronto, McLennan Physical Laboratories,
60 St.\ George Street, Toronto, Ontario, Canada  M5S 3H8}
e-mail:\ barnaby@cita.utoronto.ca}

\date{January, 2009}

\begin{abstract} 
In a variety of models the motion of the inflaton may trigger the production of some non-inflaton particles during inflation,
for example via parametric resonance or a phase transition.  Such models have attracted interest recently for a variety of reasons,
including the possibility of slowing the motion of the inflaton on a steep potential.  In this review we show that interactions between the produced
particles and the inflaton condensate can lead to a qualitatively new mechanism for generating cosmological fluctuations from inflation.
We illustrate this effect using a simple prototype model $g^2 (\phi-\phi_0)^2\chi^2$ for the interaction between the inflaton, $\phi$, and iso-inflaton, $\chi$.
Such interactions are quite natural in a variety of inflation models from supersymmetry and string theory.  Using both lattice field theory simulations and analytical 
calculations, we study the quantum production of $\chi$ particles and their subsequent rescatterings off the condensate $\phi(t)$, which generates bremsstrahlung 
radiation of light inflaton fluctuations $\delta\phi$.  This mechanism leads to observable features in the primordial power spectrum.  We derive observational constraints
on such features and discuss their implications for popular models of inflation, including brane/axion monodromy.  Inflationary particle production also leads 
to a very novel kind of nongaussian signature which may be observable in future missions.  We argue that this mechanism provides a simple and well-motivated option to generate large nongaussianity,
without fine-tuning the inflationary trajectory or appealing to re-summation of an infinite series of high dimension operators.
\hspace{3mm}
\begin{center}
\emph{This work is dedicated to the memory of Lev Kofman.}
\end{center}
\end{abstract}

\maketitle

\begin{widetext}
\tableofcontents\vspace{5mm}
\end{widetext}

\section{Introduction} 

In recent years the inflationary paradigm has become a cornerstone of modern cosmology.  In the simplest scenario 
the observed cosmological perturbations are seeded by the quantum vacuum fluctuations of the inflaton field \cite{fluct}.
This mechanism predicts a nearly scale invariant spectrum of adiabatic primordial fluctuations, consistent
with recent observational data \cite{WMAP7}.  
In addition to this standard mechanism, there are also several alternatives for generating cosmological
perturbations from inflation; examples include modulated fluctuations 
\cite{modulated1,modulated2} 
and the curvaton mechanism \cite{curvaton}.  
These various scenarios all lead to similar predictions for the  power spectrum.  On the other hand, nongaussian
statistics (such as the bispectrum) provide a powerful tool to observationally discriminate between different
mechanisms for generating the curvature perturbation.  
In this review, which is based on \cite{ir,ppcons,pptheory,ppNG}, we will present a \emph{qualitatively new} mechanism for generating
cosmological perturbations during inflation.  We discuss in detail the predictions of this new scenario for both the spectrum and nongaussianity
of the primordial curvature fluctuations, showing how this new mechanism may be observationally distinguished from previous approaches.

\subsection{Nongaussianity from Inflation}

The possibility to discriminate between various inflationary scenarios has lead to a recent surge of interest in computing and 
measuring nongaussian statistics.  Although single field, slow roll models are known to produce negligible nongaussianity \cite{riotto,maldacena,seerylidsey},
there are now a variety of scenarios available in the literature which may predict an observable signature.  Departures from gaussianity
are often parametrized in the following form
\begin{equation}
\label{local_f_NL}
  \zeta(x) = \zeta_g(x) + \frac{3}{5} f_{NL} \left[ \zeta_g^2(x)  - \langle \zeta_g^2(x)\rangle \right]
\end{equation}
where $\zeta(x)$ is the primordial curvature perturbation, $\zeta_g(x)$ is a gaussian random field and $f_{NL}$ characterizes the degree of 
nongaussianity.  The ansatz (\ref{local_f_NL}) is known as the ``local'' form of nongaussianity.  

Although the local ansatz (\ref{local_f_NL}) has received significant attention, it is certainly not the only well-motivated model for a nongaussian
curvature perturbation.  For example, the nongaussian part of $\zeta(x)$ need not be correlated with the gaussian part.  Consider a 
primordial curvature perturbation of the form
\begin{equation}
\label{uncorr}
  \zeta(x) = \zeta_g(x) + F_{NL}\left[\chi_g(x)\right]
\end{equation}
where $F_{NL}$ is some nonlinear (not necessarily quadratic) function and $\chi_g(x)$ is a gaussian field which is uncorrelated with $\zeta_g(x)$.
Both (\ref{local_f_NL}) and (\ref{uncorr}) are local in position space, however, these two types of nongaussianity will have very different observational
implications.  The uncorrelated ansatz (\ref{uncorr}) for the primordial curvature perturbation can arise, for example, in models with preheating into light fields  
\cite{preheatNG,preheatNG2}.  (See also \cite{preheatNG3,preheatNG4} for more discussion of nongaussianity from preheating and \cite{multibrid} for another model
where nongaussianity is generated at the end of inflation.)

A useful quantity to consider is the \emph{bispectrum},
$B(k_1,k_2,k_3)$, which is the 3-point correlation function of the Fourier transform of the primordial curvature perturbation
\begin{equation}
\label{bispectrum_def}
  \langle \zeta_{\bf k_1} \zeta_{\bf k_2} \zeta_{\bf k_3} \rangle  = (2\pi)^3 \delta({\bf k_1} + {\bf k_3} + {\bf k_3}) B(k_i)
\end{equation}
where $k_i \equiv |{\bf k_i}|$.
The delta function appearing in (\ref{bispectrum_def}) reflects translational invariance and ensures that $B(k_i)$ depends on three momenta ${\bf k_i}$
which form a triangle: ${\bf k_1} + {\bf k_2} + {\bf k_3} = 0$.  Rotational invariance implies that $B(k_i)$ is symmetric in its arguments.  

If we assume the ansatz (\ref{local_f_NL}) for the primordial curvature perturbation then $B(k_i)$ has a very particular dependence on momenta; 
it peaks in the squeezed limit where one of the wave-numbers is much smaller than the remaining two (e.g.\ - $k_1 \ll k_2, k_3$).
Such a bispectrum is referred to as having a \emph{squeezed} shape.  However, other shapes of bispectrum are worth considering.
A bispectrum is referred to as ``equilateral'' if it peaks when $k_1 = k_2 = k_3$ and ``flattened'' if it peaks when one of the wave-numbers 
is half the size of the remaining two (e.g.\ - $2k_1 = k_2= k_3$).  

Without assuming any specific form for the primordial curvature perturbation, such as (\ref{local_f_NL}) or (\ref{uncorr}), one may
characterize an arbitrary bispectrum (\ref{bispectrum_def}) by specifying its shape, running and size \cite{small_sound}.   
As discussed above, the shape refers to the configuration of triangle on which $B(k_i)$ is maximal (squeeze, equilateral or flattened).
The running of the bispectrum refers to how the magnitude of $B(k_i)$ depends on the overall size of the triangle.  For example,
in the case of scale invariant fluctuations, the bispectrum must scale as $B(\lambda k_1,\lambda k_2, \lambda k_3) = \lambda^{-6} B(k_1,k_2,k_3)$.
Finally, the overall size of the bispectrum is often quantified by evaluating the magnitude of $B(k_i)$ on some fixed equilateral triangle.  However, 
the skewness of the probability density function (defined later) might provide a better measure of the size of nongaussianity.

Different types of nongaussian signatures are correlated with properties of the underlying inflation model.  Let us first consider
some examples with small running:\footnote{By ``small'' here we refer to any model where the running of the bispectrum is proportional
to slow-variation parameters or arises due to loop effects.  This does \emph{not} necessarily mean that such running cannot lead to interesting 
observational signatures, see \cite{shandera,running1,running2,obs_running}.}
\begin{enumerate}
  \item A large bispectrum of local shape, along with iso-curvature effects, is associated with models where multiple fields are light (or otherwise
  dynamically important)  during inflation.  Examples include  the curvaton mechanism \cite{curvatonNG} or models with
  turning points along the inflationary trajectory \cite{turnNG}. The observational bound on local type nongaussianity, coming from the WMAP7
  \cite{WMAP7} data, is 
  $-10 < f_{NL}^{\mathrm{local}} < 74$ \cite{NGlocal_constraints} at 95\% confidence level.  When combined with Large Scale Structure (LSS) data the bound
  becomes somewhat stronger: $-1 < f_{NL}^{\mathrm{local}} < 65$ \cite{NGlocal_LSS}.
  \item A large local bispectrum \emph{without} any iso-curvature fluctuations can \emph{only} be produced by nonlocal inflation models \cite{NLNG}.  
  For any single-field inflation model described by a local low-energy effective field theory, the results of \cite{consistency} imply that the ratio of
  the 3-point correlation function to the square of the 2-point function must be of order the spectral tilt, in the squeezed limit.  
  Hence, it has been argued that a large squeezed bispectrum must be associated with the presence of multiple light degrees of freedom, and hence 
  iso-curvature effects.  However, in \cite{NLNG} it was shown single field nonlocal inflation models can
  produce a large squeezed bispectrum in the regime where the underlying scale of nonlocality is much larger than the Hubble scale during inflation.  
  Such constructions evade the no-go theorem of \cite{consistency} precisely because they violate the usual assumption of cluster decomposition.  
  Moreover, models of this type are \emph{not} subsumed by the general analysis of \cite{EFT} since nonlocal field theories with infinitely 
  many derivatives 
  cannot be obtained in the regime of low-energy effective field theory.  It is nevertheless sensible to study such constructions since they may
  be derived from ultra-violet (UV) complete frameworks, such as string field theory or $p$-adic string theory.  See \cite{NLmath} for details concerning 
  the underlying consistency of nonlocal field theories and see 
  \cite{NLreview} for a succinct review of nonlocal cosmology. 
  \item A large equilateral bispectrum is typically associated with a small sound speed for the inflaton perturbations \cite{small_sound}, 
  such as in Dirac-Born-Infeld (DBI) inflation models \cite{DBI}.  However, such a signature may also be obtained in multi-field gelaton \cite{geltron}
  or trapped 
  inflation \cite{trapped} models.   The observational bound on equilateral type nongaussianity is 
  $-125 < f_{NL}^{\mathrm{equil}} < 435$ at 95\% confidence  level \cite{NGconstraints}.
  \item A large flattened bispectrum is associated with non-vacuum initial conditions \cite{small_sound,nonBD1,nonBD2,nonBD}.\footnote{To our 
  knowledge there is no explicit computation of the observational bound on flattened nongaussianity.  In \cite{nonBD1} a template (the enfolded  
  model) was proposed.  The analysis of \cite{NGconstraints} is sufficiently general to study this shape, however, they do not explicitly
  place bounds on $f_{NL}^{\mathrm{flat}}$ but instead constrain an alternative shape (the orthogonal model) which is a superposition of flattened and equilateral
  shapes.}
\end{enumerate}

If we relax the assumption that the bispectrum is close to scale invariant then a much richer variety of nongaussian signatures is possible.
For example, in models with sharp steps in the inflaton potential \cite{chen1,chen2} the bispectrum is large only for triangles with a particular
characteristic size.  We will refer to such a signature as a \emph{localized nongaussian feature}.  Localized nongaussianities 
are not well constrained by current observation, but may be observable in future missions.  

Given the significant role that nongaussianity may play in discriminating between different models of the early universe, it is of crucial importance
to explore and classify all possible  consistent signatures for the bispectrum and other nongaussian statistics.  Indeed, in this review we will
describe a new kind of signature -- uncorrelated nongaussian features -- which is predicted in a variety of simple and well 
motivated models of inflation, but which has nevertheless
been overlooked in previous literature.

\subsection{Inflationary Particle Production}

Recently, a new mechanism for generating
cosmological perturbations during inflation was proposed \cite{ir}.  This new mechanism, dubbed \emph{infra-red (IR) cascading}, is qualitatively different
from previous proposals (such as the curvaton or modulated fluctuations) in that it does
not rely on the quantum vacuum fluctuations of some light scalar fields during inflation.  Rather, the scenario involves the production
of massive iso-curvature particles \emph{during} inflation.  
These subsequently rescatter off the slow-roll condensate to generate bremsstrahlung radiation of
light inflaton fluctuations (which induce curvature perturbations and temperature anisotropies in the usual manner).  IR cascading
can also be distinguished from previous mechanisms from the observational perspective: this new mechanism leads to novel 
features in both the spectrum and bispectrum.

In principle, IR cascading may occur in any model where non-inflaton (iso-curvature) particles are produced {during} inflation.  Models
of this type
have attracted considerable interest recently; examples have been studied where particle production occurs via parametric 
resonance \cite{ir,chung,chung2,elgaroy,sasaki,modulated_trapping,brane_brem,trapped,ppcons},
as a result of a phase transition \cite{KL,KP,BBS,preheatNG,adams,step_model,gobump,brane_annihilations}, 
or otherwise \cite{sorbo}.  Recent interest in inflationary particle production has been stimulated by various considerations:
\begin{enumerate}
  \item Particle production arises naturally in a number of microscopically realistic models of inflation, including
           examples from string theory \cite{trapped} 
           and supersymmetric (SUSY) field theory \cite{berrera}.  In particular, inflationary particle production
           is a generic feature of open string inflation models \cite{ppcons}, such as brane/axion monodromy \cite{monodromy1,monodromy2,monodromy3}.
 \item The energetic cost of producing particles during inflation has a dissipative effect on the dynamics of the inflaton.
          Particle production may therefore slow
          the motion of the inflaton, even on a steep potential.  This gives rise to a 
          new inflationary mechanism, called \emph{trapped inflation} \cite{beauty,trapped,terminal}, 
          which may circumvent
          some of the fine tuning problems associated with standard slow-roll inflation.  See \cite{trapped} for an explicit string theory realization of trapped 
          inflation and \cite{terminal} for a generalization to higher dimensional moduli spaces and enhanced symmetry loci.
          The idea of using
          dissipative dynamics to slow the motion of the inflaton is qualitatively similar to warm inflation \cite{warm} and also to the variant of natural
          inflation \cite{natural} proposed recently by Anber \& Sorbo \cite{sorbo}.
  \item Observable features in the primordial power spectrum, generated by particle 
           production and IR cascading, offer a novel example of the non-decoupling of high scale physics in the Cosmic Microwave Background (CMB)
            \cite{chung,jim}.  In the most interesting examples, the produced 
            particles are
            extremely massive for (almost) the entire history of the universe, however, their effect cannot be integrated out due to the non-adiabatic
            time dependence of the iso-inflaton 
            mode functions during particle production.  In \cite{chung} particle production during large field inflation models was
            proposed as a possible probe of Planck-scale physics.
\end{enumerate}

In this article we study in detail the impact of particle production and IR cascading on the observable primordial curvature perturbations.
In order to illustrate the basic physics we focus on a very simple and general prototype model where the inflaton, $\phi$,
and iso-inflaton, $\chi$, fields interact via the coupling
\begin{equation}
\label{int}
  \mathcal{L}_{\mathrm{int}} = -\frac{g^2}{2} (\phi-\phi_0)^2 \chi^2
\end{equation}
We expect, however, that our results will generalize in a straightforward way to more complicated models,
such as higher spin iso-inflaton fields or gauged interactions, wherein the physics of particle production and rescattering
is essentially the same.  Our result may also have implications for inflationary phase transitions, because spinodal decomposition
can be interpreted as a kind of particle production and similar bi-linear interactions will induce rescattering effects.  

Scalar field interactions of the type (\ref{int}) have also been studied recently in connection with non-equilibrium Quantum Field
Theory (QFT) \cite{nonequilibrium},
in particular with applications to the theory of preheating after inflation \cite{KLS,KLS97,FK,MT1,MT2}
and also moduli trapping \cite{beauty,terminal} at enhanced symmetry points.  Although our focus
is on particle production \emph{during} inflation (as opposed to during preheating, after inflation) some of our results nevertheless have implications
for preheating, moduli trapping and also non-equilibrium QFT more generally.  For example, in \cite{ir} analytical and numerical studies of rescattering
and IR cascading during inflation made it possible to observe, for the first time, the dynamical approach to the turbulent scaling regime that was 
discovered in \cite{B1,B2}.

Particle production during inflation in the model (\ref{int}) leads to observable features in the primordial power spectrum, $P(k)$.
A number of recent studies have claimed evidence for localized features in $P(k)$ that are incompatible with the 
simplest power-law model $P(k) \sim k^{n_s-1}$ \cite{chung2,gobump,features,features2,morefeatures1,morefeatures2,morefeatures3,features3,yokoyama1,yokoyama3,yokoyama2,hoi1,hoi,contaldi,yokoyama4}.
Although these observed features may simply be statistical anomalies (see, for example, \cite{nofeatures}) there remains the tantalizing
possibility that they represent some new physics beyond the simplest slow roll model.  Upcoming polarization data may play an important
role in distinguishing these possibilities \cite{gobump}.
In the meantime, it is interesting to determine the extent to which such features may be explained by a simple and
well motivated model such as (\ref{int}).  Moreover, because (\ref{int}) is a complete microscopic model (as opposed
to a phenomenological modification of the power spectrum) it is possible to predict a host of correlated
observables, such as features in the scalar bispectrum and tensor power spectrum.
Hence, it should be possible to robustly rule out (or confirm!) the possibility that some massive 
iso-curvature particles were produced during the observable portion of inflation.  

If detected, features from particle production and IR cascading 
will provide a rare and powerful new window into the microphysics driving inflation.  This scenario opens up the possibility of learning 
some details about how
the inflaton couples to other particles in nature, as opposed to simply reconstructing the inflaton potential along the slow roll trajectory.  Moreover, due
to the non-decoupling discussed above, features from particle production and IR cascading may probe new (beyond the standard model) physics 
at extraordinarily high energy scales.

The outline of this paper is as follows.  In section \ref{sec_overview} we provide a brief, qualitative overview of the dynamics of particle production
and IR cascading in the model (\ref{int}).  In section \ref{sec_numerical} we study in detail this same dynamics using fully nonlinear lattice field
theory simulations.  In section \ref{sec_analytical} we provide an analytical theory of particle production and IR cascading in an expanding universe.
A complimentary analytical analysis, using second order cosmological perturbation theory, is provided in section \ref{sec_metric}.  In section
\ref{sec_cons} we consider the observational constraints on inflationary particle production using a variety of data sets.  In section
\ref{sec_micro} we provide several explicit microscopic realizations of our scenario and study the implications of our observational constraints
on models of string theory inflation, in particular brane monodromy.  In section \ref{sec_ng} we quantify and characterize the nongaussianity
generated by particle production and IR cascading.  Finally, in section \ref{sec_conc}, we conclude and discuss possible future directions.

\section{Overview and Summary of the Mechanism}
\label{sec_overview}

In this section we provide a brief overview of the dynamics of particle production and IR cascading in the model (\ref{int}) and also
summarize the resulting observational signatures.  In the remainder of this article we will flesh out the details of this mechanism 
with analytical and numerical calculations.  

We consider the following model
\begin{equation}
  S = \int d^4 x \sqrt{-g} \left[ \frac{M_p^2}{2} R - \frac{1}{2}(\partial \phi)^2 - V(\phi) - \frac{1}{2}(\partial \chi)^2 - \frac{g^2}{2}(\phi-\phi_0)^2\chi^2 \right]  \label{L}
\end{equation}
where $R$ is the Ricci curvature constructed from the metric $g_{\mu\nu}$, $\phi$ is the inflaton field and $\chi$ is the iso-inflaton.  
As usual, we assume a flat FRW space-time with scale factor $a(t)$
\begin{equation}
  ds^2 \equiv g_{\mu\nu} dx^{\mu} dx^{\nu} = -dt^2 + a^2(t) d{\bf x}^2
\end{equation}
and employ the reduced Planck mass $M_p \cong 2.43\times 10^{18} \mathrm{GeV}$.  We leave the potential $V(\phi)$ driving inflation unspecified 
except to assume that it is sufficiently flat in 
the usual sense; that is $\epsilon \ll 1$, $|\eta| \ll 1$ where
\begin{equation}
\label{slow_roll}
  \epsilon \equiv \frac{M_p^2}{2} \left(\frac{V'}{V}\right)^2, \hspace{5mm}
  \eta \equiv M_p^2 \frac{V''}{V}
\end{equation}
are the usual slow roll parameters.  

Note that one might wish to supplement (\ref{L}) by its supersymmetric completion in order  to protect 
the flatness of the inflaton potential from large radiative corrections coming from loops of the $\chi$ field.  We expect that our results will 
carry over in a straightforward way to SUSY models and also to more complicated scenarios such as higher spin iso-inflaton fields and 
(possibly) inflationary phase transitions.

The coupling $\frac{g^2}{2}(\phi-\phi_0)^2\chi^2$ in (\ref{L}) is introduced to ensure that
the iso-inflaton field can become instantaneously massless at some point $\phi=\phi_0$  along the inflaton trajectory
(which we assume occurs during the observable range of $e$-foldings of inflation).
At this moment $\chi$ particles will be produced by quantum effects.  

Let us first consider the homogeneous dynamics of the inflaton field, $\phi(t)$.  Near the point $\phi = \phi_0$ we can generically expand
\begin{equation}
  \phi(t) \cong \phi_0 + v t
\end{equation}
where $v \equiv \dot{\phi}(0)$ and we have arbitrarily set the origin of time so that $t=0$ corresponds to the
moment when $\phi = \phi_0$.  (We are, of course, assuming that $\dot{\phi}(0) \not= 0$.)  The interaction (\ref{int})
induces an effective (time varying) mass for the $\chi$ 
particles of the form
\begin{equation}
\label{m_approx}
  m_\chi^2 = g^2 (\phi - \phi_0)^2 \cong k_\star^4 t^2
\end{equation}
where we have defined the characteristic scale
\begin{equation}
\label{kstar}
  k_\star = \sqrt{g |v|}
\end{equation}
It is straightforward to verify that the simple expression (\ref{m_approx}) will be a good approximation 
for $(H |t|)^{-1} \lsim \mathcal{O}(\epsilon,\eta)$ which, in most models, will be true for the entire observable $60$ $e$-foldings of 
inflation.

Note that, without needing to specify the background inflationary potential $V(\phi)$, we can write the ratio $k_\star / H$ as
\begin{equation}
\label{ratio}
  \frac{k_\star}{H} = \sqrt{\frac{g}{2\pi \mathcal{P}_\zeta^{1/2}}}
\end{equation}
where $\mathcal{P}_{\zeta}^{1/2} = 5\times 10^{-5}$ is the usual amplitude of the vacuum fluctuations from inflation.  
In this work we assume $k_\star > H$ which is easily satisfied for reasonable values of the coupling $g^2  > 10^{-7}$.
In particular, for $g^2 \sim 0.1$ we have $k_\star / H \sim 30$.

The scenario we have in mind is the following.  Inflation starts at some field value $\phi > \phi_0$ and the inflaton rolls toward
the point $\phi = \phi_0$.  Initially, the iso-inflaton field is extremely massive $m_\chi \gg H$
and hence it stays pinned in the vacuum, $\chi = 0$, and does not contribute to super-horizon curvature fluctuations.  
Eventually, at $t=0$, the inflaton rolls through the point $\phi=\phi_0$ where $m_\chi = 0$
and $\chi$ particles are produced.  To describe this burst of particle production one must solve for
the following equation for the $\chi$-particle mode functions in an expanding universe
\begin{equation}
\label{intro_chi}
  \ddot{\chi}_k + 3 H \dot{\chi}_k + \left[ \frac{k^2}{a^2} + k_\star^4 t^2  \right] \chi_k = 0
\end{equation}
Equations of this type are well-studied in the context of preheating after inflation \cite{KLS97} and moduli trapping \cite{beauty}.
The initial conditions for (\ref{intro_chi}) should be chosen to ensure that the q-number field $\chi$ is in the adiabatic vacuum in
the asymptotic past (see sections \ref{sec_numerical} and \ref{sec_analytical} for more details).
In the regime $k_\star > H$ particle production is fast compared to the expansion time and one can solve (\ref{intro_chi}) very 
accurately for the occupation number of the created $\chi$ particles
\begin{equation}
\label{n_k}
  n_k = e^{-\pi k^2 / k_\star^2} 
\end{equation}
Very quickly after the moment $t=0$, within a time $\Delta t \sim k_\star^{-1} \ll H^{-1}$,
these produced $\chi$ particles become non-relativistic ($m_\chi > H$) and their number density starts to dilute as $a^{-3}$.

\begin{figure}[htbp]
\bigskip \centerline{\epsfxsize=0.25\textwidth\epsfbox{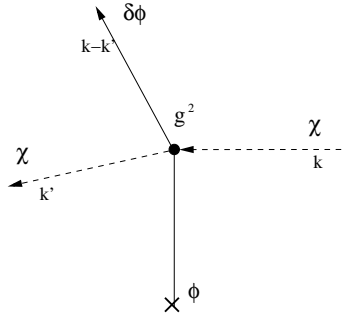}}
\caption{Rescattering diagram.
}
\label{Fig:diag}
\end{figure}

Following the initial burst of particle production there are two distinct physical effects which take place.  First, the energetic cost of producing the gas of
massive out-of-equilibrium $\chi$ particles drains energy from the inflaton condensate, forcing $\dot{\phi}$ to drop abruptly.  This velocity dip
is the result of the backreaction of the produced $\chi$ fluctuations on homogeneous condensate $\phi(t)$.  The second physical effect is
that the produced massive $\chi$ particles rescatter off the condensate via the diagram Fig.~\ref{Fig:diag} and emit  bremsstrahlung
radiation of light inflaton fluctuations (particles).  

\begin{figure}[htbp]
\bigskip \centerline{\epsfxsize=0.4\textwidth\epsfbox{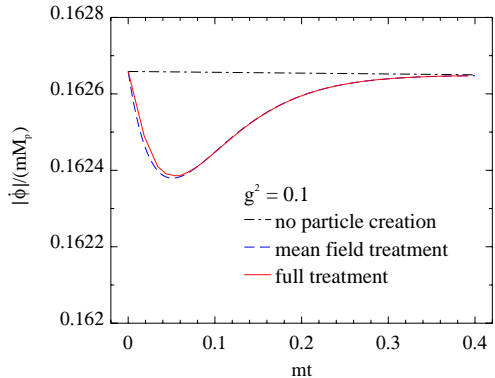}}
\caption{$|\dot\phi|/(M_pm)$ plotted against $m t$ for $g^2=0.1$ (where $m=V_{,\phi\phi}$ is the effective inflaton mass). Time $t=0$ corresponds to the 
moment when $\phi=\phi_0$ and $\chi$-particles are produced copiously. 
The solid red line is the lattice field theory result taking into account the full dynamics of rescattering and IR cascading 
while the dashed blue line is the result of a mean field theory treatment which ignores
rescattering \cite{sasaki}. The dot-dashed black line is the inflationary trajectory in the absence of particle creation.
}
\label{Fig:dotphi}
\end{figure}

Backreaction and rescattering leave distinct imprints in the observable cosmological perturbations.  Let us first discuss the impact
of backreaction.  In Fig.~\ref{Fig:dotphi} we plot the velocity dip resulting from the backreaction of the produced $\chi$
particle on the homogeneous inflaton condensate $\phi(t)$.  From this figure we see that the quantity $\ddot{\phi}/(H\dot{\phi})$ becomes large in the dip.  
This violation of slow roll is a transient effect; at late times the produced $\chi$ particles become extremely massive and their number density dilutes
as $a^{-3}$.

One can understand the temporary slowing-down of the inflaton from an analytical perspective.  Backreaction is taken into account using the mean-field equation
\begin{equation}
\label{mean}
  \ddot{\phi} + 3 H \dot{\phi} + V_{,\phi} + g^2 (\phi - \phi_0) \langle \chi^2 \rangle  = 0
\end{equation}
where the vacuum average is computed following \cite{KLS97,beauty}
\begin{equation}
  \langle \chi^2 \rangle \cong \frac{n_\chi a^{-3}}{ g |\phi - \phi_0|}
\end{equation}
In equation (\ref{mean}) we have implicitly assumed that the usual Coleman-Weinberg corrections to the inflaton potential
have \emph{already} been absorbed into $V(\phi)$, hence the vacuum average $\langle \chi^2\rangle$ should include \emph{only}
the effects of non-adiabatic particle production.
(Here $n_\chi = \int \frac{d^3 k}{(2\pi)^3} n_k \sim k_\star^3$ is the total number density of produced $\chi$
particles and the factor $a^{-3}$ reflects the usual volume dilution of non-relativistic matter.)
In Fig.~\ref{Fig:dotphi} we have plotted the solution of (\ref{mean}) along with the exact result obtained from lattice
field theory simulations, illustrating the accuracy of this simple treatment.

Using the mean field approach, one finds that the transient violation of
slow roll leads to a ``ringing pattern'' (damped oscillations) in the power spectrum $P_\phi(k) = \frac{k^3}{2\pi^2} |\delta\phi_k|^2$ 
of inflaton fluctuations \cite{sasaki}.  This ringing pattern
is localized around wave-numbers which left the horizon at the moment when particle production
occurred.  The effect is very much analogous to Fresnel diffraction at a sharp edge.

The second physical effect, rescattering, was considered for the first time in the context of inflationary particle production in \cite{ir}.
Fig.~\ref{Fig:diag} illustrates the dominant process: bremsstrahlung emission of long-wavelength $\delta\phi$ fluctuations from rescattering
of the produced $\chi$ particles off the condensate.  The time scale for such processes is set by the microscopic scale, $k_\star^{-1}$, and is thus very short compared
to the expansion time, $H^{-1}$.  Moreover, the production of inflaton fluctuations $\delta\phi$ deep in the infra-red (IR) is extremely energetically
inexpensive, since the inflaton is very nearly massless.  The combination of the short time scale for rescattering and the energetic cheapness
of radiating  IR $\delta\phi$ leads to a rapid build-up of power in long wavelength inflaton modes: IR cascading.  This effect generates a bump-like
feature in the power spectrum of inflaton fluctuations, very different from the ringing pattern associated with backreaction.  The bump-like
feature from rescattering dominates over the ringing pattern from backreaction for all values of parameters.

In \cite{ir} the model (\ref{L}) was studied using lattice field theory simulations, without neglecting any physical processes (that is
to say that full nonlinear structure of the theory, including backreaction and rescattering effects, was accounted for consistently).
However, this same dynamics can be understood analytically by solving the equation for the inflaton fluctuations
$\delta\phi$ in the approximation that all interactions are neglected, except for the diagram Fig.~\ref{Fig:diag}.  The appropriate equation
is
\begin{equation}
\label{inf_eqn}
  \delta\ddot{\phi} + 3 H \delta\dot{\phi} - \frac{\grad^2}{a^2}\delta\phi + V_{,\phi\phi}\delta \phi \cong -g^2\left[\phi(t)-\phi_0\right]\chi^2
\end{equation}
See \cite{pptheory} for a detailed analytical theory.  The solution of (\ref{inf_eqn}) may be split into two parts: the solution of the homogeneous equation and the particular solution
which is due to the source term.  Schematically we have
\begin{equation}
\label{hom+par}
  \delta\phi(t,{\bf x}) = \underbrace{\delta\phi_{\mathrm{vac}}(t,{\bf x})}_{\mathrm{homogeneous}} + \underbrace{\delta\phi_{\mathrm{resc}}(t,{\bf x})}_{\mathrm{particular}}
\end{equation}
The former contribution is the homogeneous solution which behaves as $\delta\phi_{\mathrm{vac}} \sim H / (2\pi)$ on large scales and, physically, corresponds
to the usual scale invariant vacuum fluctuations from inflation.  The particular solution, $\delta\phi_{\mathrm{resc}}$, corresponds physically to inflaton
fluctuations which are generated by rescattering.  The abrupt growth of $\chi$ inhomogeneities at $t=0$ sources the particular solution $\delta\phi_{\mathrm{resc}}$,
leading to the production of inflaton fluctuations which subsequently cross the horizon and become frozen. 

As mentioned earlier, rescattering generates a bump-like contribution to the primordial power spectrum of the curvature perturbations.
To good approximation this may be described by a simple semi-analytic fitting function
\begin{equation}
  P(k) = A_s \left(\frac{k}{k_0}\right)^{n_s-1} + A_{\mathrm{IR}} \left(\frac{\pi e}{3}\right)^{3/2}\left(\frac{k}{k_{\mathrm{IR}}}\right)^3 e^{-\frac{\pi}{2} \left(\frac{k}{k_{\mathrm{IR}}}\right)^2}
\label{P_fit}
\end{equation}
where the first term corresponds to the usual vacuum fluctuations from inflation (with amplitude $A_s$ and spectral index $n_s$) 
while the second term corresponds to the bump-like feature from particle production and IR cascading.  The amplitude of this feature ($A_{\mathrm{IR}}$) depends on $g^2$, while the location 
($k_{\mathrm{IR}}$) depends on $\phi_0$.

\begin{figure}[tbp]
\begin{center}
\includegraphics[width=3in]{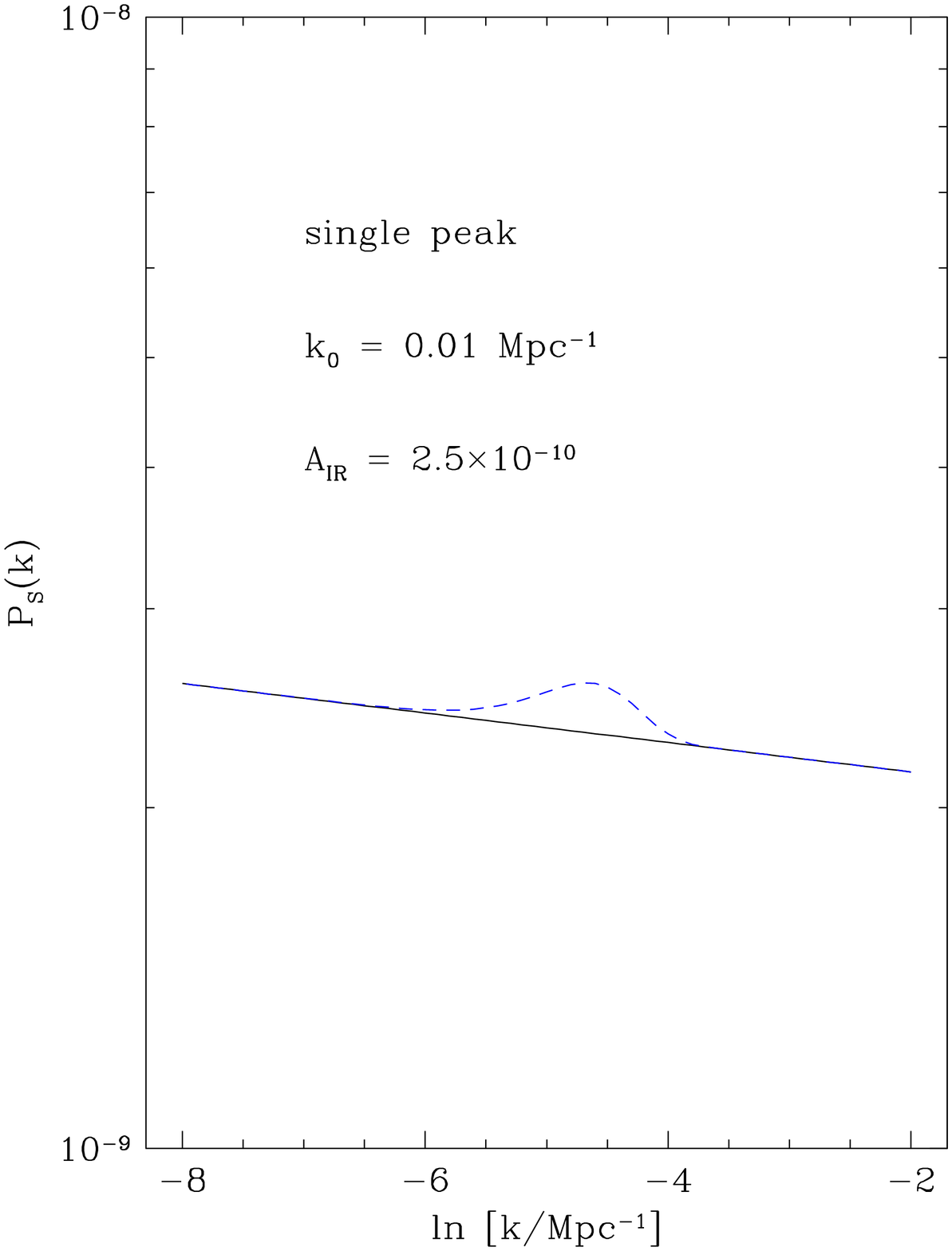}
\includegraphics[width=3in]{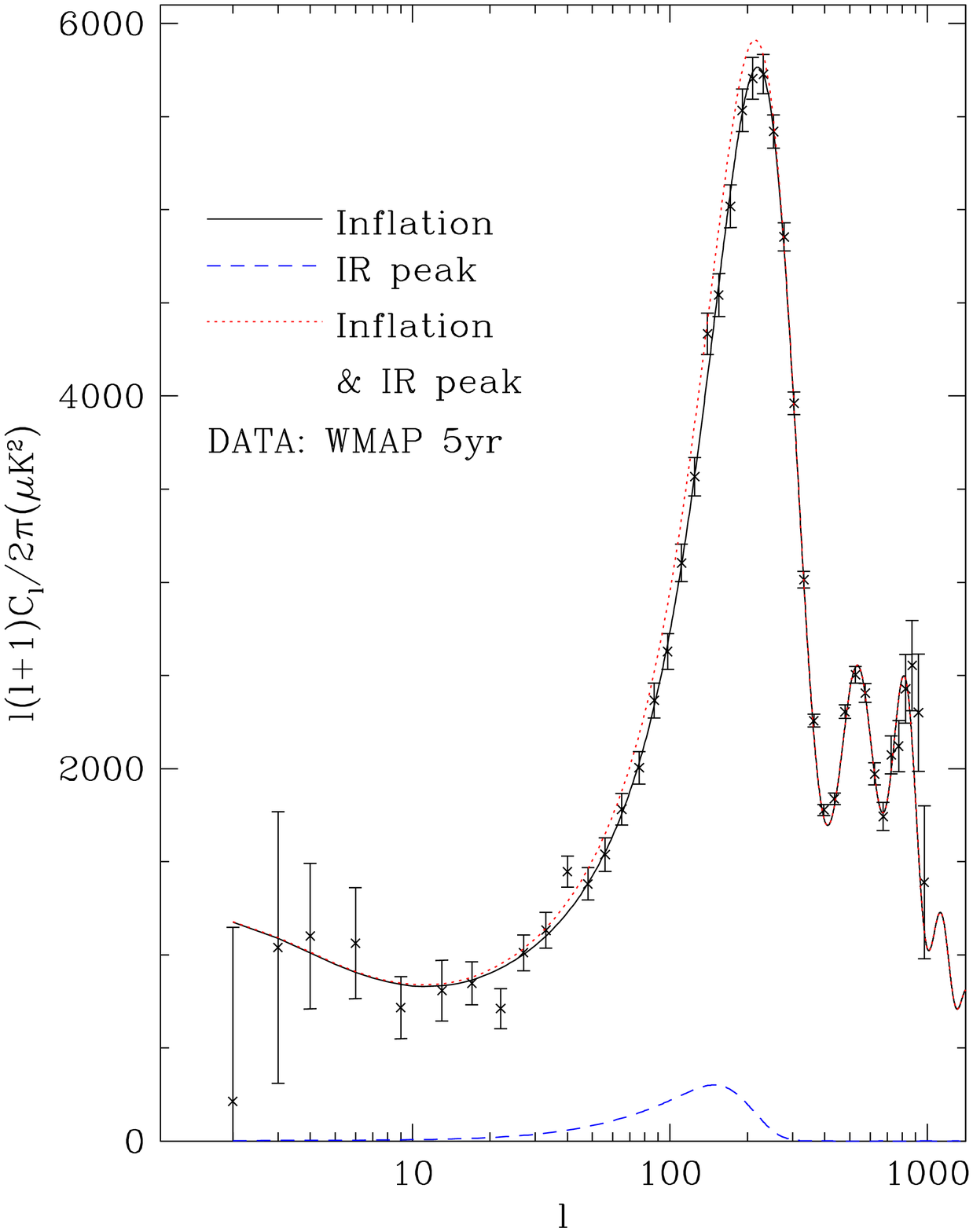}
\caption{The top panel shows a sample bump in the power spectrum with amplitude $A_{\mathrm{IR}} = 2.5\times 10^{-10}$ which corresponds 
to a coupling $g^2 \sim 0.01$.
The feature is located at $k_{\mathrm{IR}} = 0.01 \, \mathrm{Mpc}^{-1}$.  This example represents a distortion of $\mathcal{O}(10\%)$ 
as compared to the usual vacuum fluctuations 
and is consistent with the data at $2\sigma$.  The bottom panel shows the CMB angular TT power spectrum for this example, 
illustrating that the distortion shows up mostly in the first peak.}\label{Fig:sample}
\end{center}
\end{figure}

In \cite{ppcons} the simple fitting function (\ref{P_fit}) was used to place observational constraints on inflationary particle production using a
variety of cosmological data sets.  Current data are consistent with rather large spectral distortions of the type (\ref{P_fit}).
Features as large as $\mathcal{O}(10\%)$ of the usual scale-invariant fluctuations from inflation are allowed, in the case that $k_{\mathrm{IR}}$ falls within the range of scales
relevant for CMB experiments.  Such a feature corresponds to a realistic coupling $g^2 \sim 0.01$.  Even larger values of $g^2$ are allowed if the feature is localized on smaller scales.
In Fig.~\ref{Fig:sample} we have illustrated the primordial power spectrum in the model (\ref{L})
for a representative choice of parameters.  We also plot the CMB angular Temperature-Temperature (TT) power spectrum for the same parameters.

In \cite{forecast} Large Scale Structure forecast constraints were considered for the model (\ref{P_fit}).  It was shown that, for $k_{\mathrm{IR}} \lsim 0.1\, \mathrm{Mpc}^{-1}$,
the constraint on $A_{\mathrm{IR}}/A_s$ will be strengthened to the $0.5\%$ level by Planck or $0.1\%$ including also data from a Square Kilometer Array (SKA).  With a Cosmic Inflation
Probe (CIP) similar constraints could be achieved for $k_{\mathrm{IR}}$ as large as $1\, \mathrm{Mpc}^{-1}$.

The prototype model (\ref{L}) may be realized microscopically in a variety of different particle physics frameworks.
In particular, particle production is a
rather generic feature of open string inflation models \cite{ppcons} where the inflaton, $\phi$, has a geometrical interpretation as the position
of some mobile D-brane.  In this context the iso-inflaton, $\chi$, corresponds to a low-lying open string excitation which is stretched between the 
mobile inflationary brane and any other (spectator) branes which inhabit the compactification volume.  If the inflationary and spectator branes
become coincident during inflation, then the symmetry of the system is enhanced \cite{beauty} and some low-lying stretched string states will become 
instantaneously massless, mimicking the interaction (\ref{int}) (see also \cite{trapped}).  An explicit realization of this scenario is provided by brane/axion
monodromy models \cite{monodromy1,monodromy2,monodromy3}.  Our observational constraints on inflationary particle production
may be used to place bounds on parameters of the underlying string model \cite{ppcons}.

The bump-like feature in $P(k)$, illustrated in Fig.~\ref{Fig:sample}, must be associated with a nongaussian feature in the 
bispectrum \cite{ir,pptheory}.  Indeed, it is evident already from inspection of equation (\ref{inf_eqn}) that the inflaton fluctuations
generated by rescattering are significantly nongaussian; the particular solution of (\ref{inf_eqn}) is bi-linear in the gaussian field $\chi$.
The nongaussian signature from IR cascading is rather novel.  The nongaussian part of $\zeta$ is uncorrelated with the gaussian part.  Moreover,
the bispectrum $B(k_i)$ is very far from scale invariant; it peaks strongly for triangles with a characteristic size $\sim k_{\mathrm{IR}}$, 
corresponding to
the location of the bump in the power spectrum (\ref{P_fit}).  The shape of the bispectrum therefore depends sensitively on the size of the triangle and is
not well-described by any of the templates that have been proposed in the literature to date.  

The magnitude of this new kind of nongaussianity may be quite large.  To quantify the effect it is useful
to introduce the probability density function (PDF), $P(\zeta)$, which is the probability that the curvature perturbation has a fluctuation of size
$\zeta$.  If we define the central moments of the PDF as
\begin{equation}
  \langle \zeta^n \rangle = \int \zeta^n P(\zeta) d\zeta
\end{equation}
then a useful measure of nongaussianity is the dimensionless skewness of the PDF, defined by
\begin{equation}
  \hat{\kappa}_3\equiv \frac{\langle \zeta^3 \rangle_c}{\langle \zeta^2 \rangle^{3/2}}
\end{equation}
where the subscript $c$ indicates that only the connected part of the correlator should be included.  The skewness $\hat{S}_3$ encodes information
about the bispectrum $B(k_i)$ integrated over all size and shape configurations and thus provides a meaningful single number to compare
the nongaussianity of inflation models which may have very different shapes or running \cite{shandera}.  

If we choose $g^2 \sim 0.01$ (which is compatible with observation for all values of $\phi_0$) then the model (\ref{L}) produces the same
value of $\hat{S}_3$ as a local model (\ref{local_f_NL}) with $f_{NL}\cong- 53$.  This large value suggests that nongaussianity from particle production
during inflation may be observable in future missions.

Depending on model parameters, the nongaussian features predicted by the model (\ref{L}) may lead to a rich variety of observable consequences
for the CMB or Large Scale Structure (LSS).  The phenomenology of this model is quite different from other constructions that have been proposed
to obtain large nongaussianity from inflation.  However, the underlying microscopic description (\ref{L}) is extremely simple and, indeed, rather generic
from the low-energy perspective.  Explicit realizations of the interaction (\ref{int}) have been obtained from string theory and SUSY.  Moreover, in order to obtain 
an observable signature it was not necessary to fine-tune the inflationary trajectory or appeal
to re-summation of an infinite series of high-dimension operators.  

\section{Numerical Study of Rescattering and IR Cascading}
\label{sec_numerical}

\subsection{HLattice Simulations}

In this section we study numerically the creation of $\delta\phi$ fluctuations by rescattering of the produced $\chi$ particles off the condensate $\phi(t)$
in the model (\ref{L}).  To this end, we have written a new lattice field theory code, HLattice \cite{HLattice}, for simulating the interactions of scalar
fields in a cosmological setting.  HLattice can be used to simulate the  dynamics of any number of interacting scalar fields with arbitrary scalar potential and metric on field space \cite{modular_preheating}.  
We solve the Klein-Gordon equations for the scalar field dynamics in an expanding FRW space-time and also solve the Friedmann equation self-consistently for the scale factor, $a(t)$.
Since the production of long wavelength $\delta\phi$ modes is so energetically inexpensive, a major requirement for successfully capturing this
effect is respecting energy conservation to very high accuracy.  HLattice conserves energy with an accuracy of order $\sim 10^{-8}$, as compared
to $10^{-3}-10^{-5}$, which has obtained using previous codes such as DEFROST \cite{defrost} or LATTICEASY \cite{latticeasy}.  A minimum accuracy 
of order $10^{-4}$ is required for the problem at hand.  

The box size of our $512^3$ simulations corresponds to
a co-moving scale which is initially $\frac{20}{2\pi} \sim 3$ times the horizon size $H^{-1}$, while $k_\star \cong 60 \sqrt{g}\, H$.  We run our
simulations for roughly 3 $e$-foldings from the initial moment $t=0$ when the $\chi$ particles are produced, although a single $e$-folding
would have been sufficient to capture the effect.  For the sake of illustration, we have chosen the standard chaotic inflation potential $V = m^2\phi^2/2$
with $m =10^{-6}\sqrt{8\pi} M_p$ for our numerical analysis.  However, our results to not depend sensitively on the choice of background inflation model.\footnote{The 
model independence of our result arises simply because all the dynamics of rescattering and IR cascading occurs within a single $e$-folding from the moment 
when $\phi=\phi_0$.  Over such a short time it will always be a good approximation to expand $\phi(t) \cong \phi_0 + v t$.  Hence the dependence on the background
dynamics arises only through $v = \dot{\phi}(0)$ which is determined by the Hubble scale and the observed amplitude of curvature perturbations.
This claim of model independence is born out by explicit analytical calculations in the next section.}
We have considered both $\phi_0 = 2 \sqrt{8\pi} M_p$ and $\phi_0 = 3.2 \sqrt{8\pi} M_p$ and also three different values of the coupling constant: $g^2 = 0.01, 0.1, 1$.  
As expected, the coupling $g^2$ determines the magnitude of the effect while $\phi_0$ simply shifts the location of the power spectrum feature.  For for this particular inflationary potential,
the choice $\phi_0 = 3.2 \sqrt{8\pi} M_p$ corresponds to putting the feature on scale slightly smaller than todays horizon.  On the other hand, $\phi_0 = 2 \sqrt{8\pi} M_p$ corresponds to 
placing the feature on scales much smaller than those probed by the CMB (we considered this case in order to be able to directly contrast our results with \cite{sasaki}).

In order to capture the quantum production of $\chi$ particles using classical lattice simulations, we start our numerical evolution 
very shortly \emph{after} particle production has occurred, when the $\chi_k$ modes are nearly adiabatic, but before any significant inflaton fluctuations
have been produced.  In practice, this corresponds to initializing the simulation at a time $t_{\mathrm{initial}} = \mathcal{O}(k_\star^{-1})$.  
The initial conditions for the modes $\chi_k(t)$ are given by the usual Bogoliubov computation \cite{KLS97,beauty}.  These are 
chosen to reproduce the occupation number $n_k = e^{-\pi k^2 / k_\star^2}$, while ensuring that the source term for the $\delta\phi$ fluctuations is 
turned on smoothly at the initial time.  As long as the initial conditions are chosen appropriately, our results are not sensitive to the choice of $t_{\mathrm{initial}}$.

At the initial time, the occupation numbers in the inflaton and iso-inflaton fluctuations are small.  However, very quickly the massive $\chi$ particles are diluted 
away by the expansion of the universe and the occupation number of the produced IR $\delta\phi$ fluctuations grows large compared to unity.  Thus, classical lattice
field theory simulations are sufficient to capture the late-time dynamics.  
 (In the next section we will provide a quantum mechanical treatment of the dynamics of particle production and IR cascading during inflation, which
will serve as an a posteriori justification for our classical lattice calculation.)

Our approach is very similar to the methodology that has been employed successfully in studies of preheating after inflation for many years \cite{defrost,latticeasy}.
In that case the initial fluctuations of the fields are chosen to reproduce the exact behaviour of the quantum correlation functions.  The occupation numbers of the fields
are small at the initial time.  However, these grow rapidly as a result of the the preheating instability and classical simulations are sufficient to capture the late-time dynamics.

\subsection{Numerical Results}

\begin{figure}[htbp]
\bigskip \centerline{\epsfxsize=0.5\textwidth\epsfbox{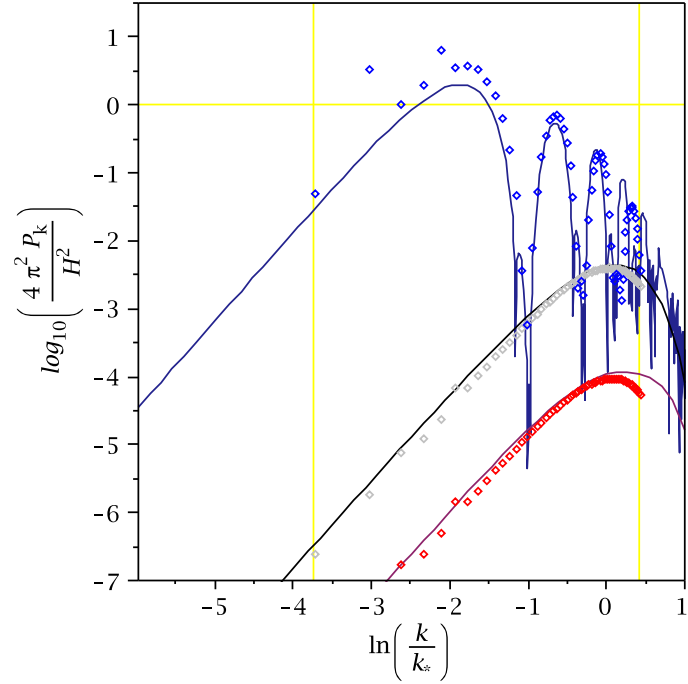}}
\caption{The power spectrum of inflaton modes induced by rescattering (normalized to the usual vacuum fluctuations) as a function of $\ln(k/k_\star)$, plotted for three representative time steps in the evolution, showing the cascading of power into the IR.
For each time step we plot the analytical result (the solid line) and the data points obtained using lattice field theory simulations (diamonds).  The time steps correspond to the following values of the scale factor: $a = 1.03, 1.04, 2.20$ (where 
$a = 1$ at the moment when $\phi = \phi_0$).  By this time the amplitude of fluctuations is saturated due to the expansion of the universe.
The vertical lines show the range of scales from our lattice simulation.
}
\label{Fig:pwr}
\end{figure}

\begin{figure}[htbp]
\bigskip \centerline{\epsfxsize=0.5\textwidth\epsfbox{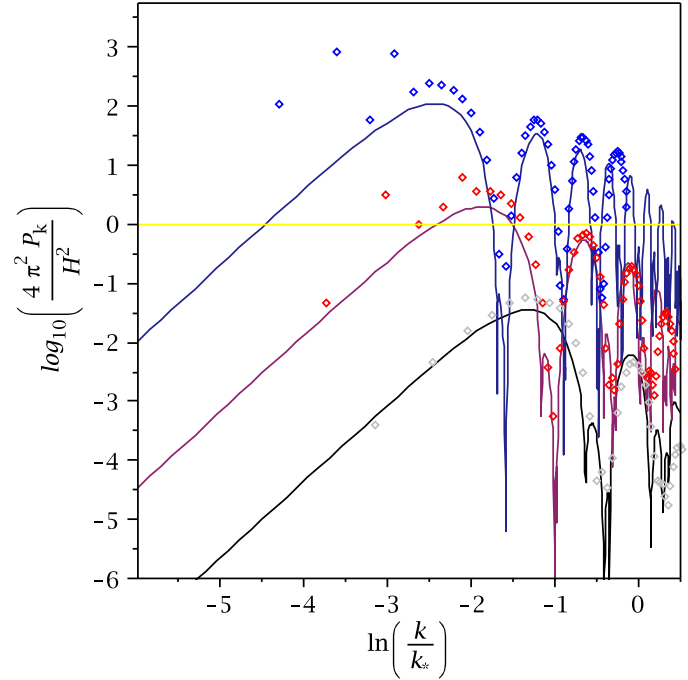}}
\caption{The dependence of the power spectrum $P_\phi$ on the coupling $g^2$.  The three curves correspond to $P_\phi$ for $g^2=0.01,0.1,1$, evaluated at a fixed value of the scale factor, $a = 2.20$. 
We see that even for small values of $g^2$ the inflaton modes induced by rescattering constitute a significant fraction of the usual vacuum fluctuations after only a single $e$-folding.
}
\label{Fig:g}
\end{figure}

\begin{figure}[htbp]
\bigskip \centerline{\epsfxsize=0.5\textwidth\epsfbox{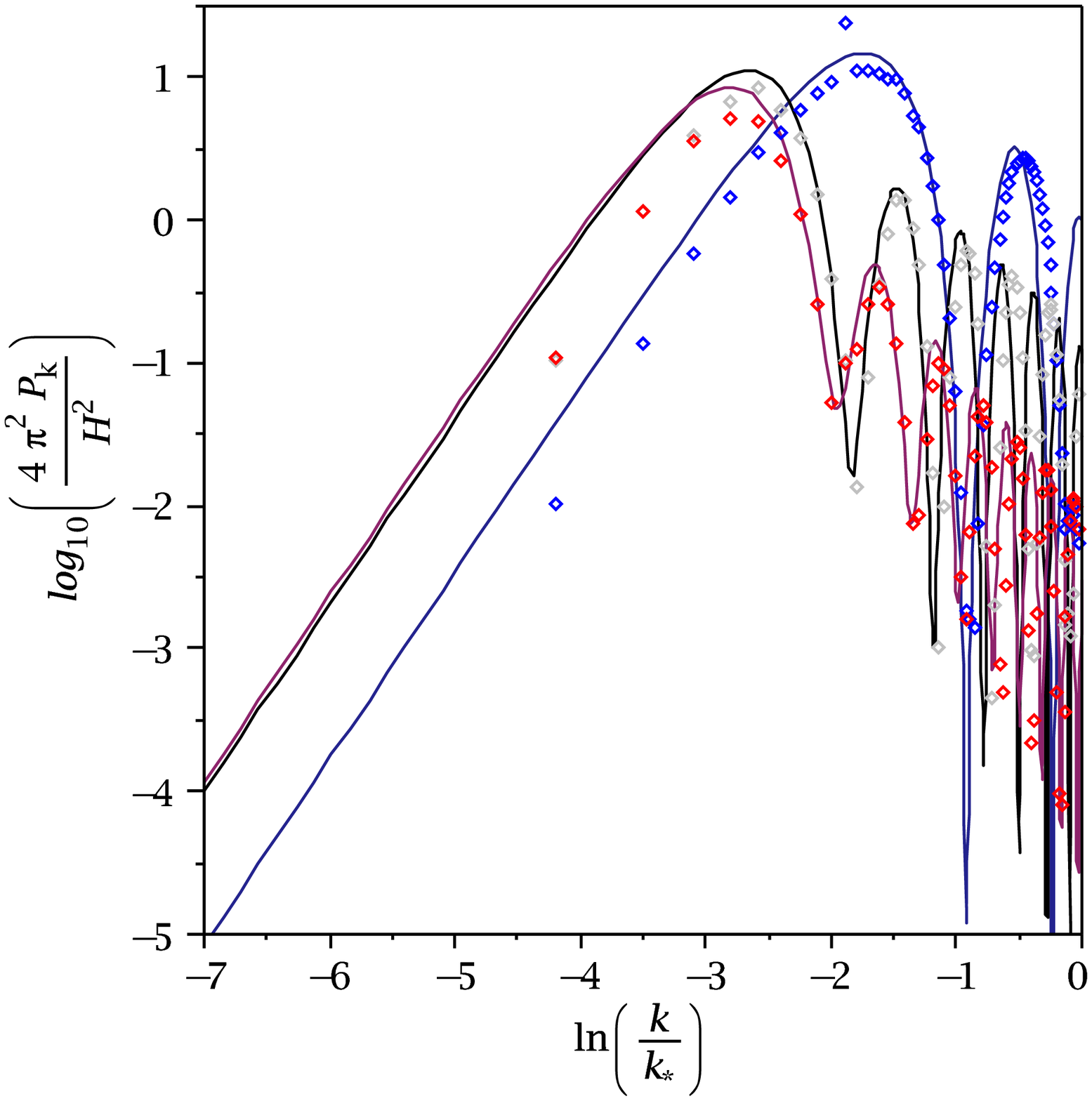}}
\caption{The power spectrum of inflaton modes induced by rescattering.
(normalized to the usual vacuum fluctuations) as a function of $\ln(k/k_\star)$, 
plotted for three representative time steps in the late-time evolution.  
This figure illustrates the final stages of IR cascading; we see the peak of the bump-like
feature slide to $k \sim e^{-3} k_\star$, at which point the associated mode functions $\delta\phi_k$
have crossed the horizon and become frozen.  At later times in the evolution the peak of the feature
and also the IR tail ($\sim k^3$) remain fixed.  Modes associated with the UV end of the spectrum
are still inside the horizon and continue to evolve as $\delta\phi_k \sim a^{-1}$, which explains the
damping of the $k>e^{-2} k_{\star}$ part of the spectrum.
For each time step we plot the analytical result (the solid line) and the data points obtained 
using lattice field theory simulations (diamonds).}
\label{Fig:pwr_late}
\end{figure}

We have studied the fully nonlinear dynamics of $\chi$ particle production and the subsequent interactions of the produced $\chi$ with the inflaton field in the model (\ref{L}), as described above.
We are interested in the power spectrum of the inflaton fluctuations
\begin{equation}
  P_\phi(k) = \frac{k^3}{2\pi^2} |\delta\phi_k|^2
\end{equation}
This contains a contribution coming from the usual quantum vacuum fluctuations from inflation that is close to the usual power-law form $k^{n_s-1}$ on large scales.  
Such a contribution would be present even in the absence of particle production and is not particularly interesting for us.  In order to isolate the effects of rescattering 
we have subtracted off this component in figures \ref{Fig:pwr}, \ref{Fig:g} and \ref{Fig:pwr_late}.  In all cases we have normalized $P_\phi$ to the amplitude of the usual vacuum fluctuations from inflation,
$H^2 / (2\pi)^2$. 

Fig.~\ref{Fig:pwr} shows time evolution of the power in the inflaton fluctuations generated by rescattering, for three different time steps early in the evolution.  
This figure illustrates how multiple rescatterings leads to a dynamical cascading of power into the IR.  To illustrate the magnitude of this effect, the horizontal yellow line
corresponds to the amplitude of the usual vacuum fluctuations from inflation.  For $g^2 \gsim 0.06$, the fluctuations from rescattering come to dominate over the vacuum 
fluctuations within a single $e$-folding.  In Fig.~\ref{Fig:g} we illustrate how the magnitude of the spectral distortion depends on the coupling, $g^2$.  
(The apparent change in the location of the feature for different values of $g^2$ arises because we are plotting the power spectrum as a function of $\ln(k / k_\star)$ and $k_\star$ depends on $g$.)

At late times, the IR portion of the power spectrum illustrated in Fig.~\ref{Fig:pwr} will remain fixed since the modes associated with
these scales have gone outside the horizon and become frozen.  On the other hand, the UV portion of this curve corresponds to modes
that are still inside the horizon, hence we expect $\delta\phi_k \sim a^{-1}$ and the UV tail of the power spectrum should damp as $a^{-2}$, due
to the Hubble expansion.  We observe precisely this behaviour in our lattice field theory simulations and this is illustrated in
Fig.~\ref{Fig:pwr_late}, which displays the dynamics of IR cascading over a much longer time scale.

Within a few $e$-foldings from the time of particle production, the entire bump-like feature from IR cascading becomes frozen outside
the horizon.  At this point the fluctuations have become classical, large-scale adiabatic density perturbations and are observable in the present epoch 
(presuming that $\phi=\phi_0$ occurs during the observable range of $e$-foldings).  In Fig.~\ref{Fig:sample} we have illustrated this bump-like feature
in both the primordial power spectrum and angular TT spectrum, for a representative choice of parameters.

\subsection{Backreaction Effects}
\label{back_sec}

As discussed previously, the production of $\chi$ fluctuations at $t=0$ back-reacts on the homogeneous $\phi(t)$ causing a transient violation of slow roll.  
We can study this backreaction numerically, by averaging the inhomogeneous field $\dot{\phi}(t,{\bf x})$ over the simulation box.
The result is plotted in Fig.~\ref{Fig:dotphi}.  We have also plotted the analytical solution of the mean-field equation (\ref{mean}), showing that this agrees with the exact numerical result.

The dynamics illustrated in Fig.~\ref{Fig:dotphi} is easy to understand physically.  The production of $\chi$ particles at $t=0$ drains kinetic energy from the condensate and hence $\dot{\phi}$ must decrease abruptly.
However, within a few $e$-foldings of the moment $t=0$, the produced iso-inflaton particles become extremely massive and are diluted by the expansion as $a^{-3}$.  At late times the inflaton velocity $\dot{\phi}$ must 
tend to the slow roll value.  Notice that the velocity $\dot{\phi}$ including backreaction effects is not changed significantly, as compared to the usual slow roll result.  This illustrates the energetic cheapness of particle production and 
IR cascading in the model (\ref{L}). 

\begin{figure}[htbp]
\bigskip \centerline{\epsfxsize=0.5\textwidth\epsfbox{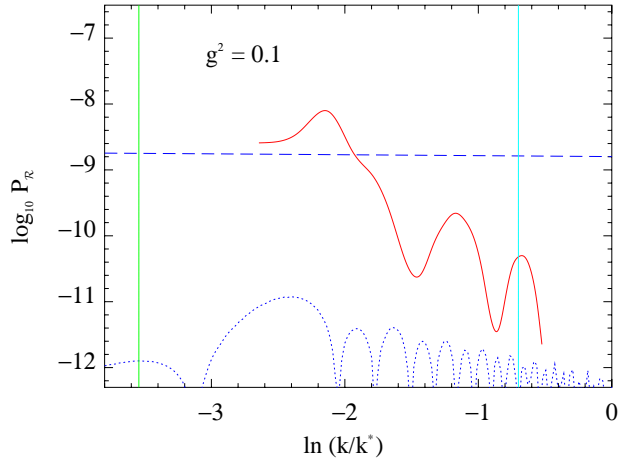}}
\caption{A comparison of curvature fluctuations from different physical effects.  The dashed blue line is the usual (nearly) scale invariant
vacuum fluctuations from inflation.  The red solid line is the bump-like feature induced by rescattering and IR cascading.  The dotted blue
line is the ringing pattern resulting from the momentary slowing-down of the inflaton (computed using the mean 
field approach of \cite{sasaki}).  The vertical lines show $aH$ at the beginning of particle production and after $\sim 3$ $e$-foldings.
This figure clearly illustrates the dominance of IR cascading over backreaction effects.  For illustration we have taken $g^2=0.1$, but the dominance is generic for all values of the coupling.
}
\label{Fig:compare}
\end{figure}

The transient violation of slow-roll illustrated in Fig.~\ref{Fig:dotphi} is expected to induce a ringing pattern in the vacuum fluctuations from inflation \cite{sasaki}.  This effect
is accounted for automatically in our HLattice simulations.  However, we would like to disentangle the effect of backreaction on the cosmological fluctuations from the 
effect of rescattering.  This will be useful in order to compare the relative importance of different physical processes, and also to guide our analytical efforts in the next section.
To this end, we consider the evolution of the curvature perturbation on co-moving hyper-surfaces, $\mathcal{R}$.  In linear theory the equation for the Fourier modes $\mathcal{R}_k$ is well known 
\begin{equation}
\label{R_k}
  \mathcal{R}''_k + 2 \frac{z'}{z} \mathcal{R}'_k + k^2 \mathcal{R}_k = 0
\end{equation}
Here the prime denotes derivatives with respect to conformal time $\tau = \int \frac{dt}{a}$ and $z \equiv a \dot{\phi} / H$.  Equation (\ref{R_k}) is only strictly valid in the absence of
entropy perturbations.  However, in our case the $\chi$ field is extremely massive $m_\chi^2 \gg H^2$ for nearly the entire duration of inflation, hence one may expect that direct iso-curvature
contributions to $\mathcal{R}$ are small.  We have solved equation (\ref{R_k}) numerically.  In order to take backreaction effects into account we 
compute the dynamics of $z(t)= a(t) \dot{\phi}(t) / H(t)$ by averaging over our HLattice simulation box.  Next, we solve equation (\ref{R_k}) given this background evolution and compute the 
power spectrum 
\begin{equation}
  P_{\mathcal{R}}(k) = \frac{k^3}{2\pi^2} |\mathcal{R}_k|^2
\end{equation}
The result is very close to the usual power-law form $k^{n_s-1}$, with small superposed oscillations resulting from the transient violation of slow roll, see Fig.~\ref{Fig:compare}.  
In order to make the ringing pattern more visible, we have subtracted off the usual (nearly) scale-invariant result which would be obtained in the absence of particle production.
For comparison, we also plot the bump-like feature from rescattering and IR cascading.  This latter contribution was obtained using the results for $P_\phi(k)$ from the previous subsection and the naive
formula $\mathcal{R} \sim \frac{H}{\dot{\phi}} \delta \phi$ (so that $P_{\mathcal{R}} \sim (2\epsilon M_p^2)^{-1} P_\phi$).  

From Fig.~\ref{Fig:compare} we see that IR cascading has a \emph{much} more significant impact on the observable curvature fluctuations than does backreaction.  Indeed, for $g^2 =0.1$ the transient violation of slow-roll yields
an order $10^{-2}$ correction to the vacuum fluctuations, while the correction from IR cascading is of order $10^{1}$.  This dominance is generic for all values of the coupling.  Thus, in developing an analytical
theory of particle production during inflation, it is a very good approximation to completely ignore backreaction effects.  

\section{Analytical Formalism}
\label{sec_analytical}

In the last section, we have studied particle production, rescattering and IR cascading using nonlinear lattice field theory simulations.
In this section we will develop a detailed analytical theory, in order to understand those results from a physical perspective.
These results were first presented in \cite{pptheory}.  We consider, again, the model (\ref{L}).  The equations of motion that we wish to solve
are
\begin{eqnarray}
  -\Box \phi + V'(\phi) + g^2 (\phi-\phi_0) \chi^2 &=& 0 \label{phiKG} \\
  -\Box \chi + g^2 (\phi-\phi_0)^2\chi &=& 0 \label{chiKG}
\end{eqnarray}
where $\Box = g_{\mu\nu}\nabla^{\mu}\nabla^{\nu}$ is the covariant d'Alembertian.  It will be useful to work with conformal time
$\tau$, related to cosmic time $t$ via $a d\tau = dt$.  In terms of conformal time the metric takes the form
\begin{eqnarray}
  ds^2 &=& -dt^2 + a^2(t) d{\bf x}\cdot d{\bf x} \nonumber \\
           &=& a^2(\tau) \left[ -d\tau^2 + d{\bf x}\cdot d{\bf x}  \right]
\end{eqnarray}
We denote derivatives with respect to cosmic time as $\dot{f} \equiv \partial_t f$ and with respect to conformal time as $f' \equiv \partial_\tau f$.  
The Hubble parameter $H = \dot{a} / a$ has conformal time analogue $\sH = a' / a$.  For an inflationary (quasi-de Sitter) phase 
($H \cong \mathrm{const}$) one has
\begin{equation}
\label{conf_scale}
  a = -\frac{1}{H\tau}\frac{1}{1-\epsilon}, \hspace{3mm} \sH = -\frac{1}{\tau}\frac{1}{1-\epsilon}
\end{equation}
to leading order in the slow roll parameter $\epsilon \ll 1$.

As discussed in section \ref{sec_overview}, the motion of the homogeneous inflaton $\phi(t)$ leads to the production
of a gas of $\chi$ particles at the moment $t=0$ when $\phi=\phi_0$.  The first step in our analytical computation is
to describe this burst of particle production in an expanding universe.  Following the initial burst, both backreaction
and rescattering effects take place.  Our formalism will focus on the latter effect, which is much more important, and 
we provide only a cursory treatment of backreaction.

\subsection{Particle Production in an Expanding Universe}
\label{chi_sec}

The first step in our scenario is the quantum mechanical production of $\chi$-particles due to the motion of $\phi$.
To understand this effect we must solve the equation for the $\chi$ fluctuations in the rolling inflaton background.
Approximating $\phi \cong \phi_0 + v t$ equation (\ref{chiKG}) gives
\begin{equation}
\label{chi_expanding}
  \ddot{\chi} + 3 H \dot{\chi} - \frac{\grad^2}{a^2} \chi + k_\star^4 t^2 \chi = 0
\end{equation} 
where $k_\star \equiv \sqrt{g |v|}$.  We remind the reader that $k_\star \gg H$ for reasonable values of the coupling, see equation (\ref{ratio}).

The flat space analogue of equation (\ref{chi_expanding}) is very well understood from studies of broad band parametric resonance during preheating \cite{KLS97}
and also moduli trapping at enhanced symmetry points \cite{beauty}.  One does not expect this treatment to differ significantly in our case
since both the time scale for particle production $\Delta t$ and the characteristic wavelength of the produced fluctuations $\lambda$ are small
compared to the Hubble scale: $\Delta t \sim \lambda^{-1} \sim k_\star^{-1} \ll H^{-1}$.  Hence, we expect that the occupation number of produced
$\chi$ particles will not differ significantly from the flat-space result (\ref{n_k}), at least on scales $k \gsim H$.
Furthermore, notice that the $\chi$ field is extremely massive for most of inflation
\begin{equation}
  \frac{m_\chi^2}{H^2} \cong \frac{k_\star^4 t^2}{H^2}
\end{equation}
Since $k_\star \gg H$ it follows that $m_\chi^2 \gg H^2$,
except in a tiny interval $H |\Delta t| \sim (H / k_\star)^2$ which amounts to roughly $10^{-3}$ $e$-foldings for $g^2 \sim 0.1$.  
Therefore, we do not expect any significant fluctuations of $\chi$ to be produced on super-horizon scales $k \lsim H$.

Let us now consider the solutions of equation (\ref{chi_expanding}).  We work with conformal time $\tau$
and write the Fourier transform of the quantum field $\chi$ as
\begin{equation}
\label{chi_fourier}
  \chi(\tau,{\bf x}) = \int \frac{d^3k}{(2\pi)^{3/2}} \frac{\xi^\chi_{\bf k}(\tau)}{a(\tau)} e^{i {\bf k}\cdot {\bf x}}
\end{equation}
Note the explicit factor of $a^{-1}$ in (\ref{chi_fourier}) which is introduced to give $\xi^\chi_{\bf k}$ a canonical kinetic 
term.  The q-number valued Fourier transform $\xi^\chi_{\bf k}(\tau)$ can be written as 
\begin{equation}
\label{chi_mode}
  \xi^\chi_{\bf k}(\tau) = a_{\bf k}\, \chi_k(\tau) + a^\dagger_{-{\bf k}}\, \chi_k^\star(\tau)
\end{equation}
where the annihilation/creation operators satisfy the usual commutation relation
\begin{equation}
\label{commutator}
  \left[a_{\bf k}, a_{\bf k'}^{\dagger}\right] = \delta^{(3)}({\bf k}-{\bf k'})
\end{equation}
and the c-number valued mode functions $\chi_k$ obey the following oscillator-like equation
\begin{equation}
\label{chi_prod}
  \chi_k''(\tau) + \omega_k^2(\tau) \chi_k(\tau) = 0
\end{equation}
The time-dependent frequency is
\begin{eqnarray}
  \omega_k^2(\tau) &=& k^2 + a^2m_\chi^2(\tau) - \frac{a''}{a} \nonumber \\
                              &\cong& k^2 + \frac{1}{\tau^2}\left[ \frac{k_\star^4}{H^2} t^2(\tau) - 2  \right]  \label{omega_chi}
\end{eqnarray}
where $m_\chi^2(\tau) = g^2 (\phi-\phi_0)^2 \cong k_\star^4 t^2(\tau)$ is the time-dependent effective mass of the $\chi$ particles and
\begin{equation}
\label{cosmic_time}
  t(\tau) = \frac{1}{H} \ln\left(\frac{-1}{H \tau}\right)
\end{equation}
is the usual cosmic time variable.  We have arbitrarily set the origin of conformal time so that $\tau=-1/H$ corresponds
to the moment when $\phi=\phi_0$.  

\begin{figure}[tbp]
\begin{center}
\includegraphics[width=1.6in]{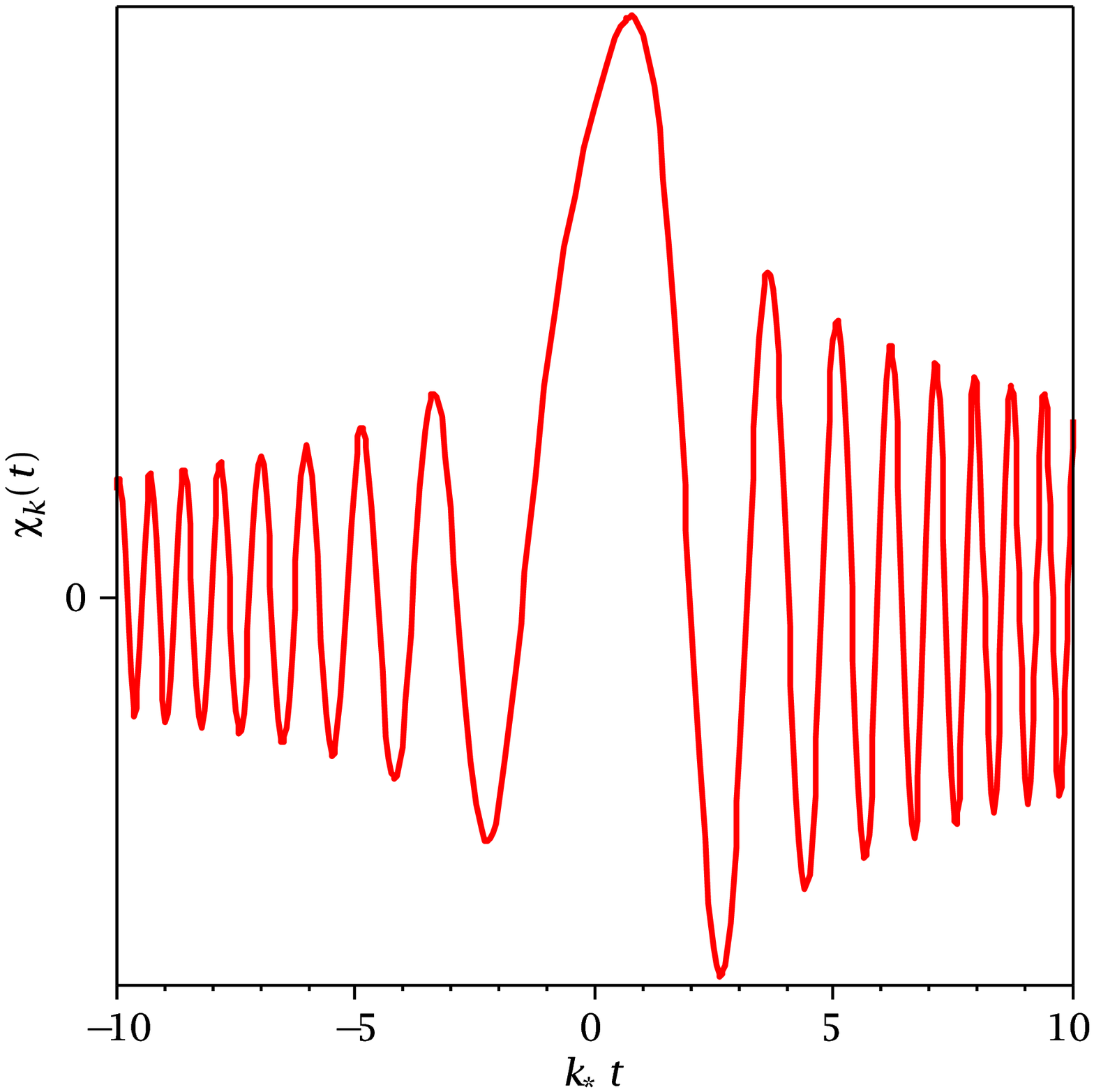}
\includegraphics[width=1.6in]{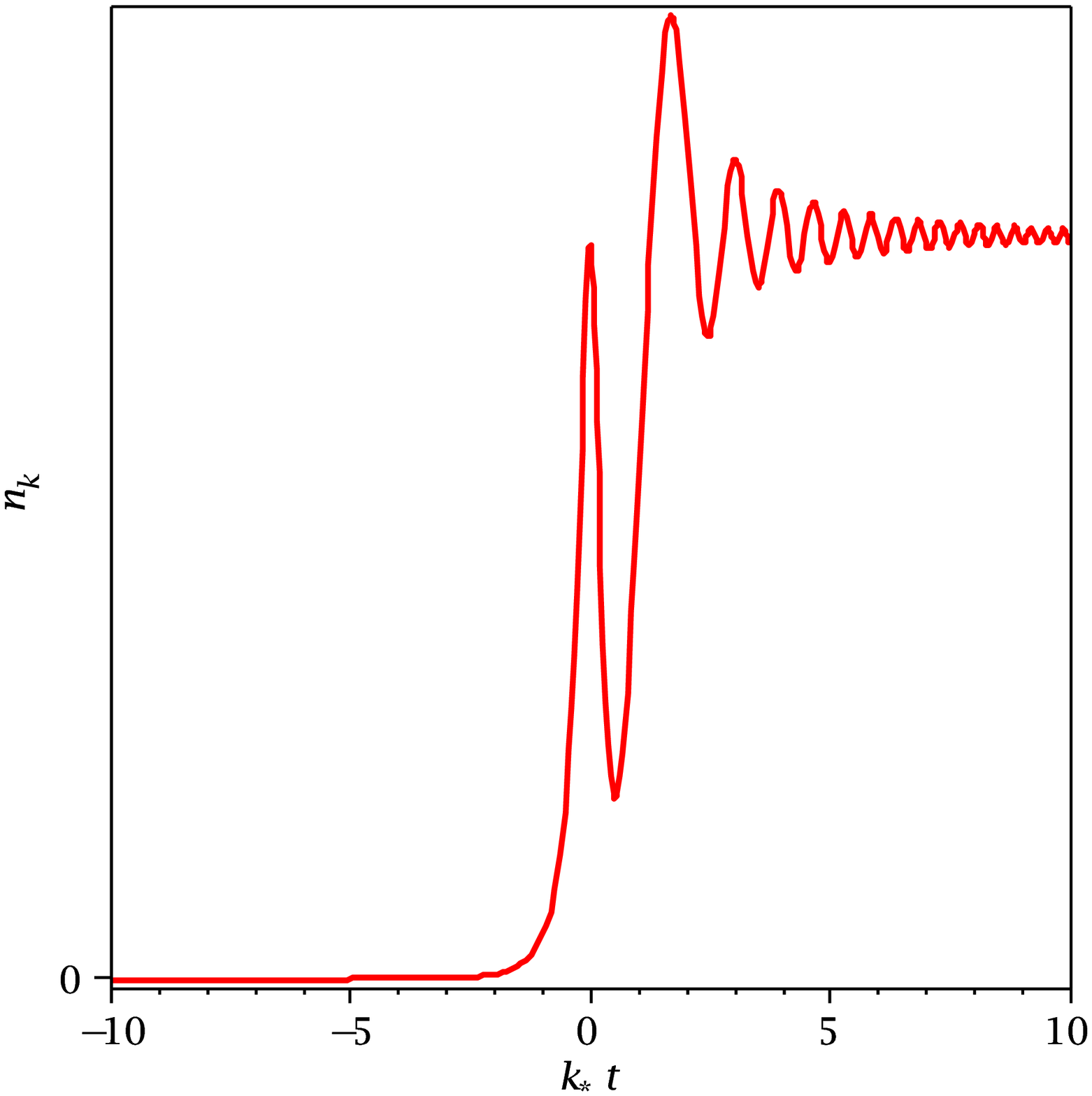}
\caption{The left panel illustrates the time dependence of the solutions $\chi_k$ of equation (\ref{chi_prod}) for a representative choice of parameters.
The oscillatory behaviour at early times represents the adiabatic initial condition.  At $t=0$ the effective frequency (\ref{omega_chi}) varies 
non-adiabatically and the fluctuations get a ``kick''.  The right panel plots the occupation number (\ref{n_k_def}) for the same mode.  No particles
are present in the adiabatic in-going regime.  This figure illustrates how the violations of adiabaticity at $t=0$ lead to production of $\chi$ particles.
}\label{Fig:chi_modes}
\end{center}
\end{figure}

In the left panel of Fig.~\ref{Fig:chi_modes} we have plotted a representative solution of (\ref{chi_prod}) in order to illustrate the qualitative behaviour
of the modes $\chi_k$.  In the right panel of Fig.~\ref{Fig:chi_modes} 
we plot the occupation number $n_k$ of particles with momentum ${\bf k}$, defined as the energy
of the mode $\frac{1}{2}|\chi'_k|^2 + \frac{1}{2}\omega_k^2 |\chi_k|^2$ divided by the energy $\omega_k$ of each particle.  Explicitly, we define
\begin{equation}
\label{n_k_def}
  n_k = \frac{\omega_k}{2}\left[ \frac{|\chi'_k|^2}{\omega_k^2} + |\chi_k|^2 \right] - \frac{1}{2}
\end{equation}
where the term $-\frac{1}{2}$ comes from extracting the zero-point energy of the linear harmonic oscillator (see \cite{KLS97} for a review).  
From the left panel of Fig.~\ref{Fig:chi_modes} we see that, near the massless point $t=0$, the fluctuations $\chi_k$ get a ``kick'' and from 
the right panel we see that the occupation number $n_k$ jumps abruptly at this same moment.

Let us now try to understand analytically the behaviour of the solutions of (\ref{chi_prod}).
At early times $t \ll -k_\star^{-1}$, the frequency $\omega_k$ varies adiabatically
\begin{equation}
\label{adiabatic_condition}
  \left|\frac{\omega'_k}{\omega_k^2}\right| \ll 1
\end{equation}
In this in-going adiabatic regime the modes $\chi_k$ are not excited and the solution of (\ref{chi_prod}) are well described
by the adiabatic solution $\chi_k(\tau) = f_k(\tau)$ where
\begin{equation}
\label{f_k}
  f_k(\tau) \equiv \frac{1}{\sqrt{2\omega_k(\tau)}} \exp\left[-i \int^\tau d\tau' \omega_k(\tau) \right]
\end{equation}
We have normalized (\ref{f_k}) to be pure positive frequency so that the state of the iso-inflaton field 
at early times corresponds to the adiabatic vacuum with no $\chi$ particles.  (Inserting (\ref{f_k}) into (\ref{n_k_def})
one finds $n_k=0$ for the adiabatic solution, as expected.)

The adiabatic solution (\ref{f_k}) ceases to be a good approximation very close to the moment when $\phi=\phi_0$, 
that is at times $|t| \lsim k_\star^{-1}$.  In this regime the adiabaticity condition (\ref{adiabatic_condition}) is violated for modes
with wave-number $H \lsim k \lsim k_\star$ and $\chi$ particles within this momentum band are produced.  During the
non-adiabatic regime we can still represent the solutions of (\ref{chi_prod}) in terms of the functions $f_k(\tau)$ as
\begin{equation}
\label{chi_bog}
  \chi_k(\tau) = \alpha_k(\tau) f_k(\tau) + \beta_k(\tau) f_k^\star(\tau)
\end{equation}
This expression affords a solution of (\ref{chi_prod}) provided the time-dependent Bogoliubov coefficients obey the following set of coupled equations
\begin{eqnarray}
  \alpha'_k(\tau) &=& \frac{\omega'_k(\tau)}{2\omega_k(\tau)} \exp\left[ +2 i \int^\tau d\tau' \omega_k(\tau') \right] \beta_k(\tau) \label{alpha_dot} \\
  \beta'_k(\tau) &=& \frac{\omega'_k(\tau)}{2\omega_k(\tau)} \exp\left[ -2 i \int^\tau d\tau' \omega_k(\tau') \right] \alpha_k(\tau) \label{beta_dot}
\end{eqnarray}
The Bogoliubov coefficients are normalized as $|\alpha_k|^2 - |\beta_k|^2 = 1$ and the assumption that no $\chi$ particles are present
in the asymptotic past\footnote{This assumption is justified since any initial excitation
of $\chi$ would have been damped out exponentially fast by the expansion of the universe.} fixes the 
initial conditions $\alpha_k = 1$, $\beta_k = 0$ for $t \rightarrow -\infty$.  This is known as the adiabatic initial condition.

From the structure of equations (\ref{alpha_dot},\ref{beta_dot}) it is clear that violations of the condition (\ref{adiabatic_condition}) near $t=0$
leads to a rapid growth in the $|\beta_k|$ coefficient.  The time variation of $\beta_k$ can be interpreted as a corresponding growth in the 
occupation number of the $\chi$ particles 
\begin{equation}
  n_k = |\beta_k|^2
\end{equation}

At late times ($t \gsim k_\star^{-1}$) adiabaticity is restored and the growth of $n_k=|\beta_k|^2$ must saturate.  By inspection of equations
(\ref{alpha_dot},\ref{beta_dot}) we can see that the Bogoliubov coefficients must tend to constant values in the out-going adiabatic regime.
Therefore, within less than an $e$-folding from the moment of particle production the solution $\chi_k$ of equation (\ref{chi_prod}) can be represented as a
simple superposition of positive frequency $f_k$ modes and negative frequency $f_k^\star$ modes.  Our goal now is to derive an analytical expression for the 
modes $\chi_k$ which is valid in this out-going adiabatic region.

Let us first study the adiabatic solution $f_k(\tau)$.  If we focus on the interesting region of phase space, $H \lsim k \lsim k_\star$, 
then the adiabatic solution (\ref{f_k}) is very well approximated by
\begin{equation}
\label{f_k_approx}
  f_k(\tau) \cong \frac{1}{a^{1/2} k_\star \sqrt{2 t(\tau)}} e^{-\frac{i}{2} k_\star^2 t^2(\tau)}
\end{equation}
where $t(\tau)$ is defined by (\ref{cosmic_time}).
It is interesting to note that equation (\ref{f_k_approx}) is identical to the analogous flat-space result \cite{ir}, except for the
factor of $a^{-1/2}$.  Taking into account also the explicit factor of $a^{-1}$ in our definition
of the Fourier transform (\ref{chi_fourier}) we recover  the expected large-scale behaviour for a massive field in de Sitter space, that is $\chi \sim a^{-3/2}$.
This dependence on the scale factor is easy to understand physically, it simply reflects the volume dilution of non-relativistic particles: 
$\rho_\chi \sim m^2_\chi \chi^2 \sim a^{-3}$.

Next, we seek an expression for the Bogoliubov coefficients $\alpha_k$, $\beta_k$ in the out-going adiabatic regime $t \gsim k_\star^{-1}$.
From (\ref{alpha_dot},\ref{beta_dot}) it is clear that the value of the Bogoliubov coefficients at late times can depends only on dynamics
during the interval $|t| \lsim k_\star^{-1}$ where the adiabaticity condition (\ref{adiabatic_condition}) is violated.  This interval
is tiny compared to the expansion time and we are justified in treating $a(\tau)$ as a constant during this phase.  Hence, it follows that the 
flat space computation of the Bogoliubov coefficients \cite{beauty,KLS97} must apply, at least for scales $k \gsim H$.  To a very
good approximation we therefore have the well-known result 
\begin{eqnarray}
  \alpha_k &\cong& \sqrt{1+e^{-\pi k^2 / k_\star^2}} \label{alpha_k} \\
  \beta_k &\cong& -i e^{-\pi k^2 / (2k_\star^2)} \label{beta_k}
\end{eqnarray}
in the out-going adiabatic regime.  Equation (\ref{beta_k}) gives the usual expression (\ref{n_k}) for the co-moving occupation number
of particles produced by a singe burst of broad-band parametric resonance:
\begin{equation}
  n_k = |\beta_k|^2 = e^{-\pi k^2 / k_\star^2}
\end{equation}

Finally, we arrive at an expression for the out-going adiabatic $\chi$ modes which is accurate
for interesting scales $k_\star \lsim k \lsim H$.  Putting together the results (\ref{f_k_approx}) and (\ref{chi_bog}) along with the well-known
expressions (\ref{alpha_k},\ref{beta_k}) we arrive at
\begin{equation}
  \chi_k(\tau) \cong 
  \sqrt{1+e^{-\pi k^2 / k_\star^2}} \frac{1}{a^{1/2} k_\star \sqrt{2 t(\tau)}} e^{-\frac{i}{2} k_\star^2 t^2(\tau)}
   - i e^{-\pi k^2 /(2 k_\star^2)}  \frac{1}{a^{1/2} k_\star \sqrt{2 t(\tau)}} e^{+\frac{i}{2} k_\star^2 t^2(\tau)} \label{chi_soln}
\end{equation}
valid for $t \gsim k_\star^{-1}$.
Equation (\ref{chi_soln}) is the main result of this subsection.  We will now justify that this expression is quite sufficient for our purposes.

For modes deep in the UV, $k \gsim k_\star$, our expression (\ref{chi_soln}), is not accurate.\footnote{The expression (\ref{f_k_approx})
for the adiabatic modes $f_k$ is not valid at high momenta where $\omega_k \cong k$.}  
However, such high momentum
particles are not produced, the condition (\ref{adiabatic_condition}) is always satisfied for $k \gg k_\star$.  Note that the absence of particle
production deep in the UV is built into our expression (\ref{chi_soln}): as $k\rightarrow \infty$ this function tends to the vacuum solution 
$\chi_k \rightarrow f_k$.

Our expression (\ref{chi_soln}) is also not
valid deep in the IR, for modes $k < H$.  To justify this neglect requires somewhat more care.  Notice that, even very far from the 
massless point, $t=0$, long wavelength modes $k \ll H$ should not be thought of as particle-like.  The large-scale mode functions are not oscillatory
but rather damp exponentially fast as $\chi \sim a^{-3/2}$.  Hence, even if we started with some super-horizon fluctuations of $\chi$ at the beginning of
inflation, these would be suppressed by an exponentially small factor before the time when particle production occurs.  Any super-horizon
fluctuation generated near $t=0$ would need to be exponentially huge to overcome this damping.  However, resonant particle production 
during inflation does
\emph{not} lead to exponential growth of mode functions.\footnote{In this regard our scenario is very different from preheating
at the end of inflation.  In the latter case the inflaton passes \emph{many} times through the massless point $m_\chi = 0$ and there are, 
correspondingly, many bursts of particle production.  After many oscillations of the inflaton field, the $\chi$ particle occupation numbers
build up to become exponentially large and, averaged over many oscillations of the background, the $\chi$ mode functions grow exponentially.
However, in our case case there is only a \emph{single} burst of particle production at $t=0$.  The resulting occupation number (\ref{n_k}) is always
less than unity and the solutions of (\ref{chi_prod}) never display exponential growth.}

To verify explicitly that there is no significant effect for super-horizon fluctuations let us consider solving equation (\ref{chi_expanding}) 
neglecting gradient terms.  The equation we wish to solve, then, is
\begin{equation}
\partial_t^2(a^{3/2} \chi) + \left[ k_\star^4 t^2 - \frac{9}{4}H^2 \right] (a^{3/2}\chi) = 0
\end{equation} 
(For simplicity we take $\epsilon=0$ for this paragraph, however, this has no effect on our results.)  The solution of this equation
may be written in terms of parabolic cylinder functions $D_{\nu}(z)$ as
\begin{eqnarray}
  \chi(t,{\bf x}) \sim \frac{1}{a^{3/2}} \left(\,  C_1 \, D_{ - \frac{1}{2} + \frac{9H^2}{8k_\star^2}i }\left[(1+i)k_\star t\right]
  +  C_2 \, D_{ - \frac{1}{2} - \frac{9H^2}{8k_\star^2}i }\left[(-1+i)k_\star t\right]   \,    \right) \label{cylinder}
\end{eqnarray}
For our purposes the precise values of the coefficients $C_1$, $C_2$ are not important.  Rather, it suffices to note that for $k_\star|t| \gsim 1$
the function (\ref{cylinder}) behaves as 
\begin{equation}
\label{chi_large_damp_thing}
  \chi(t,{\bf x}) \sim |t|^{-1/2} e^{-3 H t / 2} \times \left[\mathrm{oscillatory}\right]
\end{equation}
This explicit large-scale asymptotics confirms our previous claims that the super-horizon fluctuations of $\chi$ damp to zero
exponentially fast, as $a^{-3/2} \sim e^{-3Ht/2}$.  As discussed previously, this damping is easy to understand in terms of the volume
dilution of non-relativistic particles.  We can also understand the power-law damping that appears in (\ref{chi_large_damp_thing}) from
a physical perspective.  The properly normalized modes behave as $a^{3/2} \chi \sim \omega_k^{-1/2}$ while on large scales we have
$\omega_k \sim |m_\chi| \sim |t|$.  Hence, the late-time damping factor $t^{-1/2}$ which appears in (\ref{chi_large_damp_thing}) reflects the 
fact that the $\chi$ particles become ever more massive as $\phi$ rolls away from the point $\phi_0$.

Finally, it is straightforward to see that the function (\ref{cylinder}) does not display
any exponential growth near $t=0$.  Hence, we conclude that there is no significant generation of super-horizon $\chi$ fluctuations
due to particle production.\footnote{This is strictly true only in the linearized theory.  It is possible that $\chi$ particles are generated
by nonlinear effects such as rescattering.  However, even such second order $\chi$ fluctuations will be extremely massive
compared to the Hubble scale and must therefore suffer exponential damping $a^{-3/2}$ on large scales.}

In this subsection we have seen that the quantum production of $\chi$ particles in an expanding universe proceeds very much as it does
in flat space.  This is reasonable since particle production occurs on a time scale short compared to the expansion time and involves
modes which are inside the horizon at the time of production.  

\subsection{Inflaton Fluctuations}

In section \ref{chi_sec} we studied the quantum production of $\chi$ particles which occurs when $\phi$ rolls past the
massless point $\phi=\phi_0$.  Subsequently, there are two distinct physical processes which take place: backreaction and rescattering.
As we have argued in section \ref{sec_numerical}, the former effect has a negligible impact of the observable spectrum of cosmological
perturbations.  Hence, we will not study this effect in any detail (see \cite{chung,elgaroy,sasaki} for analytical calculations).
Instead we provide a cursory treatment of backreaction in Appendix A, in order to clear up some common misconceptions.

In this subsection we study the rescattering of produced $\chi$ particle off the inflaton condensate.  
The dominant process to consider is the diagram illustrated in Fig.~\ref{Fig:diag}, corresponding to 
bremsstrahlung emission of $\delta\phi$ fluctuations (particles) in the background of the external field.  (There is also
a sub-dominant process of the type $\chi\chi\rightarrow\delta\phi\delta\phi$ which is phase space suppressed.)  Taking into
account only the rescattering diagram illustrated in Fig.~\ref{Fig:diag} is equivalent to solving the following equation for the q-number inflaton
fluctuation
\begin{equation}
\label{delta_phi}
  \left[\partial_t^2 + 3 H \partial_t - \frac{\grad^2}{a^2} + m^2 \right] \delta \phi = -g^2 \left[\phi(t)-\phi_0\right] \chi^2
\end{equation}
where we have introduced the notation $m^2 \equiv V_{,\phi\phi}$ for the inflaton effective mass. (Note that we are not assuming a background
potential of the form $m^2 \phi^2 / 2$, only that $V_{,\phi\phi}\not= 0$ in the vicinity of the point $\phi=\phi_0$.)

Equation (\ref{delta_phi}) may be derived by noting that (\ref{L}) gives an interaction of the form $g^2(\phi-\phi_0) \delta \phi\chi^2$ between
the inflaton and iso-inflaton, in the background of the external field $\phi(t)$.  Equivalently, one may construct this equation by a straightforward
iterative solution of (\ref{phiKG}).

We work in conformal time and define the q-number Fourier
transform $\xi^\phi_{\bf k}(\tau)$ of the inflaton fluctuation analogously to (\ref{chi_fourier}):
\begin{equation}
\label{phi_fourier}
  \delta\phi(\tau,{\bf x}) = \int \frac{d^3 k}{(2\pi)^{3/2}}\frac{\xi_{\bf k}^\phi(\tau)}{a(\tau)} e^{i {\bf k}\cdot {\bf x}}
\end{equation}
(To avoid potential
confusion we again draw the attention of the reader to the explicit factor $a^{-1}$ in our convention for the Fourier 
transform.)  The equation of motion (\ref{delta_phi}) now takes the form
\begin{equation}
\label{phi_mode}
 \left[\partial_\tau^2 + k^2 + a^2 m^2 - \frac{a''}{a} \right] \xi_{\bf k}^\phi(\tau)
= -g k_\star^2 a(\tau) t(\tau) \int \frac{d^3k'}{(2\pi)^{3/2}} \xi^\chi_{\bf k'}\xi^\chi_{\bf k-k'}(\tau)
\end{equation}
The solution of (\ref{phi_mode}) consists of two parts: the solution of the homogeneous equation and the particular solution which is due
to the source.  The former corresponds, physically, to the usual vacuum fluctuations from inflation.  On the other hand, the particular solution
corresponds physically to the secondary inflaton modes which are generated by rescattering.

\subsection{Homogeneous Solution and Green Function}

We consider first the homogeneous solution of (\ref{phi_mode}).  Since the homogeneous solution is a gaussian field, we may
expand the q-number Fourier transform in terms of annihilation/creation operators $b_{\bf k}$, $b_{\bf k}^\dagger$ and c-number mode
functions $\phi_k(\tau)$ as
\begin{equation}
\label{phi_annihilation}
  \xi^{\phi}_{\bf k}(\tau) = b_{\bf k}\, \phi_k(\tau) + b^\dagger_{-{\bf k}} \,\phi_k^\star(\tau)
\end{equation}
Here the inflaton annihilation/creation operators $b_{\bf k}$, $b_{\bf k}^\dagger$ obey 
\begin{equation}
  \left[b_{\bf k}, b_{\bf k'}^{\dagger}\right] = \delta^{(3)}({\bf k} - {\bf k'})
\end{equation}
and commute with the annihilation/creation operators of the $\chi$-field:
\begin{equation}
  \left[a_{\bf k}, b_{\bf k'} \right] = \left[ a_{\bf k}, b_{\bf k'}^{\dagger} \right] = 0
\end{equation}

Using (\ref{conf_scale}) and (\ref{slow_roll}) it is straightforward to see that the homogeneous inflaton mode functions obey the following equation
\begin{equation}
\label{simple_inf_eqn}
  \partial_\tau^2 \phi_k + \left[ k^2 - \frac{1}{\tau^2}\left( \nu^2 - \frac{1}{4}  \right)  \right] \phi_k = 0
\end{equation}
where we have defined
\begin{equation}
\label{nu1}
  \nu \cong \frac{3}{2} - \eta + \epsilon
\end{equation}
The properly normalized mode function solutions are well known and may be written 
in terms of the Hankel function of the first kind as
\begin{equation}
\label{phi_k}
  \phi_k(\tau) = \frac{\sqrt{\pi}}{2} \sqrt{-\tau} H_{\nu}^{(1)}(-k\tau)
\end{equation}
This solution corresponds to the usual quantum vacuum fluctuations of the inflaton field during inflation.

In passing, let us compute the power spectrum of the quantum vacuum fluctuations from inflation.  Using the solutions (\ref{phi_k})
we have
\begin{equation}
\label{P_phi_vac}
  P_\phi^{\mathrm{vac}}(k) = \frac{k^3}{2\pi^2}\left|\frac{\phi_k}{a} \right|^2 \cong \frac{H^2}{(2\pi)^2} \left(\frac{k}{a H}\right)^{n_s-1}
\end{equation}
on large scales $k \ll aH$.  The explicit factor of $a^{-2}$ in (\ref{P_phi_vac}) appears to cancel the $a^{-1}$ in our definition of the
Fourier transform (\ref{phi_fourier}).   The spectral index is
\begin{equation}
\label{n_s_no_metric}
  n_s - 1 = 3 - 2\nu \cong 2\eta - 2\epsilon
\end{equation}
using (\ref{nu1}).

Given the solution (\ref{phi_k}) of the homogeneous equation, it is now trivial to construct the retarded Green function for equation 
(\ref{phi_mode}).  This may be written in terms of the free theory mode functions (\ref{phi_k}) as
\begin{eqnarray}
  G_k(\tau-\tau') &=& i \Theta(\tau-\tau') \left[ \,\,\,\phi_k(\tau) \phi_k^\star(\tau') 
                                                 -  \phi^\star_k(\tau) \phi_k(\tau')\,\,\, \right] \nonumber \\
  &=& \frac{i \pi}{4} \Theta(\tau-\tau') \sqrt{\tau\tau'}\left[\, H_{\nu}^{(1)}(-k\tau) H_{\nu}^{(1)}(-k\tau')^{\star} 
  - \,  H_{\nu}^{(1)}(-k\tau)^{\star} H_{\nu}^{(1)}(-k\tau') \, \right] \label{green}
\end{eqnarray}

\subsection{Particular Solution: Rescattering Effects}

We now consider the particular solution of (\ref{phi_mode}).  This is readily constructed using the Green function (\ref{green})
as
\begin{equation}
 \xi_{\bf k}^\phi(\tau) =   -\frac{g k_\star^2}{(2\pi)^{3/2}} \int d\tau' d^3k' G_k(\tau-\tau') \,
  a(\tau') t(\tau') \,\xi^\chi_{\bf k'}\xi^\chi_{\bf k-k'}(\tau') \nonumber \label{xi_k_particular}
\end{equation}
Notice that this particular solution is statistically independent of the homogeneous solution (\ref{phi_annihilation}).  
In other words, the particular solution can be expanded in terms of the annihilation/creation operators $a_{\bf k}, a_{\bf k}^\dagger$ 
associated with the $\chi$ field, whereas the homogeneous solution is written in terms of the annihilation/creation operators $b_{\bf k}, b_{\bf k}^\dagger$ associated
with the inflaton vacuum fluctuations.  These two sets of operators commute with one another.

We will ultimately be interested in computing the $n$-point correlation functions of the particular solution (\ref{xi_k_particular}).  
For example, carefully carrying out the Wick contractions, the connected contribution to the 2-point function is
\begin{eqnarray}
&& \langle \xi^\phi_{\bf k_1} \xi^\phi_{\bf k_2}(\tau) \rangle = \frac{2 g^2 k_\star^4}{(2\pi)^{3}} \delta^{(3)}({\bf k_1}+{\bf k_2})
   \nonumber \\
&& \times \int d\tau' d\tau'' a(\tau')a(\tau'')t(\tau')t(\tau'')  G_{k_1}(\tau-\tau')  G_{k_2}(\tau-\tau'') \nonumber \\
&& \,\,\times \int d^3k' \chi_{k_1-k'}(\tau')\chi_{k_1-k'}^\star(\tau'')\chi_{k'}(\tau')\chi_{k'}^\star(\tau'') \label{2pt}
\end{eqnarray}
The power spectrum of $\delta\phi$ fluctuations generated by rescattering is then defined in terms of the 2-point function
in the usual manner
\begin{equation}
  \langle \xi^\phi_{\bf k}(t) \xi^\phi_{\bf k'}(\tau) \rangle \equiv
  \delta^{(3)}( {\bf k} + {\bf k'} ) \frac{2\pi^2}{k^3} a(\tau)^2 P_\phi^{\mathrm{resc}} \label{pwr_def}
\end{equation}
(The explicit factor of $a^2$ in the definition (\ref{pwr_def}) appears to cancel the factor of $a^{-1}$ in our convention
for Fourier transforms (\ref{phi_fourier}).)

The total power spectrum is simply the sum of the contribution from the vacuum fluctuations (\ref{P_phi_vac}) and the contribution from rescattering
(\ref{pwr_def}):
\begin{equation}
\label{Psum}
  P_\phi(k) = P_\phi^{\mathrm{vac}}(k) + P_\phi^{\mathrm{resc}}(k)
\end{equation}
There are no cross-terms, owing to the fact $a_{\bf k}$ and $b_{\bf k}$ commute.

\subsection{Renormalization}

We now wish to evaluate the 2-point correlator (\ref{2pt}).  
In principle, this is straightforward: first substitute the result (\ref{chi_soln}) 
for the $\chi_k$ modes and the result (\ref{green}) for the Green function into (\ref{2pt}), next evaluate the integrals.
However, there is a subtlety. 
The resulting power spectrum is formally infinite.  Moreover, the 2-point correlation function (\ref{2pt}) receives contributions 
from two distinct effects.  There is a contribution from particle production, which we are interested in.  However, there is also 
a contribution coming from quantum vacuum fluctuations of the $\chi$ field interacting non-linearly with the inflaton.  The latter
contribution would be present even in the absence of particle production, when $\alpha_k = 1$, $\beta_k=0$.

In order to isolate the effects of particle production on the inflaton fluctuations, we would like to subtract off the contribution to
the 2-point correlation function (\ref{2pt}) which is coming from the quantum vacuum fluctuations of $\chi$.  This subtraction
also has the effect of rendering the power spectrum (\ref{pwr_def}) finite, since it extracts the usual UV divergent contribution 
associated with the Minkowski-space vacuum fluctuations.

As a step towards renormalizing the 2-point correlation function of inflaton fluctuations from rescattering (\ref{2pt}), let us first
consider the simpler problem of renormalizing the 2-point function of the gaussian field $\chi$.
We defined the renormalized 2-point function in momentum space as follows:
\begin{equation}
  \langle \xi^\chi_{k_1}(t_1) \xi^\chi_{k_2}(t_2) \rangle_{\mathrm{ren}} =  
  \langle \xi^\chi_{k_1}(t_1) \xi^\chi_{k_2}(t_2) \rangle \nonumber - \langle \xi^\chi_{k_1}(t_1) \xi^\chi_{k_2}(t_2) \rangle_{\mathrm{in}} \label{ren_rule}
\end{equation}
In (\ref{ren_rule}) the quantity $\langle \xi^\chi_{k_1}(t_1) \xi^\chi_{k_2}(t_2) \rangle_{\mathrm{in}}$ is the contribution which would be present 
even in the absence of particle production, computed by simply taking the solution (\ref{chi_bog}) with $\alpha_k=1$, $\beta_k=0$.  Explicitly, we
have
\begin{equation}
  \langle \xi^\chi_{k_1}(t_1) \xi^\chi_{k_2}(t_2) \rangle_{\mathrm{in}} = \delta^{(3)}({\bf k_1} + {\bf k_2}) f_{k_1}(t_1) f_{k_2}^\star(t_2)
\end{equation}
where $f_k$ are the adiabatic solutions (\ref{f_k}).

To see the impact of this subtraction, let us consider the renormalized variance for the iso-inflaton field, $\langle\chi^2\rangle$.  Employing the 
prescription (\ref{ren_rule}) we have
\begin{eqnarray}
  \langle \chi^2(\tau,{\bf x}) \rangle_{\mathrm{ren}} &=& 
    \int \frac{d^3 k}{(2\pi)^3 a^2(\tau)}\left[  |\chi_k(\tau)|^2 - \frac{1}{2\omega_k(\tau)}   \right] \nonumber \\
    &=& \langle \chi^2(\tau,{\bf x}) \rangle - \delta_M \label{ren_var}
\end{eqnarray}
where $\delta_M$ is the contribution from the Coleman-Weinberg potential.  This proves that our prescription reproduces the scheme advocated
in \cite{beauty}.  The renormalized variance (\ref{ren_var}) is finite and may be computed explicitly using our solutions (\ref{chi_soln}).  
We find
\begin{equation}
\label{ren_var_explicit}
  \langle\chi^2(t,{\bf x}) \rangle_{\mathrm{ren}} \cong \frac{n_\chi a^{-3}}{g |\phi-\phi_0|}
\end{equation}
where
\begin{equation}
  n_\chi \equiv \int \frac{d^3k}{(2\pi)^{3}} n_k \sim k_\star^3
\end{equation}
is the total co-moving number density of produced $\chi$ particles.  The result (\ref{ren_var_explicit}) was employed in \cite{sasaki} to quantify the
effect of backreaction on the inflaton condensate in the mean field treatment (\ref{mean}).  Hence, the renormalization scheme (\ref{ren_rule}) was implicit
in that calculation also.

At the level of the 2-point function, our renormalization scheme is tantamount to assuming that Coleman-Weinberg corrections are already absorbed into 
the definition of the inflaton potential, $V(\phi)$.  In general, such corrections might steepen $V(\phi)$ and spoil slow-roll inflation.  Here, we assume 
that this problem has already been dealt with, either by fine-tuning the bare inflaton potential or else by including extended SUSY (which can minimize 
dangerous corrections).  See also \cite{beauty} for a related discussion.  Note also that our renormalization procedure is equivalent to the quasi-particle
normal ordering scheme described in \cite{russian_text}.

Having established a scheme for remormalizing the 2-point function of the gaussian field $\chi$, it is now straightforward to 
consider higher order correlation functions.  We simply re-write the 4-point function as a product of 2-point functions 
using Wick's theorem.  Next, each Wick contraction is renormalized as (\ref{ren_rule}).  Applying this prescription to (\ref{2pt}) 
amounts to
\begin{eqnarray}
  \langle \xi^\phi_{k_1}(\tau)\xi^\phi_{k_2}(\tau) \rangle_{\mathrm{ren}} 
  &=& \frac{2 g^2 k_\star^4}{(2\pi)^3} \delta^{(3)}({\bf k_1}+{\bf k_2}) \nonumber \\
  &&\hspace{-28mm}\times \int d\tau'd\tau''t(\tau') t(\tau'') a(\tau')a(\tau'') G_{k_1}(\tau-\tau')  G_{k_2}(\tau-\tau'') \nonumber \\
  &&\hspace{-28mm} \times \int d^3k' \left[ \chi_{k_1-k'}(\tau')\chi_{k_1-k'}^\star(\tau'') - f_{k_1-k'}(\tau')f_{k_1-k'}^\star(\tau'')\right] 
               \left[\chi_{k'}(\tau')\chi_{k'}^\star(\tau'')- f_{k'}(t')f_{k'}^\star(\tau'')\right]  \label{pwr_phi_ren}
\end{eqnarray}
where $f_k(t)$ are the adiabatic solutions defined in (\ref{f_k}).

\subsection{Power Spectrum}

We are now in a position to compute the renormalized power spectrum of inflation fluctuations generated by rescattering, $P_\phi^{\mathrm{resc}}(k)$.
We renormalize the 2-point correlator of the inflaton fluctuations generated by rescatter according to (\ref{pwr_phi_ren}) and extract the power spectrum
by comparison to (\ref{pwr_def}).  We have relegated the technical details to appendix B and here we simply state the final result
\begin{eqnarray}
  P_\phi^{\mathrm{resc}}(k) &=& \frac{g^2 k^3 k_\star}{16\pi^5} \left[ \,\,\,\,\,\,\,\, \frac{e^{-\pi k^2 / (2k_\star^2)}}{2\sqrt{2}}\left( I_2(k,\tau)^2 + |I_1(k,\tau)|^2   \right)   \right. \nonumber \\
 &+& \left[ e^{-\pi k^2/(4k_\star^2)} + \frac{1}{2\sqrt{2}}e^{-3\pi k^2/(8k_\star^2)} \right]\left( I_2(k,\tau)^2 - \mathrm{Re}\left[I_1(k,\tau) \right] \right)   \nonumber \\
  &+& \left.  \left[ \frac{8\sqrt{2}}{3\sqrt{3}} e^{-\pi k^2 / (3k_\star^2)} + \frac{4\sqrt{2}}{5\sqrt{5}} e^{-3 \pi k^2/(5k_\star^2)} \right]
              \mathrm{Im}\left[ I_1(k,\tau)I_2(k,\tau) \right]\,\,\,\,\,\,\,\, \right] \label{full_pwr_result}
\end{eqnarray}
where the functions $I_1$, $I_2$ are the curved space generalization of the characteristic integrals defined in \cite{ir}.  Explicitly we have
\begin{eqnarray}
  I_1(k,\tau) &=& \frac{1}{a(\tau)}\int d\tau' G_k(\tau-\tau') e^{i k_\star^2 t^2(\tau')} \\
  I_2(k,\tau) &=& \frac{1}{a(\tau)}\int d\tau' G_k(\tau-\tau') 
\end{eqnarray}
The characteristic integral $I_2$ can be evaluated analytically, however, the resulting expression is not particularly enlightening.  Evaluation
of the integral $I_1$ requires numerical methods.  More details in Appendix B.  Equation (\ref{full_pwr_result}) is the main result of this section. 

\subsection{Comparison to Lattice Field Theory Simulations}

In section \ref{sec_numerical} the results of our analytical formalism were plotted alongside the out-put of fully nonlinear HLattice simulations.  
It is evident from Figs.~\ref{Fig:pwr}, \ref{Fig:g} and \ref{Fig:pwr_late} that the agreement between these approaches is extremely good, even very late into the
evolution and in the regime $g^2 \sim 1$.  The consistency of perturbative quantum field theory analytics and nonlinear classical lattice simulations 
provides a highly nontrivial check on our calculation.

\subsection{The Bispectrum}

So far, we have shown how to compute analytically the power spectrum generated by particle production, rescattering and IR 
cascading during  inflation in the model (\ref{L}).  We found that IR cascading leads to a bump-like contribution to the primordial power
spectrum of the inflaton fluctuations.  However, this same dynamics must also have a nontrivial impact on nongaussian statistics,
such as the bispectrum.  Indeed, it is already evident from our previous analysis that the inflaton fluctuations generated by rescattering
may be significantly nongaussian.  From the expression (\ref{xi_k_particular}) we see that the particular solution (due to rescattering) is bi-linear is the gaussian field $\chi$.

We define the bispectrum of the inflaton field fluctuations in terms of the three point correlation function as
\begin{equation}
  \langle \xi_{\bf k_1}^\phi \xi_{\bf k_2}^\phi \xi_{\bf k_3}^\phi (\tau) \rangle  = (2\pi)^3 a^3(\tau) \delta({\bf k_1} + {\bf k_2} + {\bf k_3})
  B_\phi(k_i)
\label{B_phi}
\end{equation}
The factor $a^3$ appears in (\ref{B_phi}) to cancel the explicit factors of $a^{-1}$ in our convention (\ref{phi_fourier}) for the Fourier transform.
It is well-known that the nongaussianity associated with the usual quantum vacuum fluctuations of the inflaton is negligible 
\cite{riotto,maldacena,seerylidsey}, therefore, when evaluating the bispectrum (\ref{B_phi}) we consider only the particular solution 
(\ref{xi_k_particular}) which is due to rescattering.  Carefully
carrying out the Wick contractions, we find the following result for the renormalized 3-point function
\begin{eqnarray}
  &&  \langle \xi_{\bf k_1}^\phi \xi_{\bf k_2}^\phi \xi_{\bf k_3}^\phi(\tau) \rangle_{\mathrm{ren}} = \frac{4 g^3 k_\star^6}{(2\pi)^{9/2}} \, \delta({\bf k_1} + {\bf k_2} + {\bf k_3}) \,
		\prod_{i=1}^{3} \int d\tau_i t(\tau_i)a(\tau_i) G_{k_i}(\tau-\tau_i) \times \nonumber \\
  &&  \int d^3p \left[ \chi_{k_1-p}(\tau_1)\chi_{k_1-p}^\star(\tau_2) - f_{k_1-p}(\tau_1)f_{k_1-p}^\star(\tau_2)   \right] 
			\left[   \chi_{k_3+p}(\tau_2)\chi_{k_3+p}^\star(\tau_3) - f_{k_3+p}(\tau_2)f_{k_3+p}^\star(\tau_3)    \right]
 \left[   \chi_{p}(\tau_1)\chi_{p}^\star(\tau_3) - f_{p}(\tau_1)f_{p}^\star(\tau_3)    \right] \nonumber \\ 
  && \,\,\,\,\,\,\,\,\,\,\,\,  + \,\,\,\,\, (k_2 \leftrightarrow k_3)\label{bispectrum_soln}
\end{eqnarray}
where the modes $\chi_k$ are defined by (\ref{chi_bog}) and $f_k$ are the adiabatic solutions (\ref{f_k}).  On the last line of (\ref{bispectrum_soln})
we have labeled schematically terms which are identical to the preceding three lines, only with $k_2$ and $k_3$ interchanged.  One may verify
that this expression is symmetric under interchange of the momenta $k_i$ by changing dummy variables of integration.

It is now straightforward (but tedious) to plug the expressions (\ref{f_k_approx}) and (\ref{chi_soln}) into (\ref{bispectrum_soln}) and evaluate 
the integrals.  This computation is tractable analytically because the time and phase-space integrals decouple.
The bispectrum is then extracted by comparison to (\ref{B_phi}).  This computation is carried out in detail in \cite{pptheory} where we have shown
that $B_\phi(k_i)$ peaks only over triangles with a characteristic size, corresponding to the location of the bump in the power spectrum.
This result is easy to understand on physical grounds, all the dynamics of rescattering
and IR cascading take place over a very short time near the moment $t=0$.  Hence, the effect of this dynamics on the primordial fluctuations must
be limited to scales leaving the horizon near the time when particle production occurs.

We will provide a cursory discussion of the bispectrum $B_\phi(k_i)$
in section \ref{sec_ng} when we discuss nongaussianity from particle production during inflation.

\subsection{Inclusion of a Bare Iso-Inflaton Mass}

In passing, it may be interesting to consider how the analysis of this section is modified in the case that our prototype action (\ref{L}) 
is supplemented by a bare mass term for the iso-inflaton field of the form $\delta\mathcal{L} = - \frac{1}{2}\mu^2 \chi^2$.  Thus, in place of (\ref{L}) 
suppose we consider the model
\begin{equation}
\mathcal{L} = -\frac{1}{2}(\partial\phi)^2 - V(\phi) - \frac{1}{2}(\partial\chi)^2 - \frac{1}{2}\mu^2 \chi^2 - \frac{g^2}{2}(\phi-\phi_0)^2\chi^2 \label{L2}
\end{equation}
Now the $\chi$ particles do not become massless at the point $\phi=\phi_0$, but rather the effective mass-squared
\begin{equation}
  m_\chi^2 = \mu^2 + g^2 (\phi-\phi_0)^2
\end{equation}
reaches a minimum value $\mu^2$ (which we assume to be positive).  Such a correction may arise due to a variety of effects and will reduce the impact 
of particle production and IR cascading on the observable cosmological fluctuations.    

In \cite{pptheory} we have carefully calculated the spectrum and bispectrum for the model (\ref{L2}) and shown explicitly how these differ from the analogous
results in the case $\mu^2=0$.  Rather than repeat this analysis, let us instead briefly consider how the additional bare mass term in (\ref{L2}) alters the 
dynamics of particle production from a more heuristic perspective.  The iso-inflaton fluctuations now obey the equation
\begin{equation}
  \ddot{\chi} + 3 H \dot{\chi} - \frac{\grad^2}{a^2} \chi + \left[ \mu^2 + k_\star^4 t^2 \right] \chi = 0
\end{equation}
rather than (\ref{chi_expanding}).  This equation was solved in \cite{beauty} in the regime where particle production is fast compared to the expansion 
time.\footnote{In the opposite regime, which corresponds to a fine-tuned coupling $g^2 \ll 10^{-7}$, the iso-inflaton will be light for a significant fraction
of inflation.  In that case the theory (\ref{L}) must be considered as a multi-field inflation model and one can no longer consistently assume $\langle \chi \rangle = 0$.
In other words, relaxing the assumption of fast particle production significantly changes the scenario under consideration and we do not pursue this possibility any further.}
The occupation number of produced $\chi$ particles is 
\begin{equation}
  n_k = r \, e^{- \pi k^2 / k_\star^2} 
\end{equation}
which differs from our previous result (\ref{n_k}) by the suppression factor
\begin{equation}
 r \equiv e^{-\pi \mu^2 / k_\star^2}  \leq 1
\end{equation}
Therefore, the effect of the inclusion of a bare mass for the iso-inflaton is to suppress the number density of produced $\chi$ particles by an amount $r$.  
This suppression reflects the reduced phase space of produced particles: the adiabaticity condition  $|\omega'_k / \omega_k^2| \ll 1$ is violated only for 
modes with $k < \sqrt{k_\star^2 -\mu^2}$.

The reduction of $n_k$ translates into a suppression for the $n$-point correlation functions of the iso-inflaton.  For example, the renormalized variance $\langle \chi^2 \rangle_{\mathrm{ren}} \propto \int d^3k\, n_k$ 
is suppressed by a factor of $r$.  The power spectrum of inflaton fluctuations generated by rescattering, $P_\phi^{\mathrm{resc}}$, is proportional to the 4-point correlator of $\chi$ and hence picks up a 
suppression factor of order $r^2$.  Similarly, the bispectrum $B_\phi$ is proportional to the 6-point correlator of $\chi$ and must be reduced by a factor of order $r^3$.  
The condition 
\begin{equation}
\label{small_mu}
  \mu^2 \ll k_\star^2
\end{equation}
is equivalent to $r \cong 1$ and ensures that the addition of a bare iso-inflaton mass will have a negligible impact on any observable.

For models obtained from string theory or super-gravity (SUGRA), it is natural to expect $\mu$ of order the Hubble scale during inflation \cite{false_vac,susy_break,rapid_roll}.  In the context of SUGRA, 
the finite energy density driving inflation breaks SUSY and induces soft scalar potentials with curvature of order $V''_{\mathrm{soft}} \sim \mu^2 \sim H^2$ \cite{susy_break}.  In the case of string theory, 
many scalars are conformally coupled to gravity \cite{rapid_roll} through an interaction of the form 
$\delta\mathcal{L} = - \frac{1}{12} R \chi^2$ where the Ricci scalar is $R \sim H^2$ during inflation.  More generally, any non-minimal coupling $\delta\mathcal{L} = -\frac{\xi}{2} R \chi^2$ between gravity and the iso-inflaton will induce a contribution 
of order $H$ to the effective mass of $\chi$, as long as $\xi = \mathcal{O}(1)$.  In all models where $\mu^2 \sim H^2$ the condition (\ref{small_mu}) is satisfied for reasonable values of the coupling $g^2 > 10^{-7}$, see equation
(\ref{ratio}).  Thus, we expect that corrections of the form $\delta\mathcal{L} = - \frac{1}{2}\mu^2 \chi^2$ will not alter our results in a wide variety of well-motivated models.  This heuristic expectation
was verified explicitly in \cite{pptheory}.

\section{Cosmological Perturbation Theory}
\label{sec_metric}

In section \ref{sec_analytical} we developed an analytical theory of particle production and IR cascading during inflation which is in very good
agreement with nonlinear lattice field theory simulations.  However, this formalism suffers from a neglect of metric perturbations and, consequently, we
were unable to rigorously discuss the gauge invariant curvature perturbation $\zeta$.\footnote{This variable is related to the quantity $\mathcal{R}$ defined
in subsection \ref{back_sec} as $\zeta \cong -\mathcal{R}$ on large scales.}  Hence, the reader may be concerned about gauge ambiguities in our results.  In this 
section we address such concerns, showing that metric perturbations may be incorporated in a straightforward manner and that their consistent inclusion does not change our results in any significant
way.  We will do so by showing explicitly that, with appropriate choice of gauge, equations (\ref{delta_phi}) and (\ref{chi_expanding}) for the fluctuations of the inflaton
and iso-inflaton still hold, to first approximation.  We will also go beyond our previous analysis by explicitly showing that in this same gauge the spectrum of the curvature fluctuations, $P_\zeta$,
is trivially related to the spectrum of inflaton fluctuations, $P_\phi$.  

To render the analysis tractable we would like to take full advantage of the results derived in the last section.  To do so, we employ the Seery
et al.\ formalism for working directly with the field equations \cite{seery} and make considerable use of results derived by
Malik in \cite{malik1,malik2}. (Note that our notations differ somewhat from those employed by Malik.
The reader is therefore urged to take care in comparing our formulae.)

We expand the inflaton and iso-inflaton fields up to second order in perturbation theory as
\begin{eqnarray}
  \phi(\tau,{\bf x}) &=& \phi(\tau) + \delta_1\phi(\tau,{\bf x}) + \frac{1}{2} \delta_2\phi(\tau,{\bf x}) \\
  \chi(\tau,{\bf x}) &=& \delta_1\chi(\tau,{\bf x}) + \frac{1}{2} \delta_2\chi(\tau,{\bf x})
\end{eqnarray}
The perturbations are defined to average to zero $\langle \delta_n\phi\rangle =\langle\delta_n\chi\rangle = 0$ so that 
$\langle \phi(t,{\bf x}) \rangle = \phi(t)$ and $\langle \chi(t,{\bf x}) \rangle = 0$.  (The condition $\langle\chi\rangle = 0$ is 
ensured by the fact that $m_\chi \gg H$ for nearly the entire duration of inflation.)

We employ the flat slicing and threading throughout this section.  With this gauge choice the perturbed metric takes the form
\begin{eqnarray}
  g_{00} &=& -a^2 (1 + 2\psi_1 + \psi_2) \\
  g_{0i} &=& a^2 \partial_i \left[ B_1 + \frac{1}{2} B_2  \right] \\
  g_{ij} &=& a^2 \delta_{ij}
\end{eqnarray}
so that spatial hyper-surfaces are flat.  Note also that in this gauge the field perturbations $\delta_n\phi$, $\delta_n\chi$
coincide with the Sasaki-Mukhanov variables \cite{SMvariable} at  both first and second order.

This perturbative approach, of course, neglects the momentary slow-down of the inflaton background due to backreaction.
However, we have already shown in subsection \ref{back_sec} that backreaction has a tiny impact on the observable cosmological perturbations 
(see also \cite{ir}).  

\subsection{Gaussian Perturbations}

In \cite{malik1} Malik has derived closed-form evolution equations for the field perturbations $\delta_n \phi$, $\delta_n \chi$ at both first ($n=1$) and 
second ($n=2$) order in perturbation theory.  Let us first study the gaussian perturbations.  The closed-form Klein-Gordon equation for $\delta_1\phi$ derived in \cite{malik1}
can be written as
\begin{equation}
 \delta_1 \phi'' + 2\sH \delta_1 \phi' - \grad^2 \delta_1\phi + \left[ a^2 m^2  - 3 \left(\frac{\phi'}{M_p}\right)^2   \right]\delta_1 \phi = 0  \label{d1phi}
\end{equation}
Following our previous analysis we expand
the first-order perturbation in terms of annihilation/creation operators as
\begin{equation}
  \delta_1 \phi(t,{\bf x}) = \int \frac{d^3k}{(2\pi)^{3/2}} \left[ b_{\bf k} \frac{\delta_1\phi_k(\tau)}{a(\tau)} e^{i {\bf k}\cdot {\bf x}} + \mathrm{h.c.}  \right]
\end{equation}
where $\mathrm{h.c.}$ denotes the Hermitian conjugate of the preceding term and we draw the attention of the reader to the 
the explicit factor of $a^{-1}$ in our definition of the Fourier transform.  Working to leading order in slow roll parameters we have
\begin{equation}
\label{d1phi_k}
  \delta_1\phi_k'' + \left[k^2 + \frac{1}{\tau^2}\left( -2 + 3\eta - 9\epsilon  \right)\right]\delta_1\phi_k = 0
\end{equation}
This equation coincides exactly with (\ref{simple_inf_eqn}) and the properly normalized solutions again take the form (\ref{phi_k}).
The only difference is that the order of the Hankel function, $\nu$, is now given by
\begin{equation}
\label{nu2}
  \nu \cong \frac{3}{2} - \eta + 3 \epsilon
\end{equation}
rather  than by equation (\ref{nu1}).  The power spectrum of the gaussian fluctuations is, again, given by (\ref{P_phi_vac}).  The correction
to the order of the Hankel function $\nu$ translates into a correction to the spectral index: instead of (\ref{n_s_no_metric}) we now have
\begin{equation}
  n_s - 1 = 2\eta - 6\epsilon
\end{equation}
which is precisely the standard result \cite{riotto_rev}.

Thus, as far as the quantum vacuum fluctuations of the inflaton are concerned, the only impact of consistently including metric perturbations
is an $\mathcal{O}(\epsilon)$ correction to the spectral index $n_s$.

Let us now turn our attention to the first order fluctuations of the iso-inflaton.  The closed-form Klein-Gordon equation for $\delta_1\chi$ derived 
in \cite{malik1} can be written as
\begin{equation}
  \delta_1 \chi'' + 2\sH \delta_1 \chi' - \grad^2 \delta_1\chi + a^2 k_\star^4 t^2(\tau) \delta_1 \phi = 0
\end{equation}
This coincides \emph{exactly} with equation (\ref{chi_expanding}), which we have already solved.  The fact that linear perturbations of $\chi$
do not couple to the metric fluctuations follows from the condition $\langle \chi \rangle  = 0$.

\subsection{Nongaussian Perturbations}

Now let us consider now the second order perturbation equations. The closed-form Klein-Gordon equation for $\delta_2\phi$ derived in \cite{malik1}
can be written as
\begin{equation}
  \delta_2 \phi'' + 2\sH \delta_2 \phi' - \grad^2 \delta_2\phi + \left[ a^2 m^2  - 3 \left(\frac{\phi'}{M_p}\right)^2   \right]\delta_2 \phi  = J(\tau,{\bf x}) \label{d2phi}
\end{equation}
As usual, the left-hand-side is identical to the first order equation (\ref{d1phi}) while the source term $J$ is constructed from a bi-linear 
combination of the first order quantities $\delta_1\phi$ and $\delta_1\chi$.  In order to solve equation (\ref{d2phi}) we require explicit
expressions for the Green function $G_k$ and the source term $J$.  The Green function is trivial for the case at hand; it is still given
by our previous result (\ref{nu1}), provided one takes into account the fact that the order of the Hankel functions $\nu$ is now given by (\ref{nu2}), 
rather than (\ref{nu1}).  In other words, the Green function for the nongaussian perturbations (\ref{d2phi}) differs from the result obtained neglecting 
metric perturbations only by $\mathcal{O}(\epsilon)$ corrections.

Next, we would like to consider the source term, $J$, appearing in (\ref{d2phi}).  Schematically, we can split the source into contributions bi-linear in 
the gaussian inflaton fluctuation $\delta_1\phi$ and contributions bi-linear in the iso-inflaton $\delta_1\chi$:
\begin{equation}
\label{Jsplit}
  J = J_{\phi} + J_{\chi}
\end{equation}
The contribution $J_\phi$ would be present even in the absence of the iso-inflaton.  These correspond, physically, to the usual nongaussian corrections
to the inflaton vacuum fluctuations coming from self-interactions.  This contribution to the source is well-studied in the literature and is known to contribute
negligibly to the bispectrum \cite{seery}.  Thus, in what follows, we will ignore $J_\phi$.

On the other hand, the contribution $J_\chi$ appearing in (\ref{Jsplit}) depends only on the iso-inflaton fluctuations $\delta_1\chi$.  This contribution can be understood, physically,
as generating nongaussian inflaton fluctuations $\delta_2\phi$ by rescattering of the produced $\chi$ particles off the condensate.  Hence, the contribution $J_\chi$ may source
large nongaussianity and is most interesting for us.  It is straightforward to compute $J_\chi$ explicitly for our model using the general results of \cite{malik1}.  We find
\begin{eqnarray}
  J_\chi &=& -2 a^2 g^2 (\phi-\phi_0) (\delta_1\chi)^2 \nonumber \\
     && \pm \frac{\sqrt{2\epsilon}}{M_p} \left[ \,\,\,\,   -a^2 g^2 (\phi-\phi_0)^2 (\delta_1\chi)^2 - \frac{1}{2} (\grad\delta_1 \chi)^2   - \frac{1}{2}(\delta_1\chi')^2     \right. \nonumber \\
     && \left. \,\,\,\, + \nabla^{-2}\left(   \partial_i(\delta_1\chi)\grad^2\partial^i(\delta_1\chi)  + \grad^2(\delta_1\chi)\grad^2(\delta_1\chi)  
 + \delta_1\chi' \grad^2 \delta_1\chi + (\grad\delta_1\chi')^2 \,\,\,\,  \right) \,\,\,\, \right] \label{J} 
\end{eqnarray}
where the upper sign is for $\phi' > 0$, the lower sign is for $\phi' < 0$.  
Notice that the contributions to $J_\chi$ on the third and fourth line of (\ref{J}) contain the inverse spatial Laplacian $\nabla^{-2}$ and are thus nonlocal.  
These terms all contain at least as many gradients as inverse gradients and hence the large
scale limit is well-defined.  In \cite{vernizzi} is was argued that these terms nearly always contribute negligibly to the curvature perturbation on large 
scales.

Let us now examine the structure of the iso-inflaton source $J_\chi$, equation (\ref{J}).  The first line of (\ref{J}) goes like $a^2g^2(\phi-\phi_0)(\delta_1\chi)^2$.  
This coincides exactly with the source term in equation (\ref{delta_phi}) which was already studied in section \ref{sec_analytical}.  On the other hand, the terms on 
the second, third and fourth lines of (\ref{J}) are new.  These represent corrections to IR cascading which arise due to the consistent inclusion of metric perturbations.
We will now argue that these ``extra'' terms are negligible as compared to the first line.  
If we denote the energy density in gaussian iso-inflaton fluctuations as $\rho_\chi \sim m_\chi^2 (\delta_1\chi)^2$ then, by inspection, we see that
the first line of (\ref{J}) is parametrically of order $\rho_\chi / |\phi-\phi_0|$ while the remaining terms are or order $\sqrt{\epsilon} \rho_\chi / M_p$.
Hence, we expect the first term to dominate for the field values $\phi \cong \phi_0$ which are relevant for IR cascading.  This suggests that the dominant
contribution to $J_\chi$ is the term which we have already taken into account in section \ref{sec_analytical}.

Let us now make this argument more quantitative.  
Inspection reveals that the only ``new'' contribution to (\ref{J}) which has any chance of competing with the ``old'' term $a^2g^2(\phi-\phi_0)(\delta_1\chi)^2$ 
is the one proportional to $\sqrt{\epsilon} a^2 g^2 (\phi-\phi_0)^2 (\delta_1\chi)^2 / M_p$ (the first term on the second line).  This new correction has the
possibility of becoming significant because it grows after particle production, as $\phi$ rolls away from $\phi_0$.  This growth, which reflects
the fact that the energy density in the $\chi$ particles increases as they become more massive, cannot persist indefinitely.  Within a few
$e$-foldings of particle production the iso-inflaton source term must behave as $J_\chi \sim a^{-3}$, corresponding to the volume dilution of non-relativistic
particles.  Hence, in order to justify the analysis of section \ref{sec_analytical} we must check that the term 
\begin{equation}
  J_{\mathrm{new}} \sim \frac{\sqrt{\epsilon}}{M_p} a^2 g^2 (\phi-\phi_0)^2 (\delta_1\chi)^2
\end{equation}
does not dominate over the term which we have already considered
\begin{equation}
  J_{\mathrm{old}} \sim a^2 g^2 (\phi-\phi_0) (\delta_1\chi)^2
\end{equation}
during the relevant time $H \Delta t = \mathcal{O}(1)$ after particle production.  It is straightforward to show that
\begin{equation}
  \frac{J_{\mathrm{old}}}{J_{\mathrm{new}}} \sim \frac{M_p}{\sqrt{\epsilon}} \frac{1}{\phi-\phi_0}
  \sim \frac{M_p H}{\dot{\phi}\sqrt{\epsilon}}\,\frac{1}{N} \sim \frac{1}{\epsilon}\, \frac{1}{N}
\end{equation}
where $N=Ht$ is the number of $e$-foldings elapsed from particle production to the time when IR cascading has completed.  Hence, 
$N = \mathcal{O}(1)$ and we conclude that the second, third and fourth lines of (\ref{J}) are (at least) slow roll suppressed as compared 
to the first line.  

In summary, we have shown that consistent inclusion of metric perturbations yields corrections to the inflaton fluctuations $\delta\phi$
which fall into two classes:
\begin{enumerate}
  \item  Slow-roll suppressed corrections to the inflaton vacuum fluctuations $\delta_1\phi$ (these amount to changing the definition of $\nu$
            in the solution (\ref{phi_k})).  These corrections have two physical effects.  First, they yield an $\mathcal{O}(\epsilon)$ correction to the
            spectral index. Second, they modify the propagator $G_k$ by an $\mathcal{O}(\epsilon)$ correction. 
  \item Corrections to the source $J$ for the nongaussian inflaton perturbation $\delta_2\phi$.  These corrections
           are the second, third and fourth lines of (\ref{J})) which, as we have seen, are slow roll suppressed.
\end{enumerate}
It should be clear that neither of these corrections alters our previous analysis in any significant way.

\subsection{Correlators}

So far, we have shown that a consistent inclusion of metric perturbations does not significantly alter our previous results for the field perturbations.
Specifically, $\delta_1\chi$ is identical to our previous solution of equation (\ref{chi_expanding}) for the iso-inflaton, while $\delta_1\phi$ coincides 
with the homogeneous solution of equation (\ref{delta_phi}), up to slow-roll corrections.  At second order in perturbation theory, we have seen that
\[
  \delta_2 \phi = \int d^4 x' G(x-x') J_\chi(x') + \mathcal{O}\left[(\delta_1\phi)^2\right]
\]
To leading order in slow roll $J_\chi \cong -2 a^2 g^2 (\phi-\phi_0)(\delta_1\chi)^2$ and the first term coincides with our previous result for the particular 
solution of  equation (\ref{delta_phi}).  The terms of order $(\delta_1\phi)^2$ represent nongaussian corrections to the vacuum fluctuations from inflation 
(coming from self-interactions of $\delta\phi$ and the nonlinearity of gravity). These would be present even in the absence of particle production, and are known 
to have a negligible impact on the spectrum and bispectrum \cite{seery}.

We are ultimately interested in the connected $n$-point correlation functions of $\delta\phi$.  For example, the 2-point function $\langle (\delta\phi)^2 \rangle$ get a contribution
of the form $\langle (\delta_1\phi)^2 \rangle$ which gives the usual nearly scale invariant large-scale power spectrum from inflation.  The cross term 
$\langle\delta_1\phi \delta_2\phi \rangle$ is of order $\langle (\delta_1\phi)^4 \rangle$ and represents a negligible ``loop'' correction to the scale-invariant
spectrum from inflation.  (The cross term does not involve the iso-inflaton since $\delta_1\phi$ and $\delta_1\chi$ are statistically independent.)
Finally, there is a contribution $\langle (\delta_2\phi)^2 \rangle$ which involves terms of order $\langle \chi^4 \rangle$ coming from rescattering and terms of order
$\langle (\delta_1\phi)^4 \rangle$ which represent (more) loop corrections to the scale-invariant spectrum from inflation.  Thus, we can schematically write
\[
  P_\phi(k) = P_\phi^{\mathrm{vac}}(k) \left[1 + (\mathrm{loops})\right] + P_\phi^{\mathrm{resc}}(k)
\]
Here $P_\phi^{\mathrm{vac}} \sim k^{n_s-1}$ is the usual nearly scale invariant spectrum from inflation and $P_\phi^{\mathrm{resc}}$ is the bump-like contribution
from rescattering and IR cascading which we have studied in the previous section.  The ``loop'' corrections to $P^{\mathrm{vac}}_k(k)$ have been studied in detail
in the literature (see, for example, \cite{loop_corrections}) and are known to be negligible in most models.

We can also make a similar schematic decomposition of the bispectrum by considering the structure of the 3-point correlator $\langle (\delta\phi)^3 \rangle$.  Following
our previous line of reasoning, it is clear that the dominant contribution comes from rescattering and is of order $\langle \chi^6 \rangle$.  The terms involving 
$\langle(\delta_1\phi)^3\rangle$, on the other hand, represent the usual nongaussianity generated during single field slow roll inflation and are known to be small \cite{seery}.

\subsection{The Curvature Perturbation}

Ultimately one wishes to compute not the field perturbations $\delta_n\phi$, $\delta_n\chi$, but rather the gauge invariant curvature fluctuation, $\zeta$.  We expand this in perturbation
theory in the usual manner
\begin{equation}
  \zeta  = \zeta_1 + \frac{1}{2}\zeta_2
\end{equation}
In \cite{malik2} Malik has derived expressions for the large scale curvature perturbation 
in terms of the Sasaki-Mukhanov variables at both first and second order in perturbation theory.  
We remind the reader that in the flat slicing (which we employ) the Sasaki-Mukhanov variable for each field simply coincides with the field
perturbation (\emph{i.e.}~- $Q_\phi = \delta \phi$ and $Q_\chi = \delta\chi$). 

At first order in perturbation theory the iso-inflaton does not contribute to the curvature perturbation (since $\langle\chi\rangle = 0$) and we have
\begin{equation} 
\label{zeta1}
  \zeta_1 = -\frac{\sH}{\phi'}\delta_1\phi
\end{equation}

At second order in perturbation theory the expression for the curvature perturbation is more involved.  Using the results of \cite{malik2} and working to leading order in slow roll parameters we find
\begin{eqnarray}
  \zeta_2 &\cong& -\frac{\sH}{\phi'}\left[ \delta_2\phi - \frac{\delta_2\phi'}{3\sH}  \right] \nonumber \\
          &+& \frac{1}{3(\phi')^2}\left[ (\delta_1\chi')^2 + a^2 g^2 v^2 t^2(\tau)(\delta_1\chi)^2   \right] \nonumber \\
          &+& \frac{1}{3(\phi')^2}\left[ (\delta_1\phi')^2 + a^2 m^2 (\delta_1\phi)^2   \right]   \label{zeta2}
\end{eqnarray}
Let us discuss the various contributions to this equation.  The third line contributes to the nongaussianity of the vacuum fluctuations 
during inflation.  These terms are known to be negligible \cite{riotto,maldacena,seerylidsey,seery} and, indeed, one may explicitly verify
that (\ref{zeta2}) would predict $f_{NL} \sim \mathcal{O}(\epsilon,\eta)$ in the absence of particle production.  

Next, we consider the
second line of (\ref{zeta2}).  This represents the direct contribution of the gaussian fluctuations $\delta_1\chi$ to the curvature perturbation.  
This contribution is tiny since the $\chi$ particles are extremely massive for nearly the entire duration of inflation and hence $\delta_1\chi \sim a^{-3/2}$
(see also \cite{trapped} for a related discussion).  The smallness of this contribution to $\zeta$ can be understood physically by noting that the 
super-horizon iso-curvature fluctuations in our model are negligible.

Finally, let us consider the contribution on the first line of (\ref{zeta2}).  This contribution is the most interesting.  
To make contact with observations we must compute the curvature perturbation at late times and on large scales.
In section \ref{sec_analytical} we have already shown that $\delta_n\phi$ is constant on large scales and at late times for both $n=1$ and $n=2$.
This is the expected result: the curvature fluctuations are frozen far outside the horizon  and in the absence of entropy perturbations 
\cite{Sconserved}.\footnote{Note that, in some cases, the curvature fluctuations may evolve significantly after horizon exit \cite{liddle,sasaki2}.
This is a concern in models where there are significant violations of slow-roll.  In section \ref{back_sec} we have already shown that the transient
violation of slow roll has a negligible effect on the curvature fluctuations in our model.  Hence, the result $\zeta_n \sim \delta_n\phi \sim \mathrm{const}$ far outside
the horizon is consistent with previous studies.}  Hence $\delta_2\phi'$ is completely negligible and the first term on the first line of (\ref{zeta2}) must dominate over the second term.
We conclude that ,at late times and on large scales, the second order curvature perturbation is very well approximated by
\begin{equation}
\label{zeta2_approx}
  \zeta_2 \cong -\frac{\sH}{\phi'}\delta_2\phi + \cdots
\end{equation}

In summary, we have shown that the power spectrum of curvature fluctuations from inflation in the model (\ref{L}) is trivially related to the power spectrum
of inflaton fluctuations
\begin{equation}
\label{P_zeta_prop}
  P_\zeta(k) \cong \frac{H^2}{\dot{\phi}^2} P_\phi(k) = \frac{1}{2\epsilon M_p^2} P_\phi(k)
\end{equation}
at both first and second order in cosmological perturbation theory.  This relation is valid at late times and for scales far outside the horizon.  The curvature spectrum
(\ref{P_zeta_prop}) may be written as
\begin{equation}
\label{P_zeta_decomp}
  P_\zeta(k) = P_\zeta^{\mathrm{vac}}(k) \left[ 1 + (\mathrm{loops}) \right] + P_\zeta^{\mathrm{resc}}(k)
\end{equation}
The power spectrum of the inflaton vacuum fluctuations agrees with the usual result obtained in linear theory \cite{riotto_rev} 
\begin{equation}
  P_{\zeta}^{\mathrm{vac}}(k) \cong \frac{H^2}{8\pi^2 \epsilon M_p^2}\left(\frac{k}{aH}\right)^{2\eta-6\epsilon}
\end{equation}
In (\ref{P_zeta_decomp}) we have schematically labeled the corrections arising from the second line of (\ref{zeta2}) and the source $J_\phi$ as ``loop''.
These are nongaussian corrections to the inflaton vacuum fluctuations arising from self-interactions of the inflaton and also the nonlinearity of gravity.  Such corrections are negligible.
The most interesting contribution to the power spectrum (\ref{P_zeta_decomp}) is due to rescattering, $P_\zeta^{\mathrm{resc}}(k)$.  This quantity is proportional to our previous result (\ref{full_pwr_result}).

In passing, notice that the bispectrum $B_\phi$ (defined by (\ref{B_phi})) of inflaton fluctuations will differ from the bispectrum $B$ of the curvature fluctuations (defined by (\ref{bispectrum_def})) only by a simple
re-scaling:
\begin{equation}
\label{B_zeta}
  B(k_i) \cong -\left(\frac{H}{\dot{\phi}}\right)^3 B_\phi(k_i) = -\frac{1}{(2\epsilon)^{3/2} M_p^3} B_\phi(k_i)
\end{equation}
The dominant contribution to $B_\phi$ comes from rescattering effects $\langle\delta_2\phi^3 \rangle \sim \langle \delta_1\chi^6 \rangle$.

The analysis of this section justifies our neglect of metric fluctuations in section \ref{sec_analytical}.

\section{Observational Constraints on Particle Production During Inflation}
\label{sec_cons}

In the previous sections of this review, we have demonstrated that particle production and IR cascading in the model (\ref{int}) leads
to a bump-like feature in the primordial power spectrum.  We now aim to determine the extent to which such a spectral distortion is 
compatible with current cosmological data.  The results of this section first appearing in the paper \cite{ppcons}.

One motivation for this study is to determine whether features generated by particle production during inflation can explain some of the 
anomalies in the observed primordial 
power spectrum, $P(k)$.  A number of different studies have hinted at the possible presence of some localized features 
in the power spectrum 
\cite{chung2,gobump,features,features2,morefeatures1,morefeatures2,morefeatures3,features3,yokoyama1,yokoyama3,yokoyama2,hoi1,hoi,contaldi,yokoyama4},
 which are not compatible with the simplest power law $P(k) \sim k^{n_s - 1}$
model.  Although such glitches may simply be statistical anomalies \cite{nofeatures}, there is also
the tantalizing possibility that they represent a signature of primordial physics beyond the simplest slow roll inflation scenario.  Forthcoming
polarization data may play a crucial role in distinguishing between these possibilities \cite{gobump}.  However, in the meantime, it is interesting to 
determine
the extent to which such features may be explained by microscopically realistic inflation models, such as (\ref{L}).

To answer this question we provide a simple semi-analytic
fitting function that accurately captures the shape of the feature generated by particle production and IR cascading
during inflation.  Next, we confront this modified power spectrum with a variety
of observational data sets.  We find no evidence for a detection, however, we note that observations are consistent with
relatively large spectral distortions of the type predicted by the model (\ref{int}).  If the feature is located on scales relevant for Cosmic Microwave Background
(CMB) experiments then its amplitude may be as large as $\mathcal{O}(10\%)$ of the usual scale-invariant fluctuations, corresponding to $g^2 \sim 0.01$.
Our results translate into a $\phi_0$-dependent bound on the coupling $g^2$ which is crucial in order to determine whether the nongaussian signal associated with particle production and IR 
cascading is detectable in future missions.

We also consider the more complicated features which result from multiple bursts of particle production and IR cascading.  Such features are
a prediction of a number of string theory inflation models, including brane/axion monodromy \cite{monodromy1,monodromy2,monodromy3}. 
For appropriate choice of the spacing between the features, we find that the constraint on $g^2$ in this scenario is even weaker than the single-bump
case.

\subsection{A Simple Parametrization of the Power Spectrum\label{sec:param}}

\begin{figure}[htbp]
\bigskip \centerline{\epsfxsize=0.35\textwidth\epsfbox{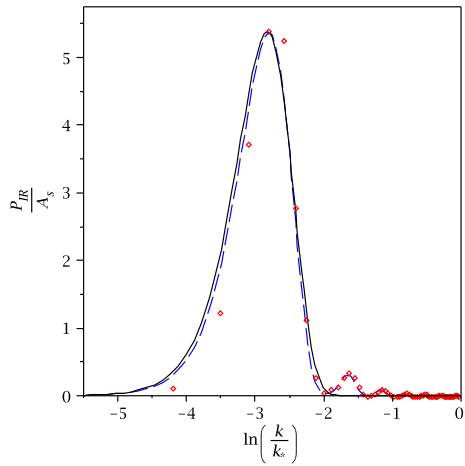}}
\caption{The bump-like features generated by IR cascading.  We plot the feature power spectrum $P_{\mathrm{bump}}(k)$ 
      obtained from fully nonlinear lattice field theory
      simulations (the red points) and also the result of an analytical calculation (the dashed blue curve) using the formalism described in
      \cite{pptheory}.  We also superpose the fitting function $\sim k^3 e^{-\pi k^2 / (2 k_\star^2)}$ (the solid black curve) to illustrate the accuracy
      of this simple formula.
}
\label{Fig:BumpFit}
\end{figure}

In \cite{ir} it was shown that particle production and IR cascading during inflation
in the model (\ref{int}) generates a bump-like contribution to the primordial
power spectrum.  As shown in Fig.~\ref{Fig:BumpFit}, this feature can be fit with a very simple function
$P_{\mathrm{bump}} \sim k^3 e^{-\pi k^2 / (2 k_\star^2)}$.
The bump-like contribution from IR cascading is complimentary to the usual
(nearly) scale-invariant contribution to the primordial power spectrum $P_{\mathrm{vac}} \sim k^{n_s-1}$ coming from
the quantum vacuum fluctuations of the inflaton.  The total, observable, power spectrum in the model
(\ref{int}) is simply the superposition of these two contributions: $P(k) \sim k^{n_s-1} +  k^3 e^{-\pi k^2 / (2 k_\star^2)}$.  (See equation (\ref{Psum}).)
This simple formula can be motivated from analytical considerations \cite{ir} and provides a good fit to lattice field theory simulations near the peak of the feature
and also in the IR tail.\footnote{This fitting formula does \emph{not} capture the small oscillatory structure in the UV tail of the feature (see Fig.~\ref{Fig:BumpFit}) which 
does not concern us since that region is not phenomenologically interesting.}

It is straightforward to generalize this discussion to allow for multiple bursts of particle production during inflation.  
Suppose there are multiple points $\phi=\phi_i$ ($i=1,\cdots,n$) along the 
inflationary trajectory where new degrees of 
freedom $\chi_i$ become massless:
\begin{equation}
\label{multiple}
  \mathcal{L}_{\mathrm{int}} = -\sum_{i=0}^{n} \frac{g_i^2}{2}(\phi-\phi_i) \chi_i^2
\end{equation}
For each instant $t_i$ when $\phi=\phi_i$ there will be an associated burst of particle production  and subsequent rescattering of
the produced massive $\chi_i$ off the condensate $\phi(t)$.  Each of these events proceeds as described above and leads to a new
bump-like contribution to the power spectrum.  
These features simply superpose owing to that fact that each field $\chi_i$ is statistically independent (so that the cross terms involving
$\chi_i\chi_j$ with $i\not= j$ in the computation of the two-point function must vanish).  Thus, we arrive at the following parametrization of the
primordial power spectrum in models with particle production during inflation:
\begin{equation}
  P(k) = A_s \left(\frac{k}{k_0}\right)^{n_s-1} + \sum_{i=1}^n A_i \left(\frac{\pi e}{3}\right)^{3/2}\left(\frac{k}{k_i}\right)^3 e^{-\frac{\pi}{2}\left(\frac{k}{k_i}\right)^2}    \label{param}
\end{equation}
where $A_s$ is the amplitude of the usual nearly scale invariant vacuum fluctuations from inflation and $k_0$ is the pivot, which we choose
to be $k_0 = 0.002 \,\mathrm{Mpc}^{-1}$ following \cite{Hinshaw2008}.  The constants $A_i$ depend on the couplings $g_i^2$ and measure the size of the
features from particle production.  We have normalized these amplitudes so that the power in the $i$-th bump, measured at the
peak of the feature, is given by $A_i$.
The location of each feature, $k_i$, is related to the number of $e$-foldings $N$ from the end of inflation to the time when the $i$-th burst of particle
production occurs: roughly $\ln( k_i / H) \sim N_i$ where $N=N_i$ at the moment when $\phi=\phi_i$.  From a purely
phenomenological perspective the locations $k_i$ are completely arbitrary.

We compare (\ref{param}) to lattice field theory simulations in order to determine the amplitude $A_i$ in terms of model parameters.  We find
\begin{equation}
\label{Ag}
  A_i \cong 10^{-6}\, g_i^{15/4}
\end{equation}
The power-law form of this relation was determined by inspection of numerical results, however, it can also be motivated by analytical considerations.
Assuming standard chaotic inflation $V = m^2\phi^2 / 2$ we have tested this formula for $g^2 = 1, 0.1, 0.01$, taking both $\phi_0 = 2 \sqrt{8\pi} M_p$ and $\phi_0 = 3.2 \sqrt{8\pi} M_p$.
We found agreement up to factors order unity in all cases.

Theoretical consistency of our calculation of the shape of the feature bounds the coupling as $10^{-7} \lsim g^2_i \lsim 1$ \cite{ir}.
Hence, the power spectrum (\ref{param}) can be obtained from sensible microphysics only when $10^{-20} \lsim A_i \lsim 10^{-6}$.
This constraint still allows for a huge range of observational possibilities: near the upper bound the feature is considerably larger
than the vacuum fluctuations while near the lower bound the feature is completely undetectable.

Note that for each bump in (\ref{param}) the IR tail $P_{\mathrm{bump}} \rightarrow k^3$ as $k\rightarrow 0$ is similar to the feature considered by Hoi, Cline \& Holder in \cite{hoi},
consistent with causality arguments about the generation of curvature perturbations by local physics.

\subsection{Data Sets and Analysis\label{sec:method}}

The primordial power spectrum for our model is parametrized as (\ref{param}).  Our aim is to derive observational constraints on the various
model parameters $A_s$, $n_s$, $k_i$ and $A_i$ using CMB, galaxy power spectrum and weak lensing data.  To this end we use the cosmoMC
package \cite{Lewis2002} to run Markov Chain Monte Carlo (MCMC) calculations to determine the likelihood of the cosmological 
parameters, including our new parameters $A_{i}$ and $k_{i}$.  We employ the following data sets.

\begin{description}
\item{\it Cosmic Microwave Background (CMB)}

Our complete CMB data sets include WMAP-5yr \cite{Komatsu2008,Hinshaw2008}, BOOMERANG \cite{Jones2006,Piacentini2006,Montroy2006}, 
ACBAR \cite{Runyan2003, Goldstein2003,Kuo2006,Reichardt2008}, CBI \cite{Pearson2003,Readhead2004a,Readhead2004b,Sievers2007}, 
VSA \cite{Dickinson2004}, DASI \cite{Halverson2002,Leitch2005}, and MAXIMA \cite{Hanany2000}. We have included the Sunyaev-Zeldovic (SZ) 
secondary anisotropy \cite{Sunyaev1972,Sunyaev1980} for WMAP-5yr, ACBAR and CBI data sets. The SZ template is obtained from hydrodynamical 
simulation \cite{Bond2005}. Also included for theoretical calculation of CMB power spectra is the CMB lensing contribution.

\item{\it Type Ia Supernova (SN)}

We employ the Union Supernova Ia data (307 SN Ia samples) from The Supernova Cosmology Project \cite{Kowalski2008}.

\item{\it Large Scale Structure (LSS)}

The 2dF Galaxy Redshift Survey (2dFGRS) data \cite{Cole2005} and Sloan Digital Sky Survey (SDSS) Luminous Red Galaxy (LRG) data 
release 4 \cite{Tegmark2006} are utilized.

Note that we have used the likelihood code based on the non-linear
modeling by Tegmark et al.\  \cite{Tegmark2006} (marginalizing the
bias $b$ and the $Q$ parameter). However with a large bump in the linear
power spectrum, this naive treatment may be not sufficient to characterize 
the non-linear response to the feature on small scales.  Ideally, this should be
obtained from N-body simulations, however, such a study is beyond the scope of 
the current work.

There are several other caveats on our results in the high-$k$ regime.  First, we assume 
linear bias for the galaxies, which may not be entirely safe at sufficiently small scales.
Moreover, sharp features in the matter power spectrum can cause sharp features in the bias 
as a function of $k$.

Keeping in mind these caveats our constraints on small scales $k \gsim 0.1\, \mathrm{Mpc}^{-1}$ should be 
taken with a grain of salt and considered as accurate only up to factors order unity.

\item{\it Weak Lensing (WL)}

 Five WL data sets are used in this paper. The effective survey area $A_{\textrm{eff}}$ and galaxy number density $n_{\textrm{eff}}$ of each survey are
 listed in Table~\ref{tblwldata_test_yes}.

\begin{table}[htbp]
{\caption{Weak Lensing Data Sets}\label{tblwldata_test_yes}}
  \begin{center}
  \begin{tabular}{lll} 
  \hline
  \hline
  Data sets& $A_{\textrm{eff}}$  & $n_{\textrm{eff}}$ \\
   & (deg$^2$) & (arcmin$^{-2}$) \\
  \hline
  COSMOS \cite{Massey2007,Lesgourgues2007}  & 1.6 & 40\\
  CFHTLS-wide \cite{Hoekstra2006,Schimd2007} & 22 & 12 \\
  GaBODS \cite{Hoekstra2002a,Hoekstra2002b} & 13 & 12.5 \\
  RCS \cite{Hoekstra2002a,Hoekstra2002b} & 53 & 8 \\
  VIRMOS-DESCART \cite{Van-Waerbeke2005,Schimd2007} & 8.5 &  15 \\
  \hline
  \end{tabular}
  \end{center}
\end{table}

For COSMOS data we use the CosmoMC plug-in written by Julien Lesgourgues \cite{Lesgourgues2007}, modified to do numerical 
marginalization on three nuisance parameters in the original code.

For the other four weak lensing data sets we use the likelihood given by \cite{Benjamin2007}. To calculate the likelihood we have written a CosmoMC 
plug-in code, with simplified marginalization on the parameters of galaxy number density function $n(z)$. More details about this plug-in can be 
found in \cite{Amigo2008}.

As for the LSS data, for small scales $k \gsim 0.1\, \mathrm{Mpc}^{-1}$ there is the caveat that the nonlinear evolution of the power spectrum in the presence of bump-like distortions may not be treated
accurately.
\end{description}



\subsection{Observational Constraints: A Single Burst of Particle Production}\label{subsec:single}

We now present our results for the observational constraints on particle production during inflation, assuming two different scenarios.

\begin{figure}[tbp]
\begin{center}
\includegraphics[width=3.2in]{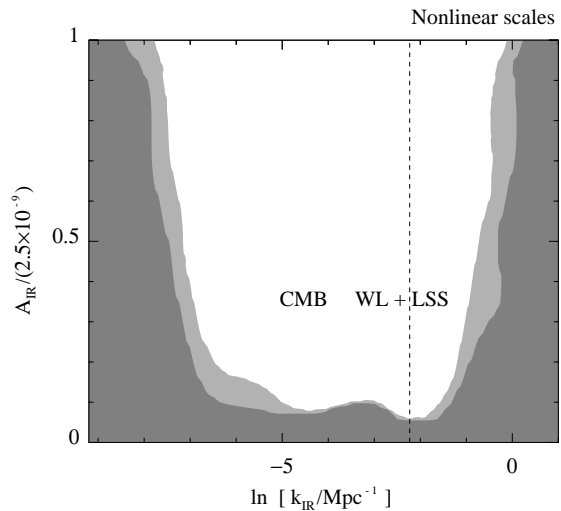}
\caption{Marginalized posterior likelihood contours for the parameters $A_{\mathrm{IR}}$ and $k_{IR}$ (the magnitude and position of the feature, respectively) in the single-bump model. Black and grey regions 
         correspond to parameter values allowed at 95.4\% and 99.7\% confidence levels, respectively.  At small scales, to the right of the dashed vertical line, our results should be taken with a grain of salt since
         the nonlinear evolution of the power spectrum may not be modeled correctly in the presence of bump-like distortions.\label{fig_2d}}
\end{center}
\end{figure}

The minimal scenario to consider is a single burst of particle production during inflation, which corresponds to taking $n=1$ in (\ref{multiple}).
The power spectrum is given by (\ref{param}) with $n=1$ and, with some abuse of notation, we denote $k_1 \equiv k_{\mathrm{IR}}$ and $A_1 \equiv A_{\mathrm{IR}}$. 
The prior we have used for $A_{\text{IR}}$ is $0\le A_{\text{IR}} \le 25\times 10^{-10}$, and for $k_{\text{IR}}$ is $-9.5\le \ln[k/\text{Mpc}^{-1}] \le 1$.  The former condition
ensures that the bump-like feature from IR cascading does not dominate over the observed scale invariant fluctuations while the latter is necessary in order to have the feature in
the observable range of scales.  In Fig.~\ref{fig_2d} we plot the marginalized posterior likelihood for the new parameters $A_{\text{IR}}$ and $k_{\text{IR}}$ describing the
magnitude and location of the bump while in Table \ref{Table_irpool} we give the best fit values for the remaining (vanilla) cosmological parameters.

\begin{table}
\begin{center} 
\caption{Constraints on the standard (``vanilla'') cosmological parameters for the single-bump model.  All errors are 95.4\% confidence level}
\label{Table_irpool}
\begin{tabular}{|c|c|}
\hline
$\Omega_bh^2$ & $0.0227^{+0.0010}_{-0.0010}$ \\
\hline
$\Omega_ch^2$ & $0.1122^{+0.0050}_{-0.0044}$ \\
\hline
$\theta$ & $1.0424^{+0.0042}_{-0.0043}$ \\
\hline
$\tau$ & $0.08^{+0.03}_{-0.03}$ \\
\hline
$n_s$ & $0.956^{+0.024}_{-0.024}$ \\
\hline
$\ln [10^{10}A_s]$ & $3.206^{+0.074}_{-0.068}$ \\
\hline
$A_{SZ}$ & $1.62^{+0.76}_{-0.74}$ \\  
\hline
$\Omega_m$ & $0.264^{+0.026}_{-0.022}$ \\
\hline
$\sigma_8$ & $0.807^{+0.034}_{-0.030}$ \\
\hline
$z_{re}$ & $10.5^{+2.5}_{-2.7}$ \\
\hline
$H_0$ & $71.6^{+2.3}_{-2.4}$ \\
\hline
\end{tabular}
\end{center}  
\end{table}

For very large scales $\lsim \mathrm{Gpc}^{-1}$, the data do not contain much information (due to cosmic variance) and hence the constraint on any modification of the power spectrum
is weak.  In this region the spectral distortion may be larger than $100\%$ of the usual scale invariant fluctuations and couplings $g^2$ order unity are allowed.  For smaller scales
$k \gsim \mathrm{Gpc}^{-1}$ the constraints are stronger and we have, very roughly, $A_{\mathrm{IR}} / A_s \lsim 0.1$ corresponding to $g^2 \lsim 0.01$.  For very small scales, $k \gsim 0.1 \, \mathrm{Mpc}^{-1}$ 
our constraints should be taken with a grain of salt since the nonlinear evolution of the power spectrum may not be modeled correctly in the presence of bump-like distortions.  At small scales nonlinear effects 
tend to wipe out features of this type (see, for example, \cite{Springel:2005nw}) and hence observational constraints for $k \gsim 0.1 \, \mathrm{Mpc}^{-1}$ may be weaker than what is presented in Fig.~\ref{fig_2d}.  
Note that in most of this nonlinear regime we find essentially no constraint on $A_{\mathrm{IR}}$, which is consistent with what would be expected in a more comprehensive treatment.

The IR cascading bump in the primordial power spectrum will be accompanied by a corresponding nongaussian feature in the bispectrum,
that will be discussed in more detail in the next section.  From the perspective of 
potentially observing this signal it is most interesting if this feature is located on scales
probed by CMB experiments.  (There is also the fascinating possibility that the nongaussianity from IR cascading could show up in the large scale
structure as in \cite{dalal,shandera,mcdonald,afshordi}.  We leave a detailed discussion to future studies.)
To get some intuition into what kinds of features in the CMB scales are still allowed by the data we focus on an example with
$A_{\text{IR}}=2.5\times 10^{-10}$ which, using (\ref{Ag}), corresponds to a reasonable coupling value $g^2 \sim 0.01$.  We take the bump to be
located at $k_{\mathrm{IR}} = 0.01\, \mathrm{Mpc}^{-1}$ and fix the remaining model parameters to $A_s=2.44\times 10^{-9}$, $n_s=0.97$ (which are compatible
with the usual values).  This sample bump in the power spectrum is illustrated in the top panel of Fig.~\ref{Fig:sample} and is consistent with the data at $2\sigma$.
In the bottom panel of Fig.~\ref{Fig:sample} we plot the associated angular CMB TT spectrum.  This example represents a surprisingly large spectral 
distortion: the total power in the feature as compared to the scale invariant vacuum fluctuations is $P_{\mathrm{bump}} / P_{\mathrm{vac}} \sim 0.1$, 
evaluated at the peak of the bump.  


\subsection{Observational Constraints: Multiple Bursts of Particle Production\label{subsec:multiple}}

Next, we consider a slightly more complicated scenario: multiple bursts of particle production leading many localized features
in the power spectrum.  For simplicity we assume that all bumps have the same magnitude $A_i \equiv A_{\mathrm{IR}}$ and we further
suppose a fixed number of $e$-foldings $\delta N$ between each burst of particle production.  This implies that the location
of the $i$-th bump will be given by $k_i = e^{(i-1)\Delta} k_1$ where $\Delta$ is a model parameter controlling the density of features.
We take the number of bursts, $n$, to be sufficiently large that the series of features extends over the whole observable range.  In
the next section we will see that these assumptions are not restrictive and that many well-motivated models are consistent with this simple set-up.  

\begin{figure}[tbp]
\begin{center}
\includegraphics[width=3.2in]{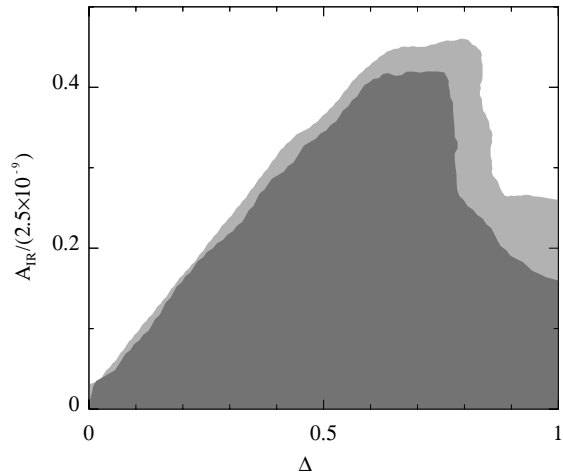}
\caption{Marginalized posterior likelihood contours for the parameters $A_{\mathrm{IR}}$ and $\Delta$ (the feature amplitude and spacing, respectively) of the multiple-bump model. Black and grey regions 
               correspond to values allowed at 95.4\% and 99.7\% confidence levels, respectively. } \label{fig_2d_multiple}
\end{center}
\end{figure}

Our multi-bump model, then, has three parameters: $A_{\mathrm{IR}}$, $k_1$ and $\Delta$.  We take the prior on the amplitude to be $A_{\mathrm{IR}} \leq 25\times 10^{-10}$ as
in section \ref{subsec:single}.  If the features are very widely spaced, $\Delta \gsim 1$, then the constraint on each bump will obviously be identical to the
results for the single-bump case presented in the section \ref{subsec:single}.  Hence the most interesting case to consider is $\Delta \lsim 1$ so that the bumps are partially overlapping.
Our prior for the density of features is therefore $0 \leq \Delta \leq 1$.  Finally, the location of the first bump
will be a historical accident in realistic models, hence we marginalize over all possible values of $k_1$ and present our constraints and 2-d likelihood plots in the space
of $A_{\mathrm{IR}}$ and $\Delta$.  This marginalized likelihood plot is presented in Fig.~\ref{fig_2d_multiple}.  In table \ref{Table_mp_close} we present the best-fit values for the vanilla cosmological
parameters.

\begin{table}
\begin{center} 
\caption{constraints on the standard (``vanilla'') cosmological parameters for the multiple-bump model.  All error bars are 95.4\% confidence level.}
\label{Table_mp_close}
\begin{tabular}{|c|c|}
\hline
$\Omega_bh^2$ & $0.0227^{+0.0009}_{-0.0009}$ \\
\hline
$\Omega_ch^2$ & $0.1126^{+0.0049}_{-0.0044}$ \\
\hline
$\theta$ & $1.0424^{+0.0039}_{-0.0043}$ \\
\hline
$\tau$ & $0.078^{+0.031}_{-0.026}$ \\
\hline
$n_s$ & $0.93^{+0.04}_{-0.17}$ \\
\hline
$\ln [10^{10}A_s]$ & $2.8^{+0.4}_{-0.9}$ \\
\hline
$A_{SZ}$ & $1.60^{+0.77}_{-0.76}$ \\
\hline
$\Omega_m$ & $0.265^{+0.026}_{-0.021}$ \\
\hline
$\sigma_8$ & $0.807^{+0.034}_{-0.030}$ \\
\hline
$z_{re}$ & $10.3^{+2.6}_{-2.5}$ \\
\hline
$H_0$ & $71.4^{+2.2}_{-2.4}$ \\
\hline
\end{tabular}
\end{center}  
\end{table}

From the likelihood plot, Fig.~\ref{fig_2d_multiple}, there is evidently a preferred value of the feature spacing, roughly $\Delta \sim 0.75$, for which the constraints are weakest.  This can be understood
as follows.  For very high density $\Delta \rightarrow 0$ the localized features from IR cascading smear together and the total power spectrum (\ref{param}) is $P(k) \sim A_s (k / k_0)^{n_s - 1} + C$ where the 
size of the constant deformation scales linearly with the density of features: $C \propto \Delta^{-1}$.  Therefore, the upper bound on the amplitude $A_{\mathrm{IR}}$ should scale linearly with $\Delta$.  Indeed, this linear
trend is very evident from Fig.~\ref{fig_2d_multiple} in the small-$\Delta$ regime.  This linear behaviour must break down at some point since as the features become infinitely widely spaced the constraint
on $A_{\mathrm{IR}}$ must go to zero.  This explains the bump in the likelihood plot, Fig.~\ref{fig_2d_multiple}, near $\Delta \sim 0.75$.

In passing, notice that the behaviour $P(k) \sim A_s (k / k_0)^{n_s - 1} + C$ for $\Delta \ll 1$ also explains why the best-fit $A_s$ in table \ref{Table_mp_close} is somewhat lower than the standard value and why the spectral tilt
$n_s-1$ is somewhat more red.

\begin{figure}[tbp]
\begin{center}
\includegraphics[width=3in]{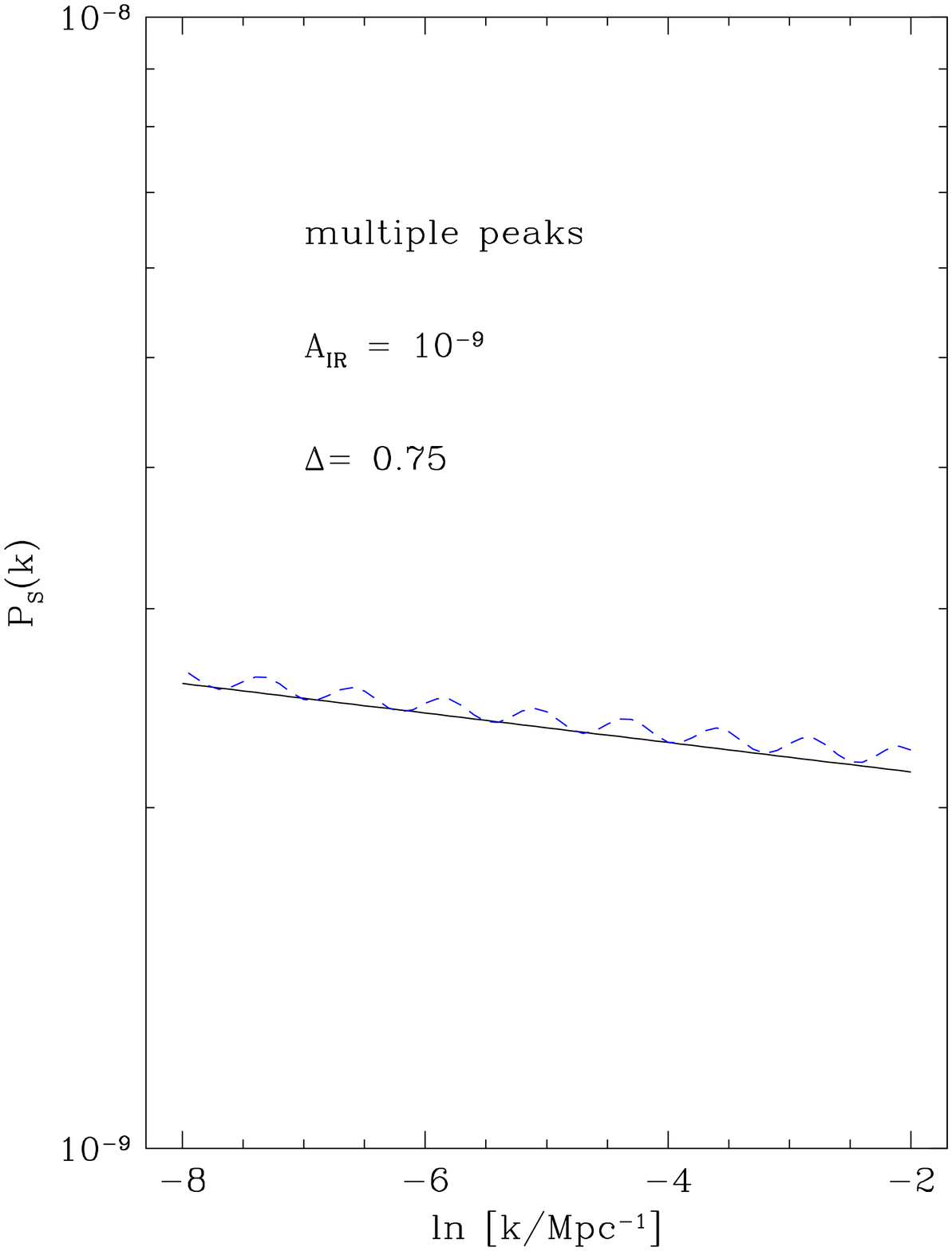}
\includegraphics[width=3in]{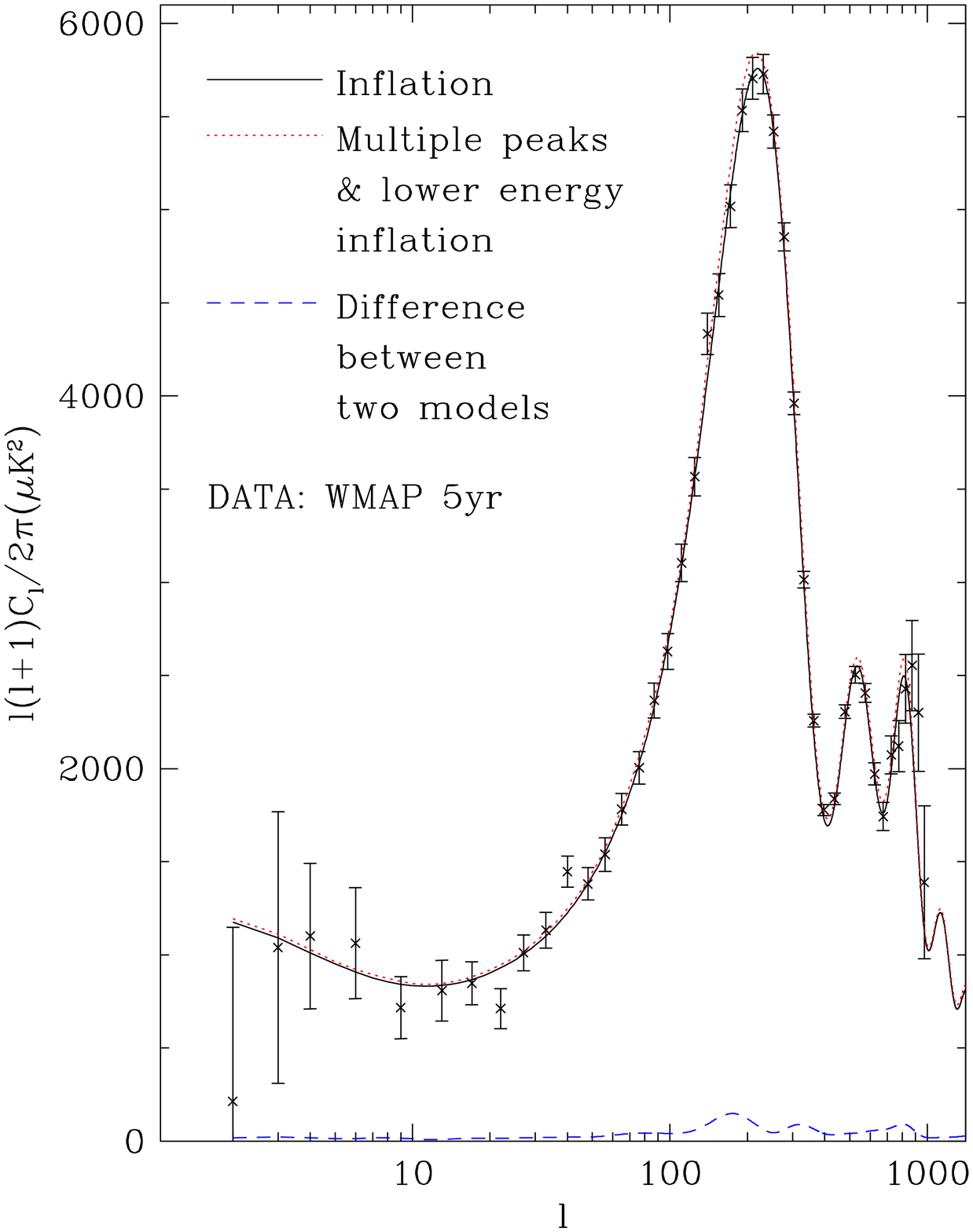}
\caption{The top panel shows a sample multiple-bump distortion with amplitude $A_{\mathrm{IR}} = 1\times 10^{-9}$ which corresponds to $g^2 \sim 0.02$.  The feature spacing is $\Delta = 0.75$.
We take the vanilla parameters to be $A_s = 1.04\times 10^{-9}$, $n_s = 0.93$ so that the scale of inflation is slightly lower than in the standard scenario and the spectral tilt is slightly redder.
The bottom panel shows the CMB angular TT power spectrum for this example.}\label{Fig:sample2}
\end{center}
\end{figure}

To get some intuition for the kinds of multi-bump distortions that are allowed by the data, we consider an example with $A_{\mathrm{IR}} = 1 \times 10^{-9}$, $\Delta = 0.75$ and fix the vanilla parameters to 
$A_s=1.04\times10^{-9}$, $n_s=0.93$.  This choice of parameters is consistent with the data at $2\sigma$ and corresponds to a reasonable coupling $g^2 \sim 0.02$.  In Fig.~\ref{Fig:sample2} we plot 
the primordial power spectrum $P(k)$ and also the CMB TT angular power spectrum for this example.

\section{Particle Physics Models}
\label{sec_micro}

In previous sections of this review we have studied in some detail the observational signatures of inflationary particle production in
the model (\ref{int}) and we have also derived observational constraints on the model parameters $g^2$ and $\phi_0$.  We now
aim to provide some explicit microscopic derivations of the model (\ref{L}) from popular theories of particle physics including string
theory and SUSY.  Our examples are meant to be illustrative, rather than exhaustive.  From the low energy perspective interactions
of the type (\ref{int}) are completely generic and hence we expect that many microscopic derivations may be possible.

\subsection{Open String Inflation Models}

String theory inflation models may be divided into two classes depending on the origin of the inflaton: closed string models and open string models.
In the former case the inflaton is typically a geometrical modulus associated with the compactification manifold (examples include racetrack
inflation \cite{racetrack}, K\"ahler modulus inflation \cite{CQ}, Roulette inflation \cite{roulette} and fibre inflation \cite{fibre}).  Reheating after
closed string inflation has been studied in \cite{closed_reheat}.  On the other hand, in the case of open string models the inflaton is typically the 
position modulus of some mobile D-brane\footnote{One notable exception is inflation driven by the open string tachyon, for example nonlocal string field theory models \cite{NLNG,NLmath,NLreview}.}
living in the compactification manifold (examples include brane inflation \cite{brane} such as the warped KKLMMT model \cite{KKLMMT}, 
D3/D7 inflation \cite{D3D7} and DBI inflation \cite{DBI}).  Reheating after open string inflation has been studied in \cite{open_reheat}.

In open string inflation models there may be, in addition to the mobile inflationary brane, 
some additional ``spectator'' branes.  If the mobile brane collides with any spectator brane during inflation then some of the stretched string states between these branes will become massless
at the moment when the branes are coincident \cite{beauty,trapped}, precisely mimicking the interaction (\ref{int}).  Thus, we expect particle 
production, IR cascading and the bump-like features described above to be a reasonably generic prediction of open string inflation.

\subsection{String Monodromy Models}

A concrete example of the heuristic scenario discussed in the last subsection is provided by the brane monodromy
and axion monodromy string theory inflation models proposed in \cite{monodromy1,monodromy2,monodromy3}.  
In the original brane monodromy model \cite{monodromy1} one considers type IIA string theory compactified on a nil manifold that is the
product of two twisted tori.  The metric on each of these twisted tori has the form
\begin{equation}
  \frac{ds^2}{\alpha'} = L_{u_1}^2 du_1^2 +  L_{u_2}^2 du_2^2 + L_x^2 (dx' + M u_1 du_2)^2
\end{equation}
where $x' = x- \frac{M}{2} u_1 u_2$ and $M$ is an integer flux number.  The dimensionless constants $L_{u_1}$, $L_{u_2}$ and $L_x$ 
determine the size of the compactification.

Inflation is realized by the motion of a D4-brane along the direction $u_1$ of the internal manifold.
The D4 spans our large 3-dimensions and wraps a 1-cycle along the direction $u_2$ of the internal space.  The size of this 1-cycle, in string units, is given by 
\begin{equation}
\label{LL}
  L = \sqrt{L_{u_2}^2 + L_x^2 M^2 u_1^2}
\end{equation}
Hence, the brane prefers to minimize its world-volume by moving to the location $u_1 = 0$ where this 1-cycle has minimal size.  This preference gives a potential to 
the D4-brane position which goes like $V \sim u_1$ in the large $u_1$ regime that is relevant for large field inflation.  

In \cite{trapped} it was shown that this scenario allows for the inclusion of a number of spectator branes stabilized at positions $u_{1} = i / M$ (with $i$ integer) 
along the inflationary trajectory.  As the mobile inflationary D4 rolls through these points particles (strings) will be produced and the resulting distribution of
features will look nearly identical to the simple multi-bump scenario studied in section \ref{subsec:multiple}.  To see this, let us now determine the distribution of 
features that is predicted from brane monodromy inflation.  The canonical inflaton $\phi$ can be related to the position of the mobile D4 as
\begin{equation}
\label{canonical}
  \phi = B\, u_1^{1/p}
\end{equation}
where $B$ is a constant with dimensions of mass that depends on model parameters.  Hence, the effective potential during inflation has the power-law form
\begin{equation}
\label{monodromy_pot}
  V(\phi) = \mu^{4-p} \phi^p
\end{equation}
For the simplest scenario described above one has $p = 2/3$.  However, the formulas (\ref{canonical},\ref{monodromy_pot}) still hold 
for the variant considered in \cite{monodromy1} with $p = 2/5$ as long as one replaces $u_1$ by a more complicated linear combination of coordinates.  These relations also hold for axion
monodromy models \cite{monodromy2} with $p=1$ and $u_1$ replaced by the axion, $c$, arising from a 2-form RR potential $C^{(2)}$ integrated over a 2-cycle $\Sigma_2$.  For \emph{all} models of the form (\ref{monodromy_pot})
the number of $e$-foldings $N$ from $\phi = \phi(N)$ to the end of inflation (which occurs at $\phi = p /\sqrt{2}$ when the slow roll parameter $\epsilon(\phi)= 1$) is given by
\begin{eqnarray}
  N &=& \frac{1}{2p} \frac{\phi^2(N)}{M_p^2} - \frac{p}{4} \nonumber \\
    &=& \frac{1}{2p} \frac{B^2}{M_p^2} u_1^{2/p} - \frac{p}{4} \label{monodromy_N}
\end{eqnarray}
Since the spectator branes are located at $u_1 = i / M$ the bursts of particle production must occur at times $N = N_i$ where
\begin{equation}
  N_i = \frac{1}{2p} \frac{B^2}{M_p^2} \left(\frac{i}{M}\right)^{2/p} - \frac{p}{4}
\end{equation}
The location $k=k_i$ of the $i$-th feature is defined, roughly, by the scale leaving the horizon at the moment $N=N_i$.  Hence, the distribution of features for brane/axion monodromy models is given by
\begin{equation}
\label{monodromy_distribution}
  \ln \left[\frac{k_i}{H} \right] \cong \tilde{B}^2 i^{2/p} - \frac{p}{4}
\end{equation}
with $p = 2/3$ or $p=2/5$ for brane monodromy and $p=1$ for axion monodromy.  In (\ref{monodromy_distribution}) the dimensionless number $\tilde{B}$ depends on model parameters.

Although the distribution of features (\ref{monodromy_distribution}) is not exactly the same as the evenly space distribution considered subsection \ref{subsec:multiple}, the two are essentially indistinguishable
over the range of scales which are probed observationally (corresponding to roughly 10 $e$-foldings of inflation).  The reason for this is simple: the inflaton is nearly constant during the first 10 $e$-foldings of inflation
and hence $\delta N \sim \delta \phi \sim \delta u_1$ within the observable region.  It follows that $k_i \cong e^{(i-1)\Delta} k_1$ to very good approximation for a huge class of models.  To see this more concretely in
the case at hand, let us compute $dN / du_1$  from (\ref{monodromy_N}).  It is straightforward to show that
\begin{equation}
  \frac{dN}{du_1} = \frac{1}{p^p} \frac{1}{\left[2\epsilon(\phi)\right]^{1-p/2}} \left( \frac{B}{M_p} \right)^{p}
\end{equation}
where
\begin{equation}
  \epsilon(\phi) \equiv \frac{M_p^2}{2}\left(\frac{V'}{V}\right)^2 = \frac{p^2}{2} \left(\frac{M_p}{\phi}\right)^2
\end{equation}
is the usual slow roll parameter.  Observational constraints on the running of the spectral index imply that $\epsilon(\phi)$ cannot change much over the observable 10 $e$-foldings of inflation.
Since $dN/du_1 \cong \mathrm{const}$ to very high accuracy it follows trivially that $N = N(u_1)$ is very close to linear and $k_i \cong e^{(i-1)\Delta} k_1$ as desired.

In the context of axion monodromy inflation models \cite{monodromy2} the multiple bump features discussed here will be complimentary to the oscillatory features described in \cite{monodromy3}
which result from the sinusoidal modulation of the inflaton potential by instanton effects.  If the bursts of particle production are sufficiently densely spaced, then signal from IR cascading may appear
oscillatory, however, it differs from the effect discussed in \cite{monodromy3} in both physical origin and also in functional form.

Let us now estimate the effective value of the couplings $g^2_i$ appearing in the prototype interaction (\ref{multiple}) that are predicted from the simplest brane monodromy model.  A complete calculation would involve
dimensionally reducing the DBI action describing the brane motion and requires knowledge of the full 10-dimensional geometry with the various embedded branes.  For our purposes, however, a simple heuristic estimate
for the collision of two D4-branes will suffice.  When $N$ D-branes become coincident the symmetry is enhanced from $U(1)^N$ to a $U(N)$ Yang Mill gauge theory.  The gauge coupling for this Yang Mills theory is given
by
\begin{equation}
  g_{\mathrm{YM}}^2 = \frac{g_s (2\pi)^2}{L}
\end{equation}
where $L$ is the volume of the 1-cycle that the D4 branes wrap and is given by (\ref{L}).  If the inflationary brane is at position $u_1$ and the $i$-th spectator brane is at $u_{1,i}$ then the distance between
the two branes is given by
\begin{equation}
  d^2 = \alpha'\, L_{u_1}^2 (u_1 - u_{1,i})^2
\end{equation}
The mass of the gauge bosons corresponding to the enhanced symmetry is
\begin{equation}
\label{boson_mass}
  M^2_i = g_{\mathrm{YM}}^2 \frac{d^2}{(2\pi)^2 (\alpha')^2} = \frac{g_s L_{u_1}^2\, (u_1 - u_{1,i})^2}{\alpha' \sqrt{L_{u_2}^2 + L_x^2 M^2 u_1^2}}
\end{equation}
To put this in the prototype form $M^2_i = g_i^2 (\phi - \phi_i)^2$ we must first convert to the canonical variable $\phi$ using the formula (\ref{canonical}) with $p = 2/3$ and 
\begin{equation}
  B = \frac{M^{1/2}}{6\pi^2}\frac{L_{u_1}L_x^{1/2}}{\sqrt{g_s \alpha'}}
\end{equation}
Next, we must Taylor expand the resulting equation about the minimum $\phi = \phi_i$.  We find
\begin{eqnarray}
  M_i^2 &\cong& g_i^2 (\phi- \phi_i)^2 + \cdots \\
  g_i^2 &=& \frac{16 g_s^2 \pi^4}{M L_x u_{1,i}} \frac{1}{\sqrt{L_{u_2}^2 + L_x^2 M^2 u_{1,i}^2}} \nonumber \\
        &=&  \frac{16 g_s^2 \pi^4}{ L_x i} \frac{1}{\sqrt{L_{u_2}^2 + L_x^2 i^2}} \label{g_monodromy}
\end{eqnarray}
where on the second line of (\ref{g_monodromy}) we have used the fact that $u_{1,i} = i / M$ (with $i$ integer) in the simplest models.  We see that the effective couplings $g_i^2$ become larger as the D4 unwinds during inflation.  
(The apparent divergence for $u_{1,i} = 0$ in the formula (\ref{g_monodromy}) is an artifact of the fact that the relation (\ref{canonical}) is not valid at small values of $u_1$.  This will not concern us here since inflation 
has already terminated at the point that our formulas break down.)

To compute the amplitude of the bump-like feature produced by brane monodromy inflation we should take into account also combinatorial factors.  When two branes become coincident the symmetry is enhanced from $U(1)^2$ to $U(2)$ 
so there are $2^2 - 2 = 2$ additional massless spin-$1$ fields appearing at the brane collision.  Thus, using equation (\ref{Ag}), the amplitude of the feature that will be imprinted in the CMB is
\begin{equation}
\label{Aeff}
  A_{i,\mathrm{eff}} = 2\times (2^2 - 2) \times \left[ 10^{-6}\cdot g_i^{15/4} \right]
\end{equation}
where the extra factor of $2$ counts the polarizations of the massless spin-$1$ fields.  This combinatorial enhancement can be much larger if the inflationary brane collides with a \emph{stack} of spectators.

The above discussion is predicated on the assumption that the original brane monodromy set-up 1 is supplemented by additional spectator branes.  This may seem like an unnecessary contrivance,
however, in order for this model to reheat successfully it may be \emph{necessary} to include spectator branes.  For example, with the reheating mechanism proposed in \cite{anke} semi-realistic 
particle phenomenology can be obtained by confining the standard model (SM) to a D6 brane which wraps the compact space.  In order to reheat into SM degrees of freedom we orient this brane so that its world-volume is 
parallel to the mobile (inflationary) D4.  In this case the end of inflation involves multiple oscillations of the D4 about the minimum of its potential.  At each oscillation the D4 collides
with the D6 and SM particles are produced by parametric resonance preheating \cite{KLS97}.  However, due to the periodic structure of the compactification, D4/D6 collisions will necessarily occur also \emph{during} 
inflation, leading to IR cascading features in the CMB.  

The timing of these D4/D6 collisions was computed in \cite{anke} for the minimal $p = 2/3$ brane monodromy model, assuming the same choices of parameters used in 
\cite{monodromy1}.  For this particular case there is only one collision (and hence one feature) during the first 10 $e$-foldings of inflation and the phenomenology is essentially the same as that considered in subsection 
\ref{subsec:single}.  What is the amplitude of this feature?  Assuming, again, the parameters employed in \cite{monodromy1} and noting that the first collision takes place at $i = 13$ \cite{anke} equation (\ref{g_monodromy})
gives $g_1^2 \cong 0.001$.  From (\ref{Aeff}) we find the effective amplitude of the feature to be $A_{1,\mathrm{eff}} / A_s \cong 0.01$.  This value is well within the observational bounds derived in subsection 
\ref{subsec:single}, however, it should be within reach of future missions \cite{forecast}.

We stress that the conclusions in the previous paragraph apply \emph{only} for the particular choice of model parameters employed in \cite{monodromy1}.  There exist other consistent parameter choices for which the 
simplest brane monodromy 
model predicts a much higher density of features with much larger amplitude.

Note that both brane and axion monodromy models may be used to realize trapped inflation \cite{trapped}.  Here we are restricting ourselves to 
the large-field regime where the potential $V = \mu^{4-p}\phi^p$ is flat enough to drive inflation without the need for trapping effects.
For a given choice of parameters one should verify that this classical potential dominates over the quantum corrections from particle production.

\subsection{A Supersymmetric Model}

Another microscopic realization of multiple bursts of particle production and IR cascading during inflation which does not
rely on string theory can be obtained from the so-called ``distributed mass'' model derived in \cite{berrera} with warm inflation \cite{warm}
in mind, however, the theory works equally well for our scenario.  This model is based
on $\mathcal{N} =1$ global SUSY and allows for the inclusion of multiple points along the inflationary trajectory where both scalar degrees of 
freedom and also their associated fermion super-partners become massless.  The distribution of features in this set-up is essentially arbitrary.

\section{Nongaussianity from Particle Production and IR Cascading}
\label{sec_ng}

We have seen that the dynamics of particle production, rescattering and IR cascading during inflation in the model (\ref{L}) leads to a bump-like
contribution to the primordial power spectrum of the observable cosmological fluctuations.  However, this dynamics must also have a nontrivial
impact on nongaussian statistics, such as the bispectrum.  Indeed, it is evident from the analytical formalism presented in sections \ref{sec_analytical}
and \ref{sec_metric} that the inflaton fluctuations $\delta\phi$ generated via the rescattering diagram Fig.~\ref{Fig:diag} may be significantly nongaussian.
(This is evident since the particular solution of equation (\ref{delta_phi}) is bi-linear in the Gaussian field $\chi$.)  In this section we characterize this
nongaussianity.  The results of this section appeared for the first time in \cite{ir,pptheory,ppNG}.  

\subsection{Probability Density Function and Cummulants}

As discussed in the introduction, nongaussianity is often characterized by the size, shape and running of the bispectrum $B(k_i)$.
A popular measure of the size of the nongaussianity is the magnitude of $B(k_i)$ on equilateral triangles.  This approach has the
advantage of being straightforward, however, it may give misleading results when one wishes to compare bispectra with
different shapes \cite{shellard} or scaling behaviour.  This is particularly true in models, such as (\ref{L}), where the bispectrum is large
only for triangles with a characteristic size and where the nongaussian part of $\zeta$ is uncorrelated with the gaussian part.  A more meaningful 
single number to compare between models is the skewness (defined below), which contains information about the bispectrum integrated 
over all scales (up to some UV smoothing scale) and all shape configurations.  (See also \cite{shellard2} for a related discussion and alternative methodology.)

In what follows it will be useful to define a normalized curvature perturbation
\begin{equation}
  \xi \equiv \frac{\zeta}{\sigma_\zeta}
\end{equation}
where $\zeta$ is the usual gauge invariant curvature perturbation and $\sigma_\zeta$ is the variance
\begin{equation}
\label{variance}
  \sigma_\zeta^2 = \langle \zeta^2\rangle = \int \frac{d^3 k_1}{(2\pi)^{3/2}}\frac{d^3 k_2}{(2\pi)^{3/2}} \langle \zeta_{\bf k_1} \zeta_{\bf k_2} \rangle
\end{equation}
If $\zeta$ is generated by the quantum vacuum fluctuations of the inflaton field then we have $\sigma_\zeta \cong 10^{-9/2}$.  

Let us define the probability density function (PDF), $P(\xi)$, as the probability that the (normalized) curvature perturbation has fluctuations
of size $\xi$.  The $n$-th central moment of the PDF is simply
\begin{equation}
  \langle \xi^n \rangle = \int_{-\infty}^{+\infty} \xi^n P(\xi) d\xi
\end{equation}
The $n$-th dimensionless cummulant $\hat{\kappa}_n$ is the connected $n$-point function.  For $\langle\xi\rangle =0$ the first few non-vanishing cummulants are:
\begin{eqnarray}
  \hat{\kappa}_3 &=& \langle \xi^3\rangle \\
  \hat{\kappa}_4 &=& \langle \xi^4 \rangle - 4 \langle \xi^2 \rangle^2
\end{eqnarray}
For a gaussian PDF we have $\hat{\kappa}_n = 0$ for $n = 3,4, \cdots$, hence the cummulants provide a measure of nongaussianity.  

In the case that $\zeta$ (and hence $\xi$) is a free field then the central limit theorem implies that the PDF is gaussian
\begin{equation}
\label{gauss_PDF}
  P(\xi) d\xi = \frac{d\xi}{\sqrt{2\pi}} \, e^{-\xi^2 / 2}
\end{equation}
This expression admits a simple generalization to the case where $P(\xi)$ is close to (but not exactly) gaussian.
This generalization is the Edgeworth expansion
\begin{eqnarray}
  P(\xi) d\xi &\cong&  \frac{d\xi}{\sqrt{2\pi}} \, e^{-\xi^2 / 2}\, \left[ 1 + \frac{\hat{\kappa}_3}{6} H_{3}(\xi) + \cdots \right] \label{edgeworth}
\end{eqnarray}
(See \cite{shandera} for a review and \cite{hidalgo} for an alternative derivation.)
In (\ref{edgeworth}) $H_{n}(\xi)$ denotes the Hermite polynomials of order $n$.  Note that the result (\ref{edgeworth}) is an expansion in cummulants.  This expression 
provides an accurate approximation of the true PDF provided the cummulants are well ordered in the sense that 
\begin{equation}
\label{ordering}
  1 \gg |\hat{\kappa}_3| \gg |\hat{\kappa}_4| \gg \cdots
\end{equation}
See \cite{structure,perturbative} for more discussion on the structure of the correlation functions and the validity of cosmological perturbation theory.

Finally, it remains to relate the cummulants to the correlation functions (such as the bispectrum) of the primordial curvature perturbation.
We can write the dimensionless skewness as an integral over the 3-point correlation function \cite{shandera,hidalgo}
\begin{equation}
  \hat{\kappa}_3 = \frac{1}{\sigma^3_\zeta} \int \frac{d^3k_1}{(2\pi)^{3/2}}\frac{d^3k_2}{(2\pi)^{3/2}}\frac{d^3k_3}{(2\pi)^{3/2}}
  \langle \zeta_{\bf k_1} \zeta_{\bf k_2} \zeta_{\bf k_3} \rangle_{\mathrm{c}}
  \label{skewness}
\end{equation}
A similar expression may be derived for the dimensionless kurtosis, $\hat{S}_4$.  From (\ref{skewness}) it is clear that the
skewness provides a measure of the magnitude of the bispectrum integrated over all triangles.  Hence, this provides a reasonable
single number to compare the size of nongaussianity between different models.  Moreover, the skewness (and higher cummulants)
are of observational interest since these determine the probability of rare fluctuations \cite{dalal,shandera,mcdonald,afshordi,lss1,lss2,lss3}.  
Observables such as the abundance of collapsed objects may therefore be used to constrain $\hat{\kappa}_3$, $\hat{\kappa}_4$, etc.

We have reviewed the derivation of the Edgeworth expansion 
(\ref{edgeworth}) only to illustrate that the dimensionless skewness (\ref{skewness}) encodes information about the bispectrum integrated 
over all momenta.  However, our actual calculation of the PDF in the next subsection will be fully non-perturbative and hence does \emph{not} rely on the 
validity of (\ref{edgeworth}) or (\ref{ordering}).

\subsection{Magnitude of the Nongaussianity from IR Cascading}

We have argued in the last subsection that the moments of the PDF provide a useful tool for quantifying the size
of nongaussianity and drawing comparisons between microscopic models whose bispectra may have very different shapes
and scaling properties.  Our goal now is to compute the skewness and kurtosis generated by particle production, rescattering and
IR cascading in the model (\ref{L}).  We will then compare this to the skewness that would be generated by more familiar constructions,
such as the local model $\zeta = \zeta_g + \frac{3}{5} f_{NL} \zeta_g^2$.

We extract the PDF of $\delta\phi$ from our HLattice simulations by measuring the fraction of the simulation box which contains the fluctuation field $\delta \phi$ at a particular 
value.  Notice that this approach is completely nonperturbative: it does not rely on the validity of the Edgeworth expansion, nor does it assume anything about the size
or ordering of the cummulants.  This procedure implicitly puts a IR cut-off at the box size $L$ and a UV cut-off at the lattice spacing, $\Lambda^{-1}$.  
Since the nongaussian effects in our model are strongly localized in Fourier space, our quantitative results are largely insensitive to $L$ and $\Lambda$.

\begin{figure}[htbp]
\bigskip \centerline{\epsfxsize=0.5\textwidth\epsfbox{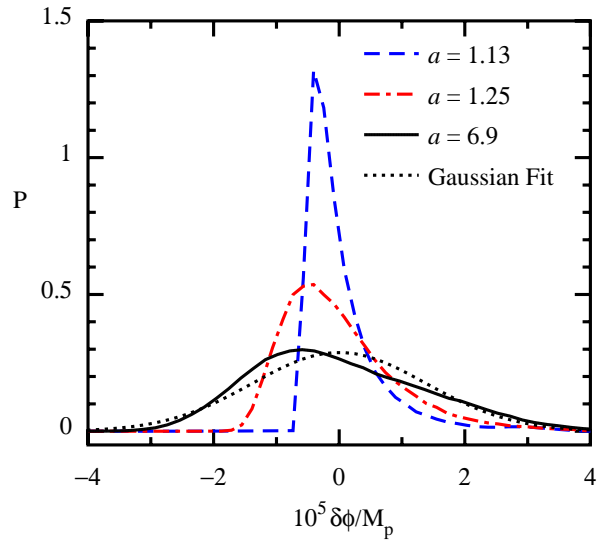}}
\caption{The PDF of the inflaton fluctuations generated by rescattering and IR cascading, at a series of different values 
of the scale factor, $a$.  The dotted black curve shows a Gaussian fit at late times and we have 
normalized the scale factor so that $a=1$ at the moment when particle production occurs.   For illustration, we have chosen $g^2=0.1$ and 
a standard chaotic inflation potential $V(\phi) = m^2\phi^2/2$.}
\label{Fig:onepi}
\end{figure}

In Fig.~\ref{Fig:onepi} we plot our numerical result for the PDF of the inflaton fluctuations generated by rescattering and IR cascading.  In order to make
the physics of inflationary particle production clear, we have subtracted off the contribution coming from the usual vacuum fluctuations of the inflaton.  That
is, the PDF in Fig.~\ref{Fig:onepi} is associated only with the contribution $\delta\phi_{\mathrm{resc}}$ in equation (\ref{hom+par}).

We can understand physically the behaviour of PDF plotted in Fig.~\ref{Fig:onepi}. 
Shortly after the initial burst of particle production the inflaton perturbations $\delta\phi$ are extremely nongaussian, due to the sudden 
appearance of the source term $J \propto \chi^2$ in the equation of motion (\ref{inf_eqn}).  Very quickly, in less than an $e$-folding, nonlinear 
interactions begin to drive the system towards gaussianity.  
A very similar behaviour has been observed in lattice simulations of out-of-equilibrium interacting scalar fields during preheating 
\cite{preheat_pdf1,preheat_pdf2}.  In the case of rescattering during preheating, the system will eventually become gaussian
when the fields thermalize.  However, in our case the universe is still inflating.
As a result, nongaussian inflaton fluctuations generated by rescattering are stretched out by the quasi-de Sitter expansion and must 
freeze once their wavelength crosses the Hubble scale.  Hence, at late times the PDF does not become completely gaussian, but rather
freezes-in with some non-trivial skewness.  Within a few $e$-foldings from the moment of particle production the time evolution of the
PDF has become completely negligible.

\begin{figure}[htbp]
\bigskip \centerline{\epsfxsize=0.5\textwidth\epsfbox{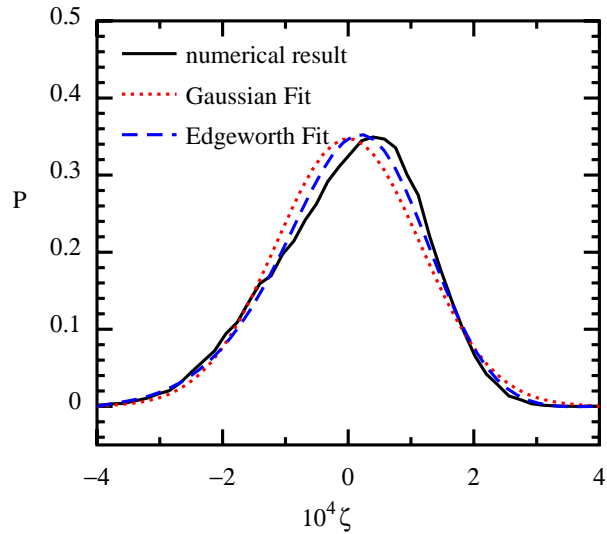}}
\caption{The PDF of the total curvature fluctuation, $\zeta$, at late times (well after all relevant modes have crossed the horizon
and frozen).  The solid black curve is the exact result from our HLattice simulations and the dotted red curve is a gaussian fit.  We have also plotted the 
leading correction to the gaussian result in the Edgeworth expansion, given explicitly by equation (24).  For illustration, we have chosen $g^2=0.1$ and 
a standard chaotic inflation potential $V(\phi) = m^2\phi^2/2$.}
\label{Fig:truePDF}
\end{figure}

In order to characterize the nongaussianity of the observable primordial fluctuations, we would like to construct the PDF for the curvature perturbation $\zeta$, 
including \emph{both} the contributions from the vacuum fluctuations of the inflaton and also from rescattering.
To this end, we construct $\zeta$ using the naive relation $\zeta = -\frac{H}{\dot{\phi}}\delta\phi$
and take into account both contributions to $\delta\phi$ in equation (\ref{hom+par}).  In Fig.~\ref{Fig:truePDF} we plot the full PDF obtained
in this manner, evaluated at very late times, well after all relevant modes have crossed the horizon and become frozen.

Given our numerical results for the PDF of the total observable curvature fluctuation, that is Fig.~\ref{Fig:truePDF}, it is straightforward to compute dimensionless 
cummulants such as the skewness ($\hat{\kappa}_3$) and kurtosis ($\hat{\kappa}_4$) for various values of the coupling 
$g^2$.  We have summarized our results in Table \ref{cummulant_table}.  Note that for $g^2=0.01$ the kurtosis $\hat{\kappa}_4$ is too small to be 
measured accurately from our HLattice simulations.

In order to give some sense of the magnitude of the nongaussianity from particle production we have also computed an 
``equivalent $f_{NL}^{\mathrm{local}}$'' defined by $5 \hat{\kappa}_3 / (18 \sigma_\zeta)$ where the variance is
$\sigma_\zeta \equiv \langle \zeta^2 \rangle^{1/2} \sim 10^{-9/2}$.  For a given $g^2$ this effective $f_{NL}^{\mathrm{local}}$
is the magnitude of $f_{NL}$ which would be necessary to reproduce the skewness $\hat{\kappa}_3$ of the IR cascading PDF using a local ansatz
$\zeta = \zeta_g + \frac{3}{5} f_{NL} \left[ \zeta_g^2 - \langle\zeta_g^2\rangle \right]$.\footnote{Our sign conventions for $f_{NL}$ are 
consistent with WMAP \cite{WMAP7}.  See \cite{shandera} for a discussion of various conventions employed in the literature.}

\begin{table}[htbp]
{\caption{Moments of the Probability Distribution Function}\label{cummulant_table}}
  \begin{center}
  \begin{tabular}{llll} 
  \hline
  $g^2$ & skewness  & kurtosis &  ``equivalent'' \\
            &  ($\hat{\kappa}_3$) & ($\hat{\kappa}_4$) &  $f_{NL}^{\mathrm{local}}$ \\
   \hline
  $1$  & $-0.51$ & $0.2$ & $-4500$ \\
  $0.1$ & $-0.49$ & $-0.1$ & $-4300$ \\
  $0.01$ & $-0.006$ & $<\mathcal{O}(10^{-3})$ & $-53$\\
  \hline
  \end{tabular}
  \end{center}
\end{table}

From Table \ref{cummulant_table} we see that IR cascading during inflation can generate significant nongaussianity.  Even taking
$g^2 = 0.01$ (which is compatible with cosmological data for any choice of $\phi_0$ \cite{ppcons}) we still obtain a skewness $\hat{\kappa}_3=-0.006$,
which is the same value that would be produced by a local model with $f_{NL} \sim -53$.  This ``equivalent'' local nongaussianity
is comparable to current observational bounds, and is well within the expected accuracy of future missions.  This suggests that nongaussian
features from particle production during inflation might be observable for reasonable values of $g^2$.  

The ``equivalent'' $f_{NL}^{\mathrm{local}}$ values presented in Table \ref{cummulant_table} must be interpreted with care.  We have included this
information only to give a heuristic sense of the magnitude of nongaussianity in our model.  It must be stressed that the PDF plotted
in Fig.~\ref{Fig:truePDF} is quite different from the analogous result for local-type nongaussianity.  For example, the value of the kurtosis
(and higher moments) are different, as is the ordering of the cummulants.  Moreover, we should emphasize that observational bounds on $f_{NL}^{\mathrm{local}}$ 
\emph{cannot} be directly applied to our model since the bispectrum in our case is uncorrelated with the vacuum fluctuations and is far from scale invariant.  
A detailed study of the detectability of nongaussianity from particle production will be the subject of a upcoming publication \cite{ppNG}.

Depending on the value of $\phi_0$, the model (\ref{int}) may lead to a variety of observable signatures.  As discussed previously, $\phi_0$
controls the location of the feature in the primordial power spectrum (\ref{P_fit}).  Nongaussian effects are also localized near the same characteristic
scale, $k_{\mathrm{IR}}$.  If $k_{\mathrm{IR}}$ corresponds to scales relevant for CMB experiments, then we predict a bump-like feature in 
the primordial power spectrum, $P_{\zeta}(k)$, and an associated feature in the bispectrum $B_{\zeta}(k_i)$ with an unusual shape (that will
be discussed the next section).  A key question is whether the nongaussian feature can be observably large in a regime where the power
spectrum feature is small enough to be compatible with current observations.  Preliminary results are encouraging: for $g^2=0.01$ the power spectrum
is consistent with all observational data \cite{ppcons} while the skewness of the PDF is rather large.  A detailed investigation will require a simple, 
separable template for the bispectrum and will be discussed in a future publication \cite{ppNG}.

On the other hand, we could imagine a scenario where the feature from IR cascading shows up on smaller scales, relevant for Large Scale Structure 
(LSS) experiments \cite{shandera,dalal,mcdonald,afshordi}.  
In this case our scenario could be probed using higher order correlations of LSS probes (such as the galaxy bispectrum) or the
abundance of collapsed objects (or voids).  The latter possibility is interesting since the cluster/void abundance is determined the tails of the PDF
and may be insensitive to the detailed shape of the bispectrum.  Quantitative predictions for observable cluster/void abudances require the PDF of the
evolved density field, smoothed on some relevant scale \cite{ppNG}, rather than the PDF of the primordial curvature perturbation (which is plotted in 
Fig.~\ref{Fig:truePDF}).  However, we can nevertheless describe the qualitative signatures which should be expected.  Our model robustly predicts
a negative skewness for both the curvature perturbation, $\zeta$, and the density field, $\delta\rho/\rho$.  Hence, we should expect a \emph{decrease}
in the abundance of the largest collapsed objects and an \emph{increase} in the abundance of the largest voids \cite{void,lss_rev}.  Owing to the localized nature
of the bispectrum feature, we expect that this effect should show up only when the density field is smoothed on a scale close to $k_{\mathrm{IR}}$.

It is worth mentioning that recent weak lensing measurement of the dark matter mass of the high-redshift galaxy cluster XMMUJ2235.3-2557 \cite{cluster} 
have been construed as a possible hint of nongaussian initial conditions \cite{lss2}.  Unfortunately, our model does not produce the correct sign of 
skewness to explain such observations.  


\subsection{Shape of the Bispectrum}

In section \ref{sec_analytical} we presented a schematic discussion of how to compute analytically the primordial bispectrum from IR cascading,
equation (\ref{B_phi}).  In \cite{pptheory} this expression is evaluated analytically.  We find that the bispectrum (\ref{B_phi}) from particle
production has a unique shape that has not been considered in previous literature.  
In order to describe this novel shape we would like to first factor out that strong
overall dependence of $B_\phi(k_i)$ on the size of the triangle.  To this end we define a ``shape function'' $S(k_i)$ in terms of the bispectrum
as follows
\begin{equation}
\label{S}
  S(k_i) = N^{-1} (k_1 k_2 k_3)^2 B(k_i)
\end{equation}
where $B(k_i)$ is related to $B_\phi(k_i)$ by equation (\ref{B_zeta}) and $N$ is a normalization factor to be discussed shortly.  
The function $S(k_i)$ has the advantage that the strong $k^6$ running of the bispectrum is extracted.  Hence, any residual scaling
behaviour displayed by $S(k_i)$ must be a result of nonlinear interactions. 

Since we expect the 3-point correlation function to be of order $P_\zeta^2(k)$ a natural choice of normalization is $N = (2\pi)^4 \mathcal{P}_\zeta^2$
where $\mathcal{P}_\zeta^{1/2} = 5\times 10^{-5}$ is the usual amplitude of the scale invariant fluctuations from inflation.  With this choice
of normalization our function $S(k_i)$ coincides with the quantity $\mathcal{G}(k_i) / (k_1 k_2 k_3)$ which was used to study localized nongaussian
features from models with steps in the inflaton potential in \cite{chen1,chen2}.  

\begin{figure}[tbp]
\begin{center}
\includegraphics[width=1.6in]{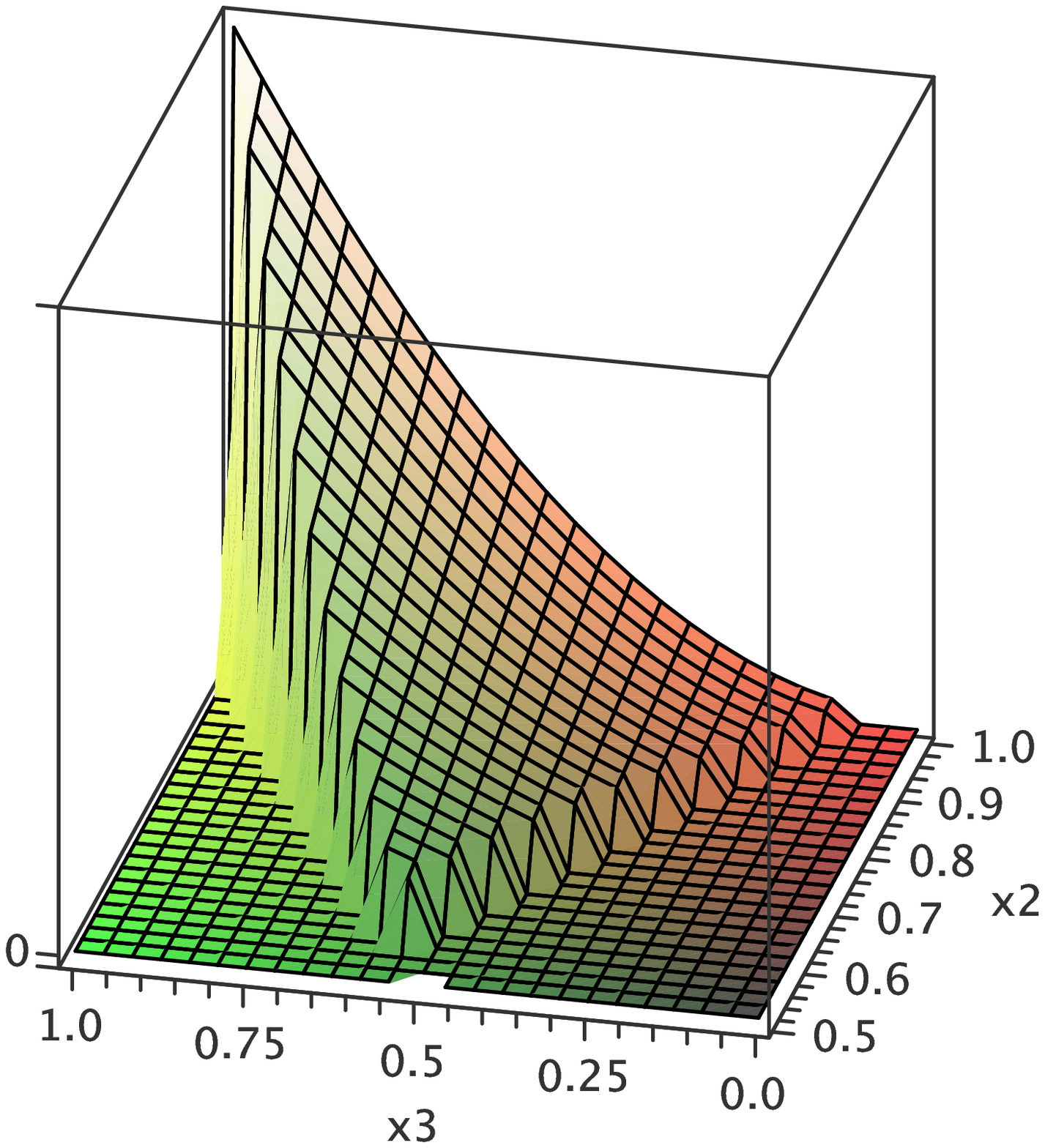}
\includegraphics[width=1.7in]{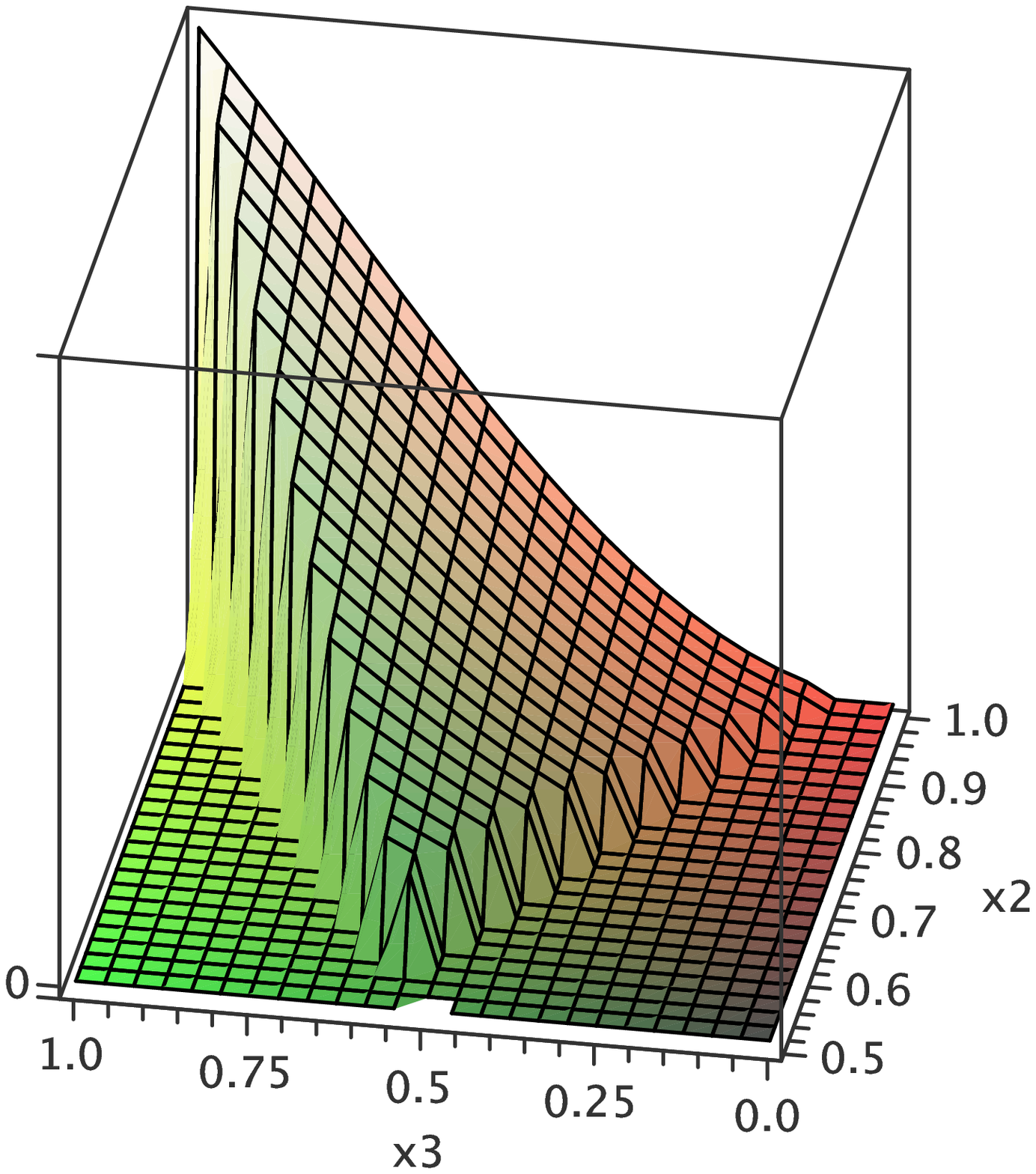}
\includegraphics[width=1.6in]{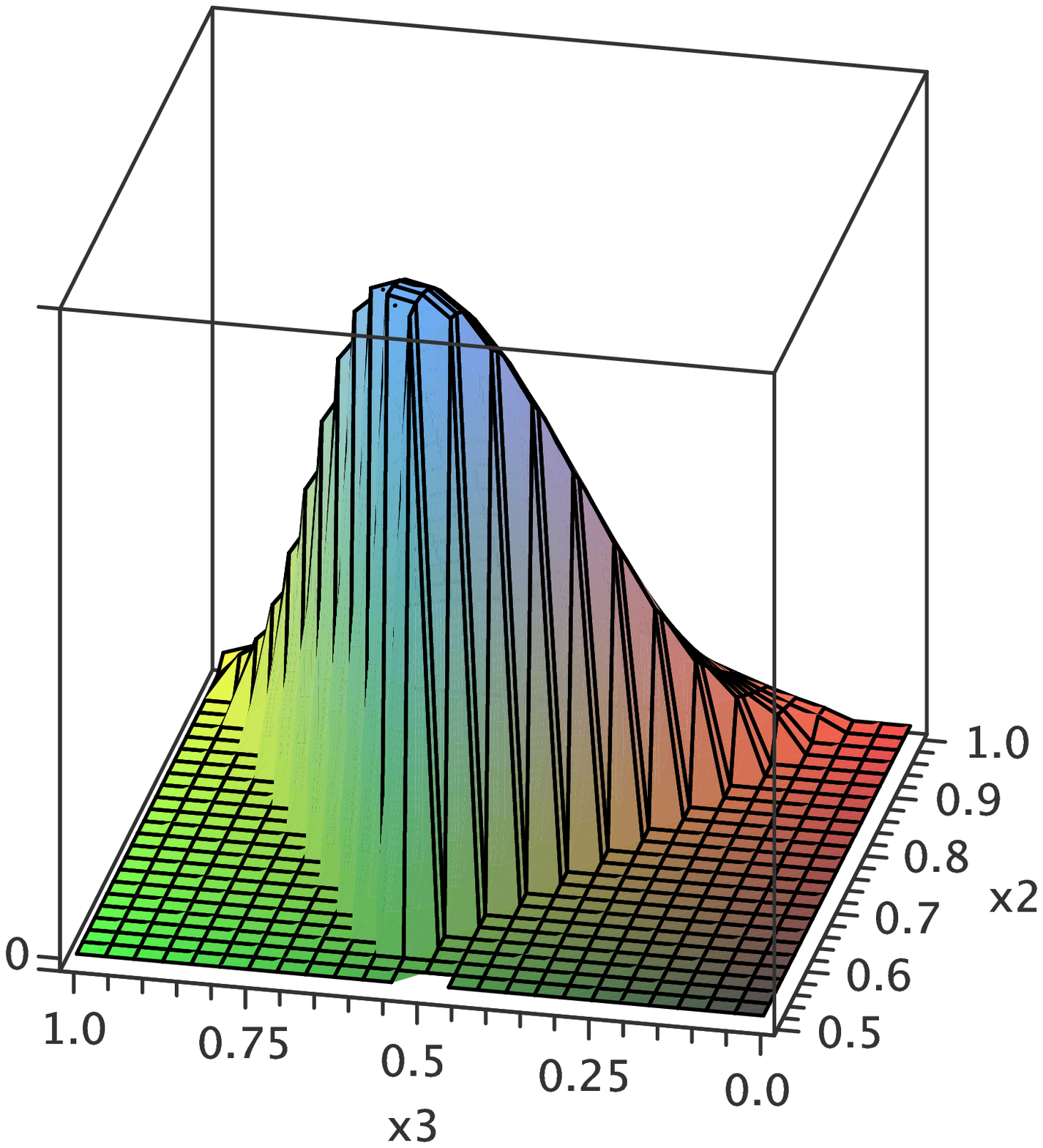}
\includegraphics[width=1.6in]{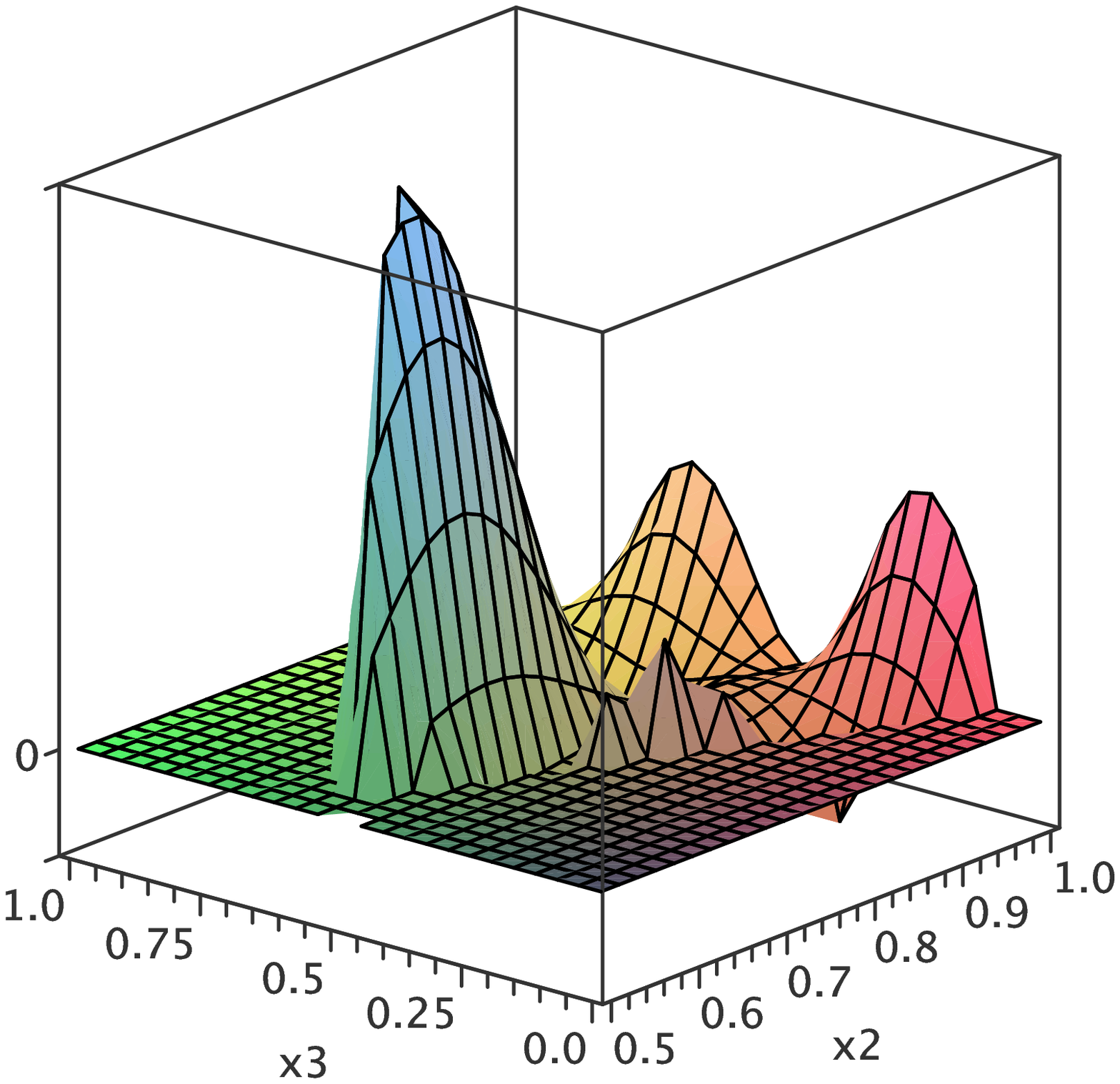}
\caption{The shape function $S(k,k x_2, k x_3)$, defined by (\ref{S}), as a function of the dimensionless quantities $x_2,x_3$ which parametrize
the shape of the triangle.  The upper left panel corresponds to $k = e^{-1}k_{\mathrm{bump}}$, the upper right panel is $k=k_{\mathrm{bump}}$,
the lower left panel is $k=e^{+1} k_{\mathrm{bump}}$ and the lower right panel is $k=e^{+2}k_{\mathrm{bump}}$.
In the IR 
($k \leq k_{\mathrm{bump}}$) 
the shape of the bispectrum is similar to the equilateral shape, however, there is also some support on 
flattened triangles near 
$k \sim e^{+1} k_{\mathrm{bump}}$.  
At larger values of $k$ the shape is unlike any other template proposed 
in the literature.
}\label{Fig:bispectrum_shape}
\end{center}
\end{figure}

Symmetry of the bispectrum under permutations of momenta implies that we can focus only on the region $k_1 \geq k_2 \geq k_3$, to avoid
counting the same configuration twice.  Moreover, the triangle inequality implies that $1-\frac{k_2}{k_1} \leq \frac{k_3}{k_1}$.  Therefore
we can completely specify the shape of the bispectrum for a given size of triangle $k$ by plotting $S(k,kx_2,kx_3)$ in the region
$x_3 \leq x_2 \leq 1$ and $1-x_2 \leq x_3$.  Because our bispectrum is very far from scale-invariant, it follows that this shape function
is sensitive to the choice of $k$.  Therefore, in Fig.~\ref{Fig:bispectrum_shape} we choose several representative choices: 
$\ln (k / k_{\mathrm{bump}}) = -1,0,1,2$.  


We see that a rich array of shapes are possible: for $k \lsim k_{\mathrm{bump}}$ the bispectrum is qualitatively similar to the equilateral model,
however, at slightly larger $k$ there is considerable support on flattened triangles also.  
Note that for $k \gsim 7.4\, k_{\mathrm{bump}}$ the shape of the bispectrum is extremely unusual and is not easily comparable to any shape
that has been proposed in previous literature.

We find that the bispectrum from IR cascading is, to a first approximation, factorisable in the sense that $B(k_1,k_2,k_3) \cong \prod_{i=1}^3 F(k_i)$ where 
$F(k) \sim C_1 k^3 e^{-C_2 k^2}$ for some constants $C_1$ and $C_2$.  This separable form is important since it permits fast algorithms for forecasting, data analysis 
and simulations \cite{factorizable}.  

In this section we have attempted to characterize the nongaussian signature associated with inflationary particle production.  This signature is
rather unique in the literature.  Our model predicts uncorrelated localized nongaussian features with a unique shape of bispectrum.
We have quantified the size of this effect by studying the dimensionless cummulants (such as the skewness) and have argued that this 
nongaussianity can be significant, depending on $g^2$.  We leave a detailed study of the observational constraints on such nongaussianities to future studies \cite{ppNG}.

\subsection{Multiple Bursts of Particle Production}

Note that our discussion of nongaussianity generalizes easily to the case where there are multiple points $\phi_i$ ($i=0, \cdots, n$) along the inflaton
trajectory where new degrees of freedom $\chi_i$ become massless, such as the model (\ref{multiple}).  Such a construction may be
quite natural in the context of brane/axion monodromy inflation \cite{trapped,monodromy1,monodromy2,monodromy3}.  In this case
the nongaussian features in $B(k_i)$ from each burst of particle production may superpose to generate a broad-band signal.  Such
a construction may (but need not) be associated with trapped inflation \cite{trapped}.  Depending
on the spacing of the points $\phi_i$ and the couplings $g_i^2$ a rich variety of nongaussianities may be possible.  
In this case that the point $\phi_i$ are sufficiently densely spaced, we expect a nongaussian signal which is close to equilateral but also
has some support on flattened triangles, consistent with the analysis of \cite{trapped,terminal}.  We leave a detailed discussion to future studies.

\section{Conclusions}
\label{sec_conc}

In this review we have considered the possibility that some non-inflaton (iso-curvature) particles were produced during 
the observable range of $e$-foldings of inflation.  Inflationary particle production might occur as a result of a phase transition, parametric
resonance or other non-adiabatic processes.  In order to illustrate the basic physics of inflationary particle production we restricted our analysis to a simple prototype
model with coupling $\frac{g^2}{2}(\phi-\phi_0)^2\chi^2$ between the inflaton $\phi$ and iso-inflaton $\chi$.  However, we expect that our qualitative 
results will apply also to SUSY models, gauged interactions, higher spin iso-inflaton fields and phase transitions.  

Models of the type we study have attracted considerable interest recently in connections with trapped inflation, trans-Planckian effects and observable features/nongaussianity
in the primordial curvature fluctuations from inflation.  Moreover, such models are quite natural from the microscopic perspective and may be
obtained in popular models of open string inflation, such as brane/axion monodromy.

We have shown that inflationary particle production in the model $\frac{g^2}{2}(\phi-\phi_0)^2\chi^2$ leads to a new mechanism for generating
cosmological fluctuations.  This mechanism is qualitatively different from previous proposals in that we do not rely on the quantum vacuum
fluctuations of light fields during inflation.  Rather, the scenario involves the production of massive $\chi$ particles during inflation, which subsequently
rescatter off the slow roll condensate $\phi(t)$ to emit bremsstrahlung radiation of light inflaton fluctuations $\delta\phi$.  We have studied this dynamics
using classical lattice field theory simulations, analytical QFT computations and also second order cosmological perturbation theory.
All of these approaches yield consistent results.  We have found that rescattering proceeds with a time scale short compared to the expansion
time.  Moreover, the emission of long-wavelength inflaton fluctuations is very energetically inexpensive.  The combination of these two effects leads to a 
rapid build-up of power in IR inflation fluctuations shortly after the moment of particle production.  This dynamical process is called IR cascading.  

Our numerical and analytical studies of rescattering and IR cascading during inflation may have relevance for trapped inflation, preheating, moduli trapping and also 
non-equilibrium QFT more generally.  For instance, we have seen that, even with a small number of out-of-equilibrium $\chi$ particles, multiple rescatterings can nevertheless
generate long-wavelength $\delta\phi$ fluctuations with huge occupations numbers.  We have also observed, for the first time, the dynamical approach to the scaling regime discovered in \cite{B1,B2}.

IR cascading during inflation leads to observable features in the primordial cosmological fluctuations.  In particular, this process generates a bump-like
contribution to the primordial scalar power spectrum.  This signature is very different from what would be obtained in a model with transient violation of 
slow roll during inflation (such as a step-like feature in $V(\phi)$), contrary to some claims in the literature.

We have studied the observational constraints on bump-like features from inflationary particle production during inflation.  We found that relatively large
distortions, of order $10\%$ of the usual scale invariant vacuum fluctuations, are compatible with current data.  We have derived observational bounds on the coupling
$g^2$ for a given $\phi_0$, which play a crucial role in determining the detectability of nongaussianity from particle production.
Our observational bounds on particle production during inflation have implications for brane/axion monodromy inflation models and other microscopic constructions.

IR cascading also has a nontrivial impact on nongaussian statistics, such as the bispectrum.  The model $\frac{g^2}{2}(\phi-\phi_0)^2\chi^2$ leads to a very novel nongaussian signature: 
uncorrelated, localized nongaussian features with a unique shape of bispectrum.  For reasonable values of the coupling, $g^2 \lsim 0.01$, this new kind of nongaussianity may be detectable
in future missions.

The nongaussian signature predicted by inflationary particle production is rather unusual, as compared to other models of inflation which are frequently studied  in the literature.  However, the underlying field theory
description is extremely simple and rather generic from the low-energy perspective.  In order to obtain observable nongaussianity it was not
necessary to fine-tune the inflaton trajectory or appeal to re-summation of an infinite series of high dimension operators.  Indeed,
the only ``tuning'' which is required for our signal to be observable is the requirement that $\phi=\phi_0$ during the observable range of $e$-foldings.

There are a variety of directions for future studies.  From the theoretical perspective, it would be interesting to explicitly generalize our results to
more complicated models with particle production during inflation (such as SUSY models, fermion $\chi$ fields and phase transitions).  There are
also a wide range of interesting phenomenological possibilities.  Varying the location of the feature we can have a variety of possible signatures for
the CMB and LSS.  We expect that IR cascading will also have implications for the spectrum of gravity waves from inflation and also primordial black
holes.  We can also imagine superposing multiple bursts of particle production to obtain an even richer variety of signatures.  We leave these possibilities 
for future investigation.

\section*{Acknowledgments}

This work has benefited considerably from the interactions with a number of people.  Thanks to
L.~Kofman and D.~Pogosian for collaboration on the original paper \cite{ir} which instigated this work.  
I am especially grateful to Z.~Huang for collaboration on the works \cite{ir,ppcons} and also for numerous 
discussions and help with numerical simulations and several figures.  Finally, thanks also to T.~Battefeld, C.~Burgess, 
J.~Cline, N.~Dalal, H.~Firouzjahi, L.~Hoi, I.~Huston, K.~Malik, P.~McDonald, A.~E.~Romano, M.~Sasaki, D.~Seery, S.~Shandera and A.~Tolley 
for helpful comments, discussions and input at various stages during the completion of this project.  

\renewcommand{\theequation}{A-\arabic{equation}}
\setcounter{equation}{0}
\section*{APPENDIX A: Backreaction Effects}
\label{appA}

\renewcommand{\thesubsection}{A.\arabic{subsection}}
\setcounter{subsection}{0}

Once $\phi$ rolls past the massless point $\phi = \phi_0$ a gas of $\chi$ particles is produced with occupation number 
given by (\ref{n_k}).  This gas costs energy, which must be drained from the condensate $\phi(t)$.  Hence,
in order for the system to conserved energy, the inflaton must slow down slightly.  As discussed in \cite{KLS97,beauty,sasaki,chung,ir},
this slowing down can be studied using the mean field equation (\ref{mean}).  Using the solutions (\ref{chi_soln}) to compute the vacuum
average $\langle\chi^2\rangle$ we find
\begin{equation}
\label{mean2}
  \ddot{\phi} + 3 H \dot{\phi} + V_{,\phi} + g \frac{k_\star^3}{(2\pi)^3} \frac{\Theta(t)}{a^3(t)} \frac{\left(\phi - \phi_0 \right)}{|\phi-\phi_0|} = 0
\end{equation}
The step function $\Theta(t)$ reflects the fact that the impact of particle production is felt only \emph{after} $\phi$ passes through $\phi_0$
and $a^{-3}$ comes from the volume dilution of non-relativistic particles.  The final term in (\ref{mean2}) may be interpreted as a 
quantum correction to the effective force due to particle production.  (Note that we have implicitly subtracted the Coleman-Weinberg
potential, which may be justified by the assumption of softly broken SUSY.)  

The solutions of (\ref{mean2}) display the expected slowing-down behaviour and have been studied analytically in \cite{beauty,sasaki,chung}.
In \cite{ir} this slowing-down was studied using inhomogeneous lattice field theory simulations and the results were found to be compatible
with the mean field treatment (see Fig.~\ref{Fig:dotphi}).
Here, we consider simple energetic arguments in order to clear up some common misconceptions.  The effective inflaton potential, including 
the effects of particle production, is
\begin{equation}
  V_{\mathrm{eff}} = V(\phi) + g \Theta(t) \frac{k_\star^3}{(2\pi)^3} \frac{|\phi-\phi_0|}{a^3}
\end{equation}
For $t \gsim k_\star^{-1}$ the $\chi$ particles are non-relativistic and their energy density is dominated by potential (rather than kinetic) energy.  Hence, we have
\begin{equation}
  \rho_\chi \cong  g \frac{k_\star^3}{(2\pi)^3} \frac{|\phi-\phi_0|}{a^3} \cong \frac{k_\star^5}{(2\pi)^3 H}\, N e^{-3 N}
\end{equation}
with $N = H t$ the number of $e$-foldings measured from the moment when $\phi = \phi_0$.  Shortly after particle production 
this energy density grows, 
corresponding to the fact that as $\phi$ moves away from $\phi_0$ the $\chi$ particles become ever more massive.  However, this
growth in the energy density cannot continue forever.  At $N=1/3$ the energy density $\rho_\chi$ peaks and at later times it decays
exponentially as $e^{-3N}$, corresponding to the volume dilution of the massive $\chi$.
Thus, the energy density in the produced $\chi$ is always bounded as 
\begin{equation}
  \rho_\chi < \frac{1}{24 \pi^3 e} \frac{k_\star^5}{H}
\end{equation}
On the short time scales relevant for particle production, $|\Delta N| \lsim 1$, it is still sensible to talk about energy conservation.  The
energy in produced $\chi$ particles must therefore be balanced by a dip in the kinetic energy of the inflaton.  (Intuitively this is to be expected
since the ``extra'' term in (\ref{mean2}) represents a force which tends to pull the inflaton back towards the point $\phi=\phi_0$.)
To get a sense of the magnitude of this effect, let us compare $\rho_\chi$ to the initial kinetic energy of the inflaton
\begin{equation}
  K_{\mathrm{in}} = \frac{\dot{\phi}^2_{\mathrm{in}}}{2} = \frac{k_\star^4}{2 g^2}
\end{equation}
Using (\ref{ratio}) it is easy to see that
\begin{equation}
  \frac{\rho_\chi}{K_{\mathrm{in}}} < 0.35 \, g^{5/2}  \ll 1
\end{equation}
Hence, the total energy density that goes into $\chi$ particles is small compared to the inflaton kinetic energy, and hence the velocity dip must
also be a small effect, roughly $\Delta \dot{\phi} / \dot{\phi} < 0.18 \, g^{5/2}$.  This simple estimate is consistent with a more quantitative
treatment \cite{ir}.

Note that, since inflation is driven by potential energy, we have $V \gg K_{\mathrm{in}} \gg \rho_\chi$.  Hence, even during particle 
production the expansion rate $H$ will still be dominated by the classical inflaton potential: $H \cong \sqrt{V/(3M_p^2)}$.  Therefore, a single burst of particle production
will \emph{not} terminate inflation.  Nor will this burst de-cohere the condensate since the occupation number of produced particles (\ref{n_k}) 
is always less than unity. (We again remind the reader that, unlike the case of broad band resonant preheating after inflation, we have 
only a single burst of particle production.)

Note that throughout this appendix we are assuming standard slow roll inflation.  In particular, our argument does not apply for trapped inflation \cite{trapped}
where the effect of backreaction on $H$ and $\phi(t)$ is much more significant.

\renewcommand{\theequation}{B-\arabic{equation}}
\setcounter{equation}{0}
\section*{APPENDIX B: Detailed Computation of $P(k)$}
\label{appB}

\renewcommand{\thesubsection}{B.\arabic{subsection}}
\setcounter{subsection}{0}

In this appendix we discuss some of the technical details associated with the computation of the renormalized power spectrum (\ref{full_pwr_result}).
First, notice that using (\ref{f_k_approx}) and (\ref{chi_soln}) we can write the quantity appearing in each renormalized Wick contraction as
\begin{eqnarray}
&& \chi_k(\tau) \chi_k^\star(\tau') - f_k(\tau) f_k^\star(\tau') \cong \nonumber \\
&& \frac{1}{k_\star^2}\frac{1}{\sqrt{a(\tau) a(\tau')}} \frac{1}{\sqrt{t(\tau)t(\tau')}} 
\left[ n_k \cos\left( \frac{k_\star^2 t^2(\tau)}{2} - \frac{k_\star^2 t^2(\tau')}{2} \right) 
+ \sqrt{n_k} \sqrt{1+n_k} \sin\left( \frac{k_\star^2 t^2(\tau)}{2} - \frac{k_\star^2 t^2(\tau')}{2} \right)  \right] \label{ren_wick_result}
\end{eqnarray}
where the occupation number $n_k$ is defined by (\ref{n_k}).
Plugging (\ref{ren_wick_result}) into (\ref{full_pwr_result}) we find
\begin{widetext}
\begin{eqnarray}
 && \hspace{-5mm} P_\phi(k) = \frac{g^2 k^3}{8 \pi^5} \left[ \,\,\,\,\,\,  \int d^3k' n_{k-k'} n_{k'}\times \int d\tau' d\tau'' \frac{G_{k}(\tau-\tau')}{a(\tau)} \frac{G_k(\tau-\tau'')}{a(\tau)}
                                                                       \cos^2\left[  \frac{k_\star^2 t^2(\tau')}{2} - \frac{k_\star^2 t^2(\tau'')}{2} \right]      \right. \nonumber \\
&+& \int d^3k' \sqrt{ n_{k-k'} n_{k'} }\sqrt{ 1 + n_{k-k'}  } \sqrt{ 1 + n_{k'}}\times
        \int d\tau' d\tau'' \frac{G_{k}(\tau-\tau')}{a(\tau)} \frac{G_k(\tau-\tau'')}{a(\tau)}
                                                                       \sin^2\left[  \frac{k_\star^2 t^2(\tau')}{2} + \frac{k_\star^2 t^2(\tau'')}{2} \right] \nonumber \\
&+& \int d^3k' \left( n_{k-k'}\sqrt{n_{k'}}\sqrt{1+n_{k'}} + n_{k'}\sqrt{n_{k-k'}}\sqrt{1+n_{k-k'}} \right) \nonumber \\
&& \,\,\,\,\,\,\,\,\,\,\,\,   \left.   \times \int d\tau' d\tau'' \frac{G_{k}(\tau-\tau')}{a(\tau)} \frac{G_k(\tau-\tau'')}{a(\tau)}
                             \cos\left[  \frac{k_\star^2 t^2(\tau')}{2} - \frac{k_\star^2 t^2(\tau'')}{2} \right]  
                             \sin\left[  \frac{k_\star^2 t^2(\tau')}{2} + \frac{k_\star^2 t^2(\tau'')}{2} \right]  \,\,\,\,\,\,\,\,\,\,\right] \label{pwr_step}
\end{eqnarray}
\end{widetext}
Notice that the time and phase space integrations in (\ref{pwr_step}) decouple.  This is the key simplification which makes an analytical
evaluation of this expression tractable.  Let us consider these integrations separately.

\subsection{Time Integrals}

All of the integrals over conformal time that appear in (\ref{pwr_step}) can be expressed in terms of two characteristic integrals which
we call $I_1$ and $I_2$.  Explicitly, these are defined as
\begin{eqnarray}
  I_1(k,\tau) &=& \frac{1}{a(\tau)}\int d\tau' G_k(\tau-\tau') e^{i k_\star^2 t^2(\tau')} \\
  I_2(k,\tau) &=& \frac{1}{a(\tau)}\int d\tau' G_k(\tau-\tau') 
\end{eqnarray}
The second characteristic integral, $I_2$, can be evaluated analytically.  However, the resulting expression is not particularly enlightening.  
Evaluation of $I_1$, on the other hand, requires numerical methods.  

Let us now show how the various integrals appearing in (\ref{pwr_step}) may be re-written in terms of $I_1$, $I_2$.  First, consider the first line
of (\ref{pwr_step}) where the following integral appears:
\begin{eqnarray}
  && \int d\tau' d\tau'' \frac{G_{k}(\tau-\tau')}{a(\tau)} \frac{G_k(\tau-\tau'')}{a(\tau)} \cos^2\left[  \frac{k_\star^2 t^2(\tau')}{2} - \frac{k_\star^2 t^2(\tau'')}{2} \right] \nonumber \\
  &=& \frac{|I_1(k,\tau)|^2}{2} + \frac{I_2(k,\tau)^2}{2} \label{tint1}
\end{eqnarray}
Next, consider the second line of (\ref{pwr_step}) where the following integral appears:
\begin{eqnarray}
  && \int d\tau' d\tau'' \frac{G_{k}(\tau-\tau')}{a(\tau)} \frac{G_k(\tau-\tau'')}{a(\tau)} \sin^2\left[  \frac{k_\star^2 t^2(\tau')}{2} + \frac{k_\star^2 t^2(\tau'')}{2} \right] \nonumber \\
  &=& -\frac{\mathrm{Re}\left[I_1(k,\tau)^2\right]}{2} + \frac{I_2(k,\tau)^2}{2} \label{tint2}
\end{eqnarray}
Finally, consider the fourth line of (\ref{pwr_step}) where the following integral appears:
\begin{eqnarray}
  && \int d\tau' d\tau'' \frac{G_{k}(\tau-\tau')}{a(\tau)} \frac{G_k(\tau-\tau'')}{a(\tau)} 
  \cos\left[  \frac{k_\star^2 t^2(\tau')}{2} - \frac{k_\star^2 t^2(\tau'')}{2} \right]\sin\left[  \frac{k_\star^2 t^2(\tau')}{2} + \frac{k_\star^2 t^2(\tau'')}{2} \right] \nonumber \\
  &=& \mathrm{Im}\left[I_1(k,\tau)I_2(k,\tau)\right] \label{tint3}
\end{eqnarray}
In the expressions (\ref{tint2}) and (\ref{tint3}) the notations $\mathrm{Re}$ and $\mathrm{Im}$ denote the real and imaginary parts, respectively.

\subsection{Phase Space Integrals}

As a warm-up to the subsequent calculation consider the following integral:
\begin{eqnarray}
   && \int d^3k' n_{k-k'}^a n_{k'}^b \nonumber \\
  && = \int d^3k' \exp\left[-a\pi |{\bf k} - {\bf k'}|^2 / k_\star^2 \right]\exp\left[-b\pi |{\bf k'}|^2 / k_\star^2 \right]  \nonumber \\
   && =  \frac{k_\star^3}{(a+b)^{3/2}} \exp\left[-\frac{ab}{a+b}\frac{\pi k^2}{k_\star^2}\right] \label{phase_identity}
\end{eqnarray}
This formula is valid when $a$, $b$ are positive real numbers. Notice that this expression is symmetric under interchange of $a$ and $b$.

The phase space integral in the first line of (\ref{pwr_step}) is computed by a trivial application of the identity (\ref{phase_identity}):
\begin{equation}
 \int d^3k' n_{k-k'}n_{k'} =  \frac{k_\star^3}{2\sqrt{2}} e^{-\pi k^2 / (2 k_\star^2)} \label{kint1}
\end{equation}
However, the remaining phase space integrals appearing in (\ref{pwr_step}) cannot be obtained exactly in closed form because they contain terms like 
$\sqrt{1+n_{k'}}$ where the gaussian factors appear under the square root.  In order to deal with such expressions, we note because $n_k \ll 1$
over most of the domain of integration, it is reasonable to replace $\sqrt{1+n_{k'}} \cong 1 + n_{k'}/2$.  
Let us now proceed in this manner.  The phase space integral on the second line of (\ref{pwr_step}) is:
\begin{eqnarray}
 && \int d^3k' \sqrt{n_{k-k'}n_{k'}}\sqrt{1+n_{k-k'}}\sqrt{1+n_{k'}} \nonumber \\
 && \cong \int d^3k'\left[  n_{k-k'}^{1/2}n_{k'}^{1/2}  + \frac{1}{2} n_{k-k'}^{3/2}n_{k'}^{1/2} + \frac{1}{2} n_{k-k'}^{1/2}n_{k'}^{3/2}  \right] \nonumber \\
 && = k_\star^3 \left[ \exp\left(-\frac{\pi k^2}{4k_\star^2}\right) + \frac{1}{2\sqrt{2}}\exp\left(-\frac{3 \pi k^2}{8k_\star^2}\right) \right] \label{kint2}
\end{eqnarray}
Finally, consider the phase space integral on the third line of (\ref{pwr_step}):
\begin{eqnarray}
 && \int d^3k' \left[ n_{k-k'} \sqrt{n_{k'}}\sqrt{1+n_{k'}}  +  n_{k'} \sqrt{n_{k-k'}}\sqrt{1+n_{k-k'}}  \right] \nonumber \\
 && \cong \int d^3k' \left[  n_{k-k'}n_{k'}^{1/2} + n_{k'}n_{k-k'}^{1/2}   + \frac{1}{2} n_{k-k'}n_{k'}^{3/2}  + \frac{1}{2} n_{k'}n_{k-k'}^{3/2}       \right] \nonumber \\
 && = k_\star^3 \left[\frac{4\sqrt{2}}{3\sqrt{3}} \exp\left(-\frac{\pi k^2}{3k_\star^2}\right) 
+ \frac{2\sqrt{2}}{5\sqrt{5}} \exp\left(-\frac{3 \pi k^2}{5k_\star^2}\right) \right] \label{kint3}
\end{eqnarray}
We have verified the formulae (\ref{kint2},\ref{kint3}) numerically.  In both cases that the numerical results agree with these 
semi-analytical expressions up to the percent level.

We can now, finally, insert the results (\ref{tint1},\ref{tint2},\ref{tint3}) and (\ref{kint1},\ref{kint2},\ref{kint3}) into the expression (\ref{pwr_step}).
Doing so, we arrive at our main analytical result, which is equation (\ref{full_pwr_result}).



\begin{thebibliography}{99}


\bibitem{fluct}
 V.~F.~Mukhanov and G.~V.~Chibisov,
``Quantum Fluctuation And 'Nonsingular' Universe,''
JETP Lett.\  {\bf 33}, 532 (1981)
[Pisma Zh.\ Eksp.\ Teor.\ Fiz.\  {\bf 33}, 549 (1981)]; 

S.~W.~Hawking,
``The Development Of Irregularities In A Single Bubble Inflationary Universe,''
Phys.\ Lett.\ B {\bf 115}, 295 (1982); 

A.~A.~Starobinsky,
``Dynamics Of Phase Transition In The New Inflationary Universe Scenario And Generation Of Perturbations,''
Phys.\ Lett.\ B {\bf 117}, 175 (1982); 
A.~H.~Guth and S.~Y.~Pi,
``Fluctuations In The New Inflationary Universe,''
Phys.\ Rev.\ Lett.\  {\bf 49}, 1110 (1982); J.~M.~Bardeen, P.~J.~Steinhardt and

M.~S.~Turner,
``Spontaneous Creation Of Almost Scale - Free Density Perturbations In An Inflationary Universe,''
Phys.\ Rev.\ D {\bf 28}, 679 (1983).

\bibitem{WMAP7}

  E.~Komatsu {\it et al.},
  ``Seven-Year Wilkinson Microwave Anisotropy Probe (WMAP) Observations:
  Cosmological Interpretation,''
  arXiv:1001.4538 [astro-ph.CO].

\bibitem{modulated1}

  L.~Kofman,
  ``Probing string theory with modulated cosmological fluctuations,''
  arXiv:astro-ph/0303614.

  F.~Bernardeau, L.~Kofman and J.~P.~Uzan,
  ``Modulated fluctuations from hybrid inflation,''
  Phys.\ Rev.\  D {\bf 70}, 083004 (2004)
  [arXiv:astro-ph/0403315].

\bibitem{modulated2}

   G.~Dvali, A.~Gruzinov and M.~Zaldarriaga,
  ``A new mechanism for generating density perturbations from inflation,''
  Phys.\ Rev.\  D {\bf 69}, 023505 (2004)
  [arXiv:astro-ph/0303591].

  G.~Dvali, A.~Gruzinov and M.~Zaldarriaga,
  ``Cosmological perturbations from inhomogeneous reheating, freezeout, and
  mass domination,''
  Phys.\ Rev.\  D {\bf 69}, 083505 (2004)
  [arXiv:astro-ph/0305548].

\bibitem{curvaton}

  D.~H.~Lyth and D.~Wands,
  ``Generating the curvature perturbation without an inflaton,''
  Phys.\ Lett.\  B {\bf 524}, 5 (2002)
  [arXiv:hep-ph/0110002].


\bibitem{ir}

  N.~Barnaby, Z.~Huang, L.~Kofman and D.~Pogosyan,
  ``Cosmological Fluctuations from Infra-Red Cascading During Inflation,''
  Phys.\ Rev.\  D {\bf 80}, 043501 (2009)
  [arXiv:0902.0615 [hep-th]].


\bibitem{ppcons}

 N.~Barnaby and Z.~Huang,
  ``Particle Production During Inflation: Observational Constraints and
  Signatures,''
  arXiv:0909.0751 [astro-ph.CO].

\bibitem{pptheory}

  N.~Barnaby,
  ``On Features and Nongaussianity from Inflationary Particle Production,''
  arXiv:1006.4615 [astro-ph.CO].

\bibitem{ppNG}

N.~Barnaby, ``Nongaussian Signatures of Inflationary Particle Production,'' \emph{work in progress}.

\bibitem{riotto}

  V.~Acquaviva, N.~Bartolo, S.~Matarrese and A.~Riotto,
  ``Second-order cosmological perturbations from inflation,''
  Nucl.\ Phys.\  B {\bf 667}, 119 (2003)
  [arXiv:astro-ph/0209156].

\bibitem{maldacena}

  J.~M.~Maldacena,
  ``Non-Gaussian features of primordial fluctuations in single field
  inflationary models,''
  JHEP {\bf 0305}, 013 (2003)
  [arXiv:astro-ph/0210603].

\bibitem{seerylidsey}

  D.~Seery and J.~E.~Lidsey,
  ``Primordial non-gaussianities in single field inflation,''
  JCAP {\bf 0506}, 003 (2005)
  [arXiv:astro-ph/0503692].

\bibitem{preheatNG}

  N.~Barnaby and J.~M.~Cline,
  ``Nongaussian and nonscale-invariant perturbations from tachyonic  preheating
  in hybrid inflation,''
  Phys.\ Rev.\  D {\bf 73}, 106012 (2006)
  [arXiv:astro-ph/0601481].

  N.~Barnaby and J.~M.~Cline,
  ``Nongaussianity from Tachyonic Preheating in Hybrid Inflation,''
  Phys.\ Rev.\  D {\bf 75}, 086004 (2007)
  [arXiv:astro-ph/0611750].

\bibitem{preheatNG2}

  J.~R.~Bond, A.~V.~Frolov, Z.~Huang and L.~Kofman,
  ``Non-Gaussian Spikes from Chaotic Billiards in Inflation Preheating,''
  Phys.\ Rev.\ Lett.\  {\bf 103}, 071301 (2009)
  [arXiv:0903.3407 [astro-ph.CO]].

\bibitem{preheatNG3}

  C.~T.~Byrnes,
  ``Constraints on generating the primordial curvature perturbation and
  non-Gaussianity from instant preheating,''
  JCAP {\bf 0901}, 011 (2009)
  [arXiv:0810.3913 [astro-ph]].

\bibitem{preheatNG4}

  D.~Mulryne, D.~Seery and D.~Wesley,
  ``Non-Gaussianity constrains hybrid inflation,''
  arXiv:0911.3550 [astro-ph.CO].

\bibitem{multibrid}

  M.~Sasaki,
  ``Multi-brid inflation and non-Gaussianity,''
  Prog.\ Theor.\ Phys.\  {\bf 120}, 159 (2008)
  [arXiv:0805.0974 [astro-ph]].

  A.~Naruko and M.~Sasaki,
  ``Large non-Gaussianity from multi-brid inflation,''
  Prog.\ Theor.\ Phys.\  {\bf 121}, 193 (2009)
  [arXiv:0807.0180 [astro-ph]].


\bibitem{small_sound}

  X.~Chen, M.~x.~Huang, S.~Kachru and G.~Shiu,
  ``Observational signatures and non-Gaussianities of general single field
  inflation,''
  JCAP {\bf 0701}, 002 (2007)
  [arXiv:hep-th/0605045].

\bibitem{shape}

  D.~Babich, P.~Creminelli and M.~Zaldarriaga,
  ``The shape of non-Gaussianities,''
  JCAP {\bf 0408}, 009 (2004)
  [arXiv:astro-ph/0405356].

\bibitem{shellard}

   J.~R.~Fergusson and E.~P.~S.~Shellard,
  ``The shape of primordial non-Gaussianity and the CMB bispectrum,''
  Phys.\ Rev.\  D {\bf 80}, 043510 (2009)
  [arXiv:0812.3413 [astro-ph]].

\bibitem{shandera}

  M.~LoVerde, A.~Miller, S.~Shandera and L.~Verde,
  ``Effects of Scale-Dependent Non-Gaussianity on Cosmological Structures,''
  JCAP {\bf 0804}, 014 (2008)
  [arXiv:0711.4126 [astro-ph]].

\bibitem{running1}

  J.~Kumar, L.~Leblond and A.~Rajaraman,
  ``Scale Dependent Local Non-Gaussianity from Loops,''
  arXiv:0909.2040 [astro-ph.CO].

\bibitem{running2}

  C.~T.~Byrnes, S.~Nurmi, G.~Tasinato and D.~Wands,
  ``Scale dependence of local $f_{NL}$,''
  arXiv:0911.2780 [astro-ph.CO].

\bibitem{obs_running}

  E.~Sefusatti, M.~Liguori, A.~P.~S.~Yadav, M.~G.~Jackson and E.~Pajer,
  ``Constraining Running Non-Gaussianity,''
  JCAP {\bf 0912}, 022 (2009)
  [arXiv:0906.0232 [astro-ph.CO]].

\bibitem{curvatonNG}

  N.~Bartolo, S.~Matarrese and A.~Riotto,
  ``On non-Gaussianity in the curvaton scenario,''
  Phys.\ Rev.\  D {\bf 69}, 043503 (2004)
  [arXiv:hep-ph/0309033].

  K.~Enqvist and S.~Nurmi,
  ``Non-gaussianity in curvaton models with nearly quadratic potential,''
  JCAP {\bf 0510}, 013 (2005)
  [arXiv:astro-ph/0508573].

  K.~A.~Malik and D.~H.~Lyth,
  ``A numerical study of non-gaussianity in the curvaton scenario,''
  JCAP {\bf 0609}, 008 (2006)
  [arXiv:astro-ph/0604387].

  M.~Sasaki, J.~Valiviita and D.~Wands,
  ``Non-gaussianity of the primordial perturbation in the curvaton model,''
  Phys.\ Rev.\  D {\bf 74}, 103003 (2006)
  [arXiv:astro-ph/0607627].

\bibitem{turnNG}

  G.~I.~Rigopoulos, E.~P.~S.~Shellard and B.~J.~W.~van Tent,
  ``Large non-Gaussianity in multiple-field inflation,''
  Phys.\ Rev.\  D {\bf 73}, 083522 (2006)
  [arXiv:astro-ph/0506704].

  F.~Vernizzi and D.~Wands,
  ``Non-Gaussianities in two-field inflation,''
  JCAP {\bf 0605}, 019 (2006)
  [arXiv:astro-ph/0603799].

 C.~T.~Byrnes, K.~Y.~Choi and L.~M.~H.~Hall,
  ``Conditions for large non-Gaussianity in two-field slow-roll inflation,''
  JCAP {\bf 0810}, 008 (2008)
  [arXiv:0807.1101 [astro-ph]].

  C.~T.~Byrnes and G.~Tasinato,
  ``Non-Gaussianity beyond slow roll in multi-field inflation,''
  JCAP {\bf 0908}, 016 (2009)
  [arXiv:0906.0767 [astro-ph.CO]].

  X.~Chen and Y.~Wang,
  ``Quasi-Single Field Inflation and Non-Gaussianities,''
  arXiv:0911.3380 [hep-th].

\bibitem{NGlocal_constraints}

  K.~M.~Smith, L.~Senatore and M.~Zaldarriaga,
  ``Optimal limits on $f_{NL}^{\mathrm{local}}$ from WMAP 5-year data,''
  JCAP {\bf 0909}, 006 (2009)
  [arXiv:0901.2572 [astro-ph]].

\bibitem{NGlocal_LSS}

  A.~Slosar, C.~Hirata, U.~Seljak, S.~Ho and N.~Padmanabhan,
  ``Constraints on local primordial non-Gaussianity from large scale
  structure,''
  JCAP {\bf 0808}, 031 (2008)
  [arXiv:0805.3580 [astro-ph]].

\bibitem{NLNG}

  N.~Barnaby, T.~Biswas and J.~M.~Cline,
  ``p-adic inflation,''
  JHEP {\bf 0704}, 056 (2007)
  [arXiv:hep-th/0612230].

  N.~Barnaby and J.~M.~Cline,
  ``Large Nongaussianity from Nonlocal Inflation,''
  JCAP {\bf 0707}, 017 (2007)
  [arXiv:0704.3426 [hep-th]].

  N.~Barnaby and J.~M.~Cline,
  ``Predictions for Nongaussianity from Nonlocal Inflation,''
  JCAP {\bf 0806}, 030 (2008)
  [arXiv:0802.3218 [hep-th]].

\bibitem{consistency}

  C.~Cheung, A.~L.~Fitzpatrick, J.~Kaplan and L.~Senatore,
  ``On the consistency relation of the 3-point function in single field
  inflation,''
  JCAP {\bf 0802}, 021 (2008)
  [arXiv:0709.0295 [hep-th]].

\bibitem{EFT}

  C.~Cheung, P.~Creminelli, A.~L.~Fitzpatrick, J.~Kaplan and L.~Senatore,
  ``The Effective Field Theory of Inflation,''
  JHEP {\bf 0803}, 014 (2008)
  [arXiv:0709.0293 [hep-th]].

\bibitem{NLmath}

  N.~Barnaby and N.~Kamran,
  ``Dynamics with Infinitely Many Derivatives: The Initial Value Problem,''
  JHEP {\bf 0802}, 008 (2008)
  [arXiv:0709.3968 [hep-th]].

  N.~Barnaby and N.~Kamran,
  ``Dynamics with Infinitely Many Derivatives: Variable Coefficient
  Equations,''
  JHEP {\bf 0812}, 022 (2008)
  [arXiv:0809.4513 [hep-th]].

  N.~Barnaby, D.~J.~Mulryne, N.~J.~Nunes and P.~Robinson,
  ``Dynamics and Stability of Light-Like Tachyon Condensation,''
  JHEP {\bf 0903}, 018 (2009)
  [arXiv:0811.0608 [hep-th]].



\bibitem{NLreview}

  N.~Barnaby,
  ``Nonlocal Inflation,''
  Can.\ J.\ Phys.\  {\bf 87}, 189 (2009)
  [arXiv:0811.0814 [hep-th]].

\bibitem{DBI}

  E.~Silverstein and D.~Tong,
  ``Scalar Speed Limits and Cosmology: Acceleration from D-cceleration,''
  Phys.\ Rev.\  D {\bf 70}, 103505 (2004)
  [arXiv:hep-th/0310221].

 M.~Alishahiha, E.~Silverstein and D.~Tong,
  ``DBI in the sky,''
  Phys.\ Rev.\  D {\bf 70}, 123505 (2004)
  [arXiv:hep-th/0404084].

\bibitem{geltron}

  A.~J.~Tolley and M.~Wyman,
  ``The Gelaton Scenario: Equilateral non-Gaussianity from multi-field
  dynamics,''
  arXiv:0910.1853 [hep-th].

\bibitem{trapped}

 D.~Green, B.~Horn, L.~Senatore and E.~Silverstein,
  ``Trapped Inflation,''
  Phys.\ Rev.\  D {\bf 80}, 063533 (2009)
  [arXiv:0902.1006 [hep-th]].

\bibitem{NGconstraints}

  L.~Senatore, K.~M.~Smith and M.~Zaldarriaga,
  ``Non-Gaussianities in Single Field Inflation and their Optimal Limits from
  the WMAP 5-year Data,''
  arXiv:0905.3746 [astro-ph.CO].

\bibitem{nonBD1}

 P.~D.~Meerburg, J.~P.~van der Schaar and P.~S.~Corasaniti,
  ``Signatures of Initial State Modifications on Bispectrum Statistics,''
  JCAP {\bf 0905}, 018 (2009)
  [arXiv:0901.4044 [hep-th]].

\bibitem{nonBD2}

  P.~D.~Meerburg, J.~P.~van der Schaar and M.~G.~Jackson,
  ``Bispectrum signatures of a modified vacuum in single field inflation with a
  small speed of sound,''
  arXiv:0910.4986 [hep-th].

\bibitem{nonBD}

  R.~Holman and A.~J.~Tolley,
  ``Enhanced Non-Gaussianity from Excited Initial States,''
  JCAP {\bf 0805}, 001 (2008)
  [arXiv:0710.1302 [hep-th]].

\bibitem{chen1}

  X.~Chen, R.~Easther and E.~A.~Lim,
  ``Large non-Gaussianities in single field inflation,''
  JCAP {\bf 0706}, 023 (2007)
  [arXiv:astro-ph/0611645].

\bibitem{chen2}

  X.~Chen, R.~Easther and E.~A.~Lim,
  ``Generation and Characterization of Large Non-Gaussianities in Single Field
  Inflation,''
  JCAP {\bf 0804}, 010 (2008)
  [arXiv:0801.3295 [astro-ph]].

\bibitem{chung}

  D.~J.~H.~Chung, E.~W.~Kolb, A.~Riotto and I.~I.~Tkachev,
  ``Probing Planckian physics: Resonant production of particles during
  inflation and features in the primordial power spectrum,''
  Phys.\ Rev.\  D {\bf 62}, 043508 (2000)
  [arXiv:hep-ph/9910437].

\bibitem{chung2}

 G.~J.~Mathews, D.~J.~H.~Chung, K.~Ichiki, T.~Kajino and M.~Orito,
  ``Constraints on resonant particle production during inflation from the
  matter and CMB power spectra,''
  Phys.\ Rev.\  D {\bf 70}, 083505 (2004)
  [arXiv:astro-ph/0406046].

\bibitem{elgaroy}

  O.~Elgaroy, S.~Hannestad and T.~Haugboelle,
  ``Observational constraints on particle production during inflation,''
  JCAP {\bf 0309}, 008 (2003)
  [arXiv:astro-ph/0306229].

\bibitem{sasaki}

  A.~E.~Romano and M.~Sasaki,
  ``Effects of particle production during inflation,''
  Phys.\ Rev.\  D {\bf 78}, 103522 (2008)
  [arXiv:0809.5142 [gr-qc]].

\bibitem{modulated_trapping}

  D.~Langlois and L.~Sorbo,
  ``Primordial perturbations and non-Gaussianities from modulated trapping,''
  arXiv:0906.1813 [astro-ph.CO].

\bibitem{brane_brem}

 P.~Brax and E.~Cluzel,
  ``Brane Bremsstrahlung in DBI Inflation,''
  arXiv:0912.0806 [hep-th].

\bibitem{KL}

  L.~A.~Kofman and A.~D.~Linde,
 ``Generation of Density Perturbations in the Inflationary Cosmology,''
  Nucl.\ Phys.\  B {\bf 282}, 555 (1987).

\bibitem{KP}

  L.~A.~Kofman and D.~Y.~Pogosian,
 ``NONFLAT PERTURBATIONS IN INFLATIONARY COSMOLOGY,''
  Phys.\ Lett.\  B {\bf 214}, 508 (1988).

\bibitem{BBS}

 D.~S.~Salopek, J.~R.~Bond and J.~M.~Bardeen,
  ``Designing Density Fluctuation Spectra in Inflation,''
  Phys.\ Rev.\  D {\bf 40}, 1753 (1989).

\bibitem{adams}

  J.~A.~Adams, B.~Cresswell and R.~Easther,
  ``Inflationary perturbations from a potential with a step,''
  Phys.\ Rev.\  D {\bf 64}, 123514 (2001)
  [arXiv:astro-ph/0102236].

\bibitem{step_model}

  P.~Hunt and S.~Sarkar,
  ``Multiple inflation and the WMAP 'glitches',''
  Phys.\ Rev.\  D {\bf 70}, 103518 (2004)
  [arXiv:astro-ph/0408138].

  P.~Hunt and S.~Sarkar,
  ``Multiple inflation and the WMAP 'glitches' II. Data analysis and
  cosmological parameter extraction,''
  Phys.\ Rev.\  D {\bf 76}, 123504 (2007)
  [arXiv:0706.2443 [astro-ph]].

\bibitem{gobump}

  M.~J.~Mortonson, C.~Dvorkin, H.~V.~Peiris and W.~Hu,
  ``Things that go bump in the CMB polarization: features from inflation versus
  reionization,''
  arXiv:0903.4920 [astro-ph.CO].

\bibitem{brane_annihilations}

  D.~Battefeld, T.~Battefeld, H.~Firouzjahi and N.~Khosravi,
  ``Brane Annihilations during Inflation,''
  arXiv:1004.1417 [hep-th].

\bibitem{sorbo}

  M.~M.~Anber and L.~Sorbo,
  ``Naturally inflating on steep potentials through electromagnetic
  dissipation,''
  arXiv:0908.4089 [hep-th].

\bibitem{berrera}

  A.~Berera and T.~W.~Kephart,
  ``Ubiquitous inflaton in string-inspired models,''
  Phys.\ Rev.\ Lett.\  {\bf 83}, 1084 (1999)
  [arXiv:hep-ph/9904410].

\bibitem{monodromy1}

 E.~Silverstein and A.~Westphal,
  ``Monodromy in the CMB: Gravity Waves and String Inflation,''
  Phys.\ Rev.\  D {\bf 78}, 106003 (2008)
  [arXiv:0803.3085 [hep-th]].

\bibitem{monodromy2}

 L.~McAllister, E.~Silverstein and A.~Westphal,
  ``Gravity Waves and Linear Inflation from Axion Monodromy,''
  arXiv:0808.0706 [hep-th].

\bibitem{monodromy3}

  R.~Flauger, L.~McAllister, E.~Pajer, A.~Westphal and G.~Xu,
  ``Oscillations in the CMB from Axion Monodromy Inflation,''
  arXiv:0907.2916 [hep-th].

\bibitem{beauty}

  L.~Kofman, A.~D.~Linde, X.~Liu, A.~Maloney, L.~McAllister and E.~Silverstein,
  ``Beauty is attractive: Moduli trapping at enhanced symmetry points,''
  JHEP {\bf 0405}, 030 (2004)
  [arXiv:hep-th/0403001].

\bibitem{terminal}

  D.~Battefeld and T.~Battefeld,
  ``A Terminal Velocity on the Landscape: Particle Production near Extra
  Species Loci in Higher Dimensions,''
  arXiv:1004.3551 [hep-th].

\bibitem{warm}
  A.~Berera,
  ``Warm Inflation,''
  Phys.\ Rev.\ Lett.\  {\bf 75}, 3218 (1995)
  [arXiv:astro-ph/9509049].

\bibitem{natural}

 K.~Freese, J.~A.~Frieman and A.~V.~Olinto,
  ``Natural inflation with pseudo - Nambu-Goldstone bosons,''
  Phys.\ Rev.\ Lett.\  {\bf 65}, 3233 (1990).

  F.~C.~Adams, J.~R.~Bond, K.~Freese, J.~A.~Frieman and A.~V.~Olinto,
  ``Natural Inflation: Particle Physics Models, Power Law Spectra For Large
  Scale Structure, And Constraints From Cobe,''
  Phys.\ Rev.\  D {\bf 47}, 426 (1993)
  [arXiv:hep-ph/9207245].

\bibitem{jim}

  C.~P.~Burgess, J.~M.~Cline, F.~Lemieux and R.~Holman,
  ``Decoupling, trans-Planckia and inflation,''
  arXiv:astro-ph/0306236.

  C.~P.~Burgess, J.~M.~Cline and R.~Holman,
  ``Effective field theories and inflation,''
  JCAP {\bf 0310}, 004 (2003)
  [arXiv:hep-th/0306079].

  C.~P.~Burgess, J.~M.~Cline, F.~Lemieux and R.~Holman,
  ``Are inflationary predictions sensitive to very high energy physics?,''
  JHEP {\bf 0302}, 048 (2003)
  [arXiv:hep-th/0210233].

\bibitem{nonequilibrium}

 J.~Berges and J.~Serreau,
  ``Progress in nonequilibrium quantum field theory,''
  arXiv:hep-ph/0302210.

  J.~Berges and J.~Serreau,
  ``Progress in nonequilibrium quantum field theory. II,''
  arXiv:hep-ph/0410330.

  J.~Berges and S.~Borsanyi,
  ``Progress in nonequilibrium quantum field theory. III,''
  Nucl.\ Phys.\  A {\bf 785}, 58 (2007)
  [arXiv:hep-ph/0610015].

\bibitem{KLS}

 L.~Kofman, A.~D.~Linde and A.~A.~Starobinsky,
  ``Reheating after inflation,''
  Phys.\ Rev.\ Lett.\  {\bf 73}, 3195 (1994)
  [arXiv:hep-th/9405187].

\bibitem{KLS97}

  L.~Kofman, A.~D.~Linde and A.~A.~Starobinsky,
 ``Towards the theory of reheating after inflation,''
  Phys.\ Rev.\  D {\bf 56}, 3258 (1997)
  [arXiv:hep-ph/9704452].

\bibitem{FK}
  
G.~N.~Felder and L.~Kofman,
  ``The development of equilibrium after preheating,''
  Phys.\ Rev.\  D {\bf 63}, 103503 (2001)
  [arXiv:hep-ph/0011160].

\bibitem{MT1}

  R.~Micha and I.~I.~Tkachev,
  ``Relativistic turbulence: A long way from preheating to equilibrium,''
  Phys.\ Rev.\ Lett.\  {\bf 90}, 121301 (2003)
  [arXiv:hep-ph/0210202].

\bibitem{MT2}

  R.~Micha and I.~I.~Tkachev,
  ``Turbulent thermalization,''
  Phys.\ Rev.\  D {\bf 70}, 043538 (2004)
  [arXiv:hep-ph/0403101].

\bibitem{B1}

  J.~Berges, A.~Rothkopf and J.~Schmidt,
  ``Non-thermal fixed points: effective weak-coupling for strongly correlated
  systems far from equilibrium,''
  Phys.\ Rev.\ Lett.\  {\bf 101}, 041603 (2008)
  [arXiv:0803.0131 [hep-ph]].

\bibitem{B2}

  J.~Berges and G.~Hoffmeister,
  ``Nonthermal fixed points and the functional renormalization group,''
  arXiv:0809.5208 [hep-th].


\bibitem{features}

  S.~Hannestad,
  ``Reconstructing the primordial power spectrum - a new algorithm,''
  JCAP {\bf 0404}, 002 (2004)
  [arXiv:astro-ph/0311491].


\bibitem{features2}

A.~Shafieloo and T.~Souradeep,
  ``Primordial power spectrum from WMAP,''
  Phys.\ Rev.\  D {\bf 70}, 043523 (2004)
  [arXiv:astro-ph/0312174].


\bibitem{morefeatures1}

  A.~Shafieloo, T.~Souradeep, P.~Manimaran, P.~K.~Panigrahi and R.~Rangarajan,
 ``Features in the Primordial Spectrum from WMAP: A Wavelet Analysis,''
  Phys.\ Rev.\  D {\bf 75}, 123502 (2007)
  [arXiv:astro-ph/0611352].


\bibitem{morefeatures2}

	L.~Covi, J.~Hamann, A.~Melchiorri, A.~Slosar and I.~Sorbera,
		``Inflation and WMAP three year data: Features have a future!,''
		Phys.\ Rev.\  D {\bf 74}, 083509 (2006)
		[arXiv:astro-ph/0606452].


\bibitem{morefeatures3}

  J.~Hamann, L.~Covi, A.~Melchiorri and A.~Slosar,
  ``New constraints on oscillations in the primordial spectrum of inflationary
  perturbations,''
  Phys.\ Rev.\  D {\bf 76}, 023503 (2007)
  [arXiv:astro-ph/0701380].


\bibitem{features3}

 P.~Mukherjee and Y.~Wang,
  ``Model-Independent Reconstruction of the Primordial Power Spectrum from WMAP
  Data,''
  Astrophys.\ J.\  {\bf 599}, 1 (2003)
  [arXiv:astro-ph/0303211].


\bibitem{yokoyama1}

  N.~Kogo, M.~Matsumiya, M.~Sasaki and J.~Yokoyama,
  ``Reconstructing the primordial spectrum from WMAP data by the cosmic
  inversion method,''
  Astrophys.\ J.\  {\bf 607}, 32 (2004)
  [arXiv:astro-ph/0309662].

\bibitem{yokoyama3}

  R.~Nagata and J.~Yokoyama,
  ``Reconstruction of the primordial fluctuation spectrum from the five-year
  WMAP data by the cosmic inversion method with band-power decorrelation
  analysis,''
  Phys.\ Rev.\  D {\bf 78}, 123002 (2008)
  [arXiv:0809.4537 [astro-ph]].

\bibitem{yokoyama2}

  R.~Nagata and J.~Yokoyama,
  ``Band-power reconstruction of the primordial fluctuation spectrum by the
  maximum likelihood reconstruction method,''
  Phys.\ Rev.\  D {\bf 79}, 043010 (2009)
  [arXiv:0812.4585 [astro-ph]].

\bibitem{hoi1}

  J.~M.~Cline and L.~Hoi,
  ``Inflationary potential reconstruction for a WMAP running power  spectrum,''
  JCAP {\bf 0606}, 007 (2006)
  [arXiv:astro-ph/0603403].

\bibitem{hoi}

  L.~Hoi, J.~M.~Cline and G.~P.~Holder,
 ``Testing the $k^3$ Component in the Primordial Perturbation Power Spectrum,''
  arXiv:0706.3887 [astro-ph].

\bibitem{contaldi}

 G.~Nicholson and C.~R.~Contaldi,
  ``Reconstruction of the Primordial Power Spectrum using Temperature and
  Polarisation Data from Multiple Experiments,''
  arXiv:0903.1106 [astro-ph.CO].

\bibitem{yokoyama4}

  K.~Ichiki, R.~Nagata and J.~Yokoyama,
  ``Cosmic Discordance: Detection of a modulation in the primordial fluctuation
  spectrum,''
  arXiv:0911.5108 [astro-ph.CO].

\bibitem{nofeatures}

 J.~Hamann, A.~Shafieloo and T.~Souradeep,
  ``Features in the primordial power spectrum? A frequentist analysis,''
  arXiv:0912.2728 [astro-ph.CO].

\bibitem{forecast}

  T.~Chantavat, C.~Gordon and J.~Silk,
  ``Large Scale Structure Forecast Constraints on Particle Production During
  Inflation,''
  arXiv:1009.5858 [astro-ph.CO].

\bibitem{HLattice}

Z.~Huang, ``HLattice: A New Code for Lattice Cosmology,'' work in progress.

\bibitem{modular_preheating}

  N.~Barnaby, J.~R.~Bond, Z.~Huang and L.~Kofman,
  ``Preheating After Modular Inflation,''
  arXiv:0909.0503 [hep-th].

\bibitem{defrost}

  A.~V.~Frolov,
  ``DEFROST: A New Code for Simulating Preheating after Inflation,''
  JCAP {\bf 0811}, 009 (2008)
  [arXiv:0809.4904 [hep-ph]].

\bibitem{latticeasy}

  G.~N.~Felder and I.~Tkachev,
  ``LATTICEEASY: A program for lattice simulations of scalar fields in an
  expanding universe,''
  [arXiv:hep-ph/0011159].

\bibitem{russian_text}

A.~A.~Grib, S.~G.~Mamayev and V.~M.~Mostepanenko, ``Vaccuum Effects in Strong Fields,'' Friedmann Laboratory Publishing, St.~Petersburg (1994).

\bibitem{riotto_rev}

   A.~Riotto,
  ``Inflation and the theory of cosmological perturbations,''
  arXiv:hep-ph/0210162.

\bibitem{false_vac}

  E.~J.~Copeland, A.~R.~Liddle, D.~H.~Lyth, E.~D.~Stewart and D.~Wands,
  ``False vacuum inflation with Einstein gravity,''
  Phys.\ Rev.\  D {\bf 49}, 6410 (1994)
  [arXiv:astro-ph/9401011].

\bibitem{susy_break}

  M.~Dine, L.~Randall and S.~D.~Thomas,
  ``Supersymmetry breaking in the early universe,''
  Phys.\ Rev.\ Lett.\  {\bf 75}, 398 (1995)
  [arXiv:hep-ph/9503303].

\bibitem{rapid_roll}

  L.~Kofman and S.~Mukohyama,
  ``Rapid roll Inflation with Conformal Coupling,''
  Phys.\ Rev.\  D {\bf 77}, 043519 (2008)
  [arXiv:0709.1952 [hep-th]].

\bibitem{seery}

  D.~Seery, K.~A.~Malik and D.~H.~Lyth,
  ``Non-gaussianity of inflationary field perturbations from the field
  equation,''
  JCAP {\bf 0803}, 014 (2008)
  [arXiv:0802.0588 [astro-ph]].


\bibitem{malik1}

  K.~A.~Malik,
  ``A not so short note on the Klein-Gordon equation at second order,''
  JCAP {\bf 0703}, 004 (2007)
  [arXiv:astro-ph/0610864].

\bibitem{malik2}

  K.~A.~Malik,
  ``Gauge-invariant perturbations at second order: Multiple scalar fields  on
  large scales,''
  JCAP {\bf 0511}, 005 (2005)
  [arXiv:astro-ph/0506532].

\bibitem{SMvariable}

  M.~Sasaki,
  ``Large Scale Quantum Fluctuations in the Inflationary Universe,''
  Prog.\ Theor.\ Phys.\  {\bf 76}, 1036 (1986).

\bibitem{vernizzi}
 
  D.~Langlois and F.~Vernizzi,
  ``Nonlinear perturbations of cosmological scalar fields,''
  JCAP {\bf 0702}, 017 (2007)
  [arXiv:astro-ph/0610064].

\bibitem{loop_corrections}

  D.~Seery,
  ``One-loop corrections to a scalar field during inflation,''
  JCAP {\bf 0711}, 025 (2007)
  [arXiv:0707.3377 [astro-ph]].

  D.~Seery,
  ``One-loop corrections to the curvature perturbation from inflation,''
  JCAP {\bf 0802}, 006 (2008)
  [arXiv:0707.3378 [astro-ph]].

\bibitem{M&W}
 
  K.~A.~Malik and D.~Wands,
  ``Evolution of second order cosmological perturbations,''
  Class.\ Quant.\ Grav.\  {\bf 21}, L65 (2004)
  [arXiv:astro-ph/0307055].

\bibitem{L&R}

  D.~H.~Lyth and Y.~Rodriguez,
  ``Non-gaussianity from the second-order cosmological perturbation,''
  Phys.\ Rev.\  D {\bf 71}, 123508 (2005)
  [arXiv:astro-ph/0502578].

\bibitem{Sconserved}

  D.~H.~Lyth, K.~A.~Malik and M.~Sasaki,
  ``A general proof of the conservation of the curvature perturbation,''
  JCAP {\bf 0505}, 004 (2005)
  [arXiv:astro-ph/0411220].


\bibitem{liddle}

  S.~M.~Leach and A.~R.~Liddle,
  ``Inflationary perturbations near horizon crossing,''
  Phys.\ Rev.\  D {\bf 63}, 043508 (2001)
  [arXiv:astro-ph/0010082].

\bibitem{sasaki2}

  S.~M.~Leach, M.~Sasaki, D.~Wands and A.~R.~Liddle,
  ``Enhancement of superhorizon scale inflationary curvature perturbations,''
  Phys.\ Rev.\  D {\bf 64}, 023512 (2001)
  [arXiv:astro-ph/0101406].


\bibitem{Hinshaw2008}
  G.~Hinshaw {\it et al.}  [WMAP Collaboration],
  ``Five-Year Wilkinson Microwave Anisotropy Probe (WMAP)
  Observations:Data Processing, Sky Maps, \& Basic Results,''
  Astrophys.\ J.\ Suppl.\  {\bf 180}, 225 (2009)
  [arXiv:0803.0732 [astro-ph]].

\bibitem{Lewis2002}
  A.~Lewis and S.~Bridle,
  ``Cosmological parameters from CMB and other data: a Monte-Carlo approach,''
  Phys.\ Rev.\  D {\bf 66}, 103511 (2002)
  [arXiv:astro-ph/0205436].

\bibitem{Komatsu2008}
  E.~Komatsu {\it et al.}  [WMAP Collaboration],
  ``Five-Year Wilkinson Microwave Anisotropy Probe (WMAP)
  Observations:Cosmological Interpretation,''
  Astrophys.\ J.\ Suppl.\  {\bf 180}, 330 (2009)
  [arXiv:0803.0547 [astro-ph]].

\bibitem{Jones2006}

W.~C.~Jones \emph{et al.}, 
``Observations of the temperature and polarization anisotropies with BOOMERANG 2003,''
New Astronomy Review, {\bf 50}, 945 (2006).


\bibitem{Piacentini2006}
  F.~Piacentini {\it et al.},
  ``A measurement of the polarization-temperature angular cross power spectrum
  of the Cosmic Microwave Background from the 2003 flight of BOOMERANG,''
  Astrophys.\ J.\  {\bf 647}, 833 (2006)
  [arXiv:astro-ph/0507507].

\bibitem{Montroy2006}
  T.~E.~Montroy {\it et al.},
  ``A Measurement of the CMB  Spectrum from the 2003 Flight of BOOMERANG,''
  Astrophys.\ J.\  {\bf 647}, 813 (2006)
  [arXiv:astro-ph/0507514].

\bibitem{Runyan2003}
  M.~C.~Runyan {\it et al.},
  ``The Arcminute Cosmology Bolometer Array Receiver,''
  Astrophys.\ J.\ Suppl.\  {\bf 149}, 265 (2003)
  [arXiv:astro-ph/0303515].

\bibitem{Goldstein2003}
  J.~H.~Goldstein {\it et al.},
  ``Estimates of Cosmological Parameters Using the CMB Angular Power Spectrum
  of ACBAR,''
  Astrophys.\ J.\  {\bf 599}, 773 (2003)
  [arXiv:astro-ph/0212517].

\bibitem{Kuo2006}
  C.~L.~Kuo {\it et al.},
  ``Improved Measurements of the CMB Power Spectrum with ACBAR,''
  Astrophys.\ J.\  {\bf 664}, 687 (2007)
  [arXiv:astro-ph/0611198].

\bibitem{Reichardt2008}
  C.~L.~Reichardt {\it et al.},
  ``High resolution CMB power spectrum from the complete ACBAR data set,''
  Astrophys.\ J.\  {\bf 694}, 1200 (2009)
  [arXiv:0801.1491 [astro-ph]].


\bibitem{Pearson2003}
  T.~J.~Pearson {\it et al.},
  ``The Anisotropy of the Microwave Background to l = 3500: Mosaic Observations
  with the Cosmic Background Imager,''
  Astrophys.\ J.\  {\bf 591}, 556 (2003)
  [arXiv:astro-ph/0205388].

\bibitem{Readhead2004a}
  A.~C.~S.~Readhead {\it et al.},
  ``Extended Mosaic Observations with the Cosmic Background Imager,''
  Astrophys.\ J.\  {\bf 609}, 498 (2004)
  [arXiv:astro-ph/0402359].

\bibitem{Readhead2004b}
  A.~C.~S.~Readhead {\it et al.},
 ``Polarization Observations with the Cosmic Background Imager,''
  arXiv:astro-ph/0409569.

\bibitem{Sievers2007}

J.~J.~Sievers \emph{et al.}, 
``Implications of the Cosmic Background Imager Polarization Data,''
Astrophys.\ J.\ {\bf 660}, 976 (2007).


\bibitem{Dickinson2004}
  C.~Dickinson {\it et al.},
  ``High sensitivity measurements of the CMB power spectrum with the extended
  Very Small Array,''
  Mon.\ Not.\ Roy.\ Astron.\ Soc.\  {\bf 353}, 732 (2004)
  [arXiv:astro-ph/0402498].

\bibitem{Halverson2002}
  N.~W.~Halverson {\it et al.},
  ``{DASI} First Results: A Measurement of the Cosmic Microwave Background
  Angular Power Spectrum,''
  Astrophys.\ J.\  {\bf 568}, 38 (2002)
  [arXiv:astro-ph/0104489].


\bibitem{Leitch2005}
  E.~M.~Leitch, J.~M.~Kovac, N.~W.~Halverson, J.~E.~Carlstrom, C.~Pryke and M.~W.~E.~Smith,
  ``DASI Three-Year Cosmic Microwave Background Polarization Results,''
  Astrophys.\ J.\  {\bf 624}, 10 (2005)
  [arXiv:astro-ph/0409357].

\bibitem{Hanany2000}
  S.~Hanany {\it et al.},
  ``MAXIMA-1: A Measurement of the Cosmic Microwave Background Anisotropy on
 angular scales of 10 arcminutes to 5 degrees,''
  Astrophys.\ J.\  {\bf 545}, L5 (2000)
  [arXiv:astro-ph/0005123].


\bibitem{Sunyaev1972}

R.~A.~Sunyaev and Y.~B.~Zeldovich, 
``The Observations of Relic Radiation as a Test of the Nature of X-Ray Radiation from the Clusters of Galaxies,''
Comments on Astrophysics and Space Physics {\bf 4}, 173 (1972).


\bibitem{Sunyaev1980}
R.~A.~Sunyaev and Y.~B.~Zeldovich, 
``Microwave background radiation as a probe of the contemporary structure and history of the universe,''
Ann.\ Rev.\ Astron.\ Astrophys.\ {\bf 18}, 537 (1980).


\bibitem{Bond2005}
  J.~R.~Bond {\it et al.},
  ``The Sunyaev-Zeldovich effect in CMB-calibrated theories applied to the
  Cosmic Background Imager anisotropy power at l > 2000,''
  Astrophys.\ J.\  {\bf 626}, 12 (2005)
  [arXiv:astro-ph/0205386].

\bibitem{Kowalski2008}
  M.~Kowalski {\it et al.}  [Supernova Cosmology Project Collaboration],
  ``Improved Cosmological Constraints from New, Old and Combined Supernova
  Datasets,''
  Astrophys.\ J.\  {\bf 686}, 749 (2008)
  [arXiv:0804.4142 [astro-ph]].

\bibitem{Cole2005}
  S.~Cole {\it et al.}  [The 2dFGRS Collaboration],
  ``The 2dF Galaxy Redshift Survey: Power-spectrum analysis of the final
  dataset and cosmological implications,''
  Mon.\ Not.\ Roy.\ Astron.\ Soc.\  {\bf 362}, 505 (2005)
  [arXiv:astro-ph/0501174].

\bibitem{Tegmark2006}
  M.~Tegmark {\it et al.}  [SDSS Collaboration],
  ``Cosmological Constraints from the SDSS Luminous Red Galaxies,''
  Phys.\ Rev.\  D {\bf 74}, 123507 (2006)
  [arXiv:astro-ph/0608632].

\bibitem{Massey2007}
  R.~Massey {\it et al.},
  ``COSMOS: 3D weak lensing and the growth of structure,''
  arXiv:astro-ph/0701480.

\bibitem{Hoekstra2006}
  H.~Hoekstra {\it et al.},
  ``First cosmic shear results from the Canada-France-Hawaii Telescope Wide
  Synoptic Legacy Survey,''
  Astrophys.\ J.\  {\bf 647}, 116 (2006)
  [arXiv:astro-ph/0511089].

\bibitem{Schimd2007}
  C.~Schimd {\it et al.},
  ``Tracking quintessence by cosmic shear: Constraints from VIRMOS-Descart  and
  CFHTLS and future prospects,''
  Astron.\ Astrophys.\  {\bf 463}, 405 (2007)
  [arXiv:astro-ph/0603158].

\bibitem{Hoekstra2002a}
  H.~Hoekstra, H.~K.~C.~Yee, M.~D.~Gladders, L.~F.~Barrientos, P.~B.~Hall and L.~Infante,
  ``A measurement of weak lensing by large scale structure in RCS fields,''
  arXiv:astro-ph/0202285.

\bibitem{Hoekstra2002b}
  H.~Hoekstra, H.~K.~C.~Yee and M.~D.~Gladders,
  ``Constraints on Omega m and sigma 8 from weak lensing in RCS fields,''
  Astrophys.\ J.\  {\bf 577}, 595 (2002)
  [arXiv:astro-ph/0204295].

\bibitem{Van-Waerbeke2005}
  L.~Van Waerbeke, Y.~Mellier and H.~Hoekstra,
  ``Dealing with systematics in cosmic shear studies: new results from the
  VIRMOS-Descart Survey,''
  Astron.\ Astrophys.\  {\bf 429}, 75 (2005)
  [arXiv:astro-ph/0406468].

\bibitem{Lesgourgues2007}
  J.~Lesgourgues, M.~Viel, M.~G.~Haehnelt and R.~Massey,
  ``A Combined analysis of Lyman-alpha forest, 3D Weak Lensing and WMAP year
  three data,''
  JCAP {\bf 0711}, 008 (2007)
  [arXiv:0705.0533 [astro-ph]].

\bibitem{Benjamin2007}
  J.~Benjamin {\it et al.},
  ``Cosmological Constraints From the 100 Square Degree Weak Lensing Survey,''
  arXiv:astro-ph/0703570.

\bibitem{Amigo2008}
  S.~De Lope Amigo, W.~Y.~Cheung, Z.~Huang and S.~P.~Ng,
  ``Cosmological Constraints on Decaying Dark Matter,''
  JCAP {\bf 0906}, 005 (2009)
  [arXiv:0812.4016 [hep-ph]].


\bibitem{Springel:2005nw}
  V.~Springel {\it et al.},
  ``Simulating the joint evolution of quasars, galaxies and their large-scale
  distribution,''
  Nature {\bf 435}, 629 (2005)
  [arXiv:astro-ph/0504097].

\bibitem{dalal}

  N.~Dalal, O.~Dore, D.~Huterer and A.~Shirokov,
  ``The imprints of primordial non-gaussianities on large-scale structure:
  scale dependent bias and abundance of virialized objects,''
  Phys.\ Rev.\  D {\bf 77}, 123514 (2008)
  [arXiv:0710.4560 [astro-ph]].

\bibitem{mcdonald}

  P.~McDonald,
  ``Primordial non-Gaussianity: large-scale structure signature in the
  perturbative bias model,''
  Phys.\ Rev.\  D {\bf 78}, 123519 (2008)
  [arXiv:0806.1061 [astro-ph]].

\bibitem{afshordi}

  N.~Afshordi and A.~J.~Tolley,
  ``Primordial non-gaussianity, statistics of collapsed objects, and the
  Integrated Sachs-Wolfe effect,''
  Phys.\ Rev.\  D {\bf 78}, 123507 (2008)
  [arXiv:0806.1046 [astro-ph]].


\bibitem{racetrack}

 J.~J.~Blanco-Pillado {\it et al.},
  ``Racetrack inflation,''
  JHEP {\bf 0411}, 063 (2004)
  [arXiv:hep-th/0406230].

\bibitem{CQ}

 J.~P.~Conlon and F.~Quevedo,
  ``Kaehler moduli inflation,''
  JHEP {\bf 0601}, 146 (2006)
  [arXiv:hep-th/0509012].

\bibitem{roulette}

  J.~R.~Bond, L.~Kofman, S.~Prokushkin and P.~M.~Vaudrevange,
  ``Roulette inflation with Kaehler moduli and their axions,''
  Phys.\ Rev.\  D {\bf 75}, 123511 (2007)
  [arXiv:hep-th/0612197].

\bibitem{fibre}

  M.~Cicoli, C.~P.~Burgess and F.~Quevedo,
  ``Fibre Inflation: Observable Gravity Waves from IIB String
  Compactifications,''
  JCAP {\bf 0903}, 013 (2009)
  [arXiv:0808.0691 [hep-th]].

\bibitem{closed_reheat}

 N.~Barnaby, J.~R.~Bond, Z.~Huang and L.~Kofman,
  ``Preheating After Modular Inflation,''
  JCAP {\bf 0912}, 021 (2009)
  [arXiv:0909.0503 [hep-th]].

  J.~Braden, L.~Kofman and N.~Barnaby,
  ``Reheating the Universe After Multi-Field Inflation,''
  JCAP {\bf 1007}, 016 (2010)
  [arXiv:1005.2196 [hep-th]].


\bibitem{brane}

  G.~R.~Dvali and S.~H.~H.~Tye,
  ``Brane inflation,''
  Phys.\ Lett.\  B {\bf 450}, 72 (1999)
  [arXiv:hep-ph/9812483].

 C.~P.~Burgess, M.~Majumdar, D.~Nolte, F.~Quevedo, G.~Rajesh and R.~J.~Zhang,
  ``The Inflationary Brane-Antibrane Universe,''
  JHEP {\bf 0107}, 047 (2001)
  [arXiv:hep-th/0105204].

\bibitem{KKLMMT}

 S.~Kachru, R.~Kallosh, A.~Linde, J.~M.~Maldacena, L.~P.~McAllister and S.~P.~Trivedi,
  ``Towards inflation in string theory,''
  JCAP {\bf 0310}, 013 (2003)
  [arXiv:hep-th/0308055].

\bibitem{D3D7}

  K.~Dasgupta, C.~Herdeiro, S.~Hirano and R.~Kallosh,
  ``D3/D7 inflationary model and M-theory,''
  Phys.\ Rev.\  D {\bf 65}, 126002 (2002)
  [arXiv:hep-th/0203019].

\bibitem{open_reheat}

  N.~Barnaby, A.~Berndsen, J.~M.~Cline and H.~Stoica,
  ``Overproduction of cosmic superstrings,''
  JHEP {\bf 0506}, 075 (2005)
  [arXiv:hep-th/0412095].

  N.~Barnaby, C.~P.~Burgess and J.~M.~Cline,
  ``Warped reheating in brane-antibrane inflation,''
  JCAP {\bf 0504}, 007 (2005)
  [arXiv:hep-th/0412040].

  N.~Barnaby,
  ``Caustic formation in tachyon effective field theories,''
  JHEP {\bf 0407}, 025 (2004)
  [arXiv:hep-th/0406120].

  N.~Barnaby and J.~M.~Cline,
  ``Creating the universe from brane-antibrane annihilation,''
  Phys.\ Rev.\  D {\bf 70}, 023506 (2004)
  [arXiv:hep-th/0403223].


\bibitem{anke}

  R.~H.~Brandenberger, A.~Knauf and L.~C.~Lorenz,
  ``Reheating in a Brane Monodromy Inflation Model,''
  JHEP {\bf 0810}, 110 (2008)
  [arXiv:0808.3936 [hep-th]].

\bibitem{jim_feat}

 J.~M.~Cline and L.~Hoi, private communication.

\bibitem{yokoyama_gw}

  R.~Saito, J.~Yokoyama and R.~Nagata,
  ``Single-field inflation, anomalous enhancement of superhorizon fluctuations,
  and non-Gaussianity in primordial black hole formation,''
  JCAP {\bf 0806}, 024 (2008)
  [arXiv:0804.3470 [astro-ph]].
 
 R.~Saito and J.~Yokoyama,
  ``Gravitational wave background as a probe of the primordial black hole
  abundance,''
  Phys.\ Rev.\ Lett.\  {\bf 102}, 161101 (2009)
  [arXiv:0812.4339 [astro-ph]].


\bibitem{shellard2}

  J.~R.~Fergusson, M.~Liguori and E.~P.~S.~Shellard,
  ``The CMB Bispectrum,''
  arXiv:1006.1642 [astro-ph.CO].

\bibitem{hidalgo}

  D.~Seery and J.~C.~Hidalgo,
  ``Non-Gaussian corrections to the probability distribution of the  curvature
  perturbation from inflation,''
  JCAP {\bf 0607}, 008 (2006)
  [arXiv:astro-ph/0604579].

\bibitem{lss1}

  A.~Slosar, C.~Hirata, U.~Seljak, S.~Ho and N.~Padmanabhan,
  ``Constraints on local primordial non-Gaussianity from large scale
  structure,''
  JCAP {\bf 0808}, 031 (2008)
  [arXiv:0805.3580 [astro-ph]].

\bibitem{lss2}

  R.~Jimenez and L.~Verde,
  ``Implications for Primordial Non-Gaussianity ($f_{NL}$) from weak lensing masses
  of high-z galaxy clusters,''
  Phys.\ Rev.\  D {\bf 80}, 127302 (2009)
  [arXiv:0909.0403 [astro-ph.CO]].

\bibitem{lss3}

  L.~Verde and S.~Matarrese,
  ``Detectability of the effect of Inflationary non-Gaussianity on halo bias,''
  Astrophys.\ J.\  {\bf 706}, L91 (2009)
  [arXiv:0909.3224 [astro-ph.CO]].

\bibitem{structure}

 S.~Shandera,
  ``The structure of correlation functions in single field inflaiton,''
  Phys.\ Rev.\  D {\bf 79}, 123518 (2009)
  [arXiv:0812.0818 [astro-ph]].

\bibitem{perturbative}

  L.~Leblond and S.~Shandera,
  ``Simple Bounds from the Perturbative Regime of Inflation,''
  JCAP {\bf 0808}, 007 (2008)
  [arXiv:0802.2290 [hep-th]].

\bibitem{preheat_pdf1}

  G.~N.~Felder and L.~Kofman,
  ``The development of equilibrium after preheating,''
  Phys.\ Rev.\  D {\bf 63}, 103503 (2001)
  [arXiv:hep-ph/0011160].

\bibitem{preheat_pdf2}

 G.~N.~Felder and O.~Navros,
  ``Inflaton fragmentation after lambda phi**4 inflation,''
  JCAP {\bf 0702}, 014 (2007)
  [arXiv:hep-ph/0701128].

\bibitem{cluster}

  M.~J.~Jee {\it et al.},
  ``Hubble Space Telescope Weak-lensing Study of the Galaxy Cluster XMMU
  J2235.3-2557 at z=1.4: A Surprisingly Massive Galaxy Cluster when the
  Universe is One-third of its Current Age,''
  Astrophys.\ J.\  {\bf 704}, 672 (2009)
  [arXiv:0908.3897 [astro-ph.CO]].

\bibitem{factorizable}

  K.~M.~Smith and M.~Zaldarriaga,
  ``Algorithms for bispectra: forecasting, optimal analysis, and simulation,''
  arXiv:astro-ph/0612571.

\bibitem{void}

  M.~Kamionkowski, L.~Verde and R.~Jimenez,
  ``The Void Abundance with Non-Gaussian Primordial Perturbations,''
  JCAP {\bf 0901}, 010 (2009)
  [arXiv:0809.0506 [astro-ph]].

\bibitem{lss_rev}

  V.~Desjacques and U.~Seljak,
  ``Primordial non-Gaussianity from the large scale structure,''
  Class.\ Quant.\ Grav.\  {\bf 27}, 124011 (2010)
  [arXiv:1003.5020 [astro-ph.CO]].

\end{thebibliography}
\end{document}